\documentclass[10pt]{iopart}
\usepackage[latin1]{inputenc}
\usepackage{pslatex}
\usepackage{float}
\usepackage{graphicx}
\usepackage{amssymb}
\IfFileExists{url.sty}{\usepackage{url}}
                      {\newcommand{\url}{\texttt}}
\usepackage[]{natbib}
\bibpunct[; ]{(}{)}{,}{a}{}{;}
\makeatletter

\providecommand{\boldsymbol}[1]{\rm{\boldmath $#1$}}
\providecommand{\tabularnewline}{\\}

\usepackage{bm}
\eqnobysec
\newcommand{\apj}{\textit{Astrophys.\ J.\ }}
\newcommand{\aap}{\textit{Astron.\ Astrophys.\ }}
\newcommand{\nat}{\textit{Nature}\ }
\newcommand{\mnras}{\textit{Mon.\ Not.\ Roy.\ Ast.\ Soc.}\ }
\newcommand{\aj}{\textit{Astron.\ J.\ }}
\newcommand{\araa}{\textit{Ann.\ Rev.\ Astron.\ Astrophys.\ }}
\newcommand{\apjs}{\textit{Astrophys. J. Supp.\ }}
\newcommand{\pasp}{\textit{Publ. Astron. Soc. Pac.\ }}
\newcommand{\kms}{km~s$^{-1}$}
\newcommand{\Ms}{M$_\odot$}

\makeatother
\begin{document}
\review{Constraints on Galaxy Structure and Evolution from the Light
of Nearby Systems}

\author{Gerald Cecil and James A Rose}
\address{
Dept.\ of Physics \& Astronomy, University of North Carolina at Chapel
Hill, NC 27599-3255 USA}
\eads{\mailto{cecil@physics.unc.edu}, \mailto{jim@physics.unc.edu}}

\begin{abstract}
We review knowledge of galaxy structures obtained by their emitted light and
in the local universe where they can be studied in great detail.
We discuss the shapes of, and stellar motions within, galaxies, compositional
clues derived from their spectra, and what luminous matter implies
about their dark matter content. Implications on the
current theory of hierarchical galaxy formation are explored.
\end{abstract}

\section{Introduction}

We have entered the age of `precision cosmology'.  Careful 
measurements of how objects of known radiant power have dimmed with distance,
the angular size of known distances at early epochs, and
the power spectrum of the matter distribution at the present epoch, have led
to a remarkably accurate cosmological model.  That is, the expansion history of the 
universe, which is governed by the density of visible and dark matter \citep{Freeman06} and by
dark energy, has been well constrained.  The tools of observational
cosmology are amazingly diverse.  Type Ia supernovae (SNe) --- the
deflagration/detonation of a white dwarf star after it accretes sufficient mass to exceed the
pressure support of degenerate electrons ---
provides a high-luminosity `standard candle' for
mapping the dimming of sources over cosmological distances, constraining
directly how the Hubble expansion parameter changes with cosmic
epoch.  Detailed maps of the power spectrum in angular scales of
temperature variations of the cosmic microwave background radiation reveal
various harmonics of density fluctuations at the time when the universe had cooled
sufficiently for electrons and protons to `recombine' into hydrogen atoms.
The scale of these fluctuations depends on the matter energy density
in the early universe, and the
angular scale observed at the present epoch depends on the expansion 
history since this recombination.
Hence, the temperature maps constrain the cosmological
world model.  Finally, the power spectrum of the 3D distribution of matter on
large scales, as traced by luminous galaxies, also constrains
both the spectrum of the initial density
fluctuations and the combination of matter and dark energy that governs
their evolution.

The recent success of observational cosmology poses both new challenges
and opportunities. Perhaps most exciting is to quantify structure growth.
In the favoured dark energy plus cold, dark matter ($\Lambda$-CDM)
model, non-baryonic dark matter (DM) is non-relativistic long before the
epoch where neutral hydrogen atoms (HI in astronomical jargon) first
form. From the minimum (Jeans) mass that can collapse by self gravity
at that time ($\gtrsim10^{6}$ solar mass, \Ms), cold dark matter
(CDM) clumps in a hierarchy of increasing mass.

This scenario has been simulated extensively into the non-linear regime, but
is largely untested by observations. Temperature anisotropies
in the cosmic microwave background radiation
probe only the \textit{start} of the linear regime of structure
formation as gravity drags on cosmic expansion. While anisotropies
support the idea that the seeds of non-linear structure have
Gaussian amplitude spectrum with random phases (i.e.,
are spatially uncorrelated), unclear is how coherent structure evolves
into the non-linear regime, and if the radiant power (luminosity
in astrophysical jargon) and mass scales and morphologies of the observed
structures are consistent with the CDM hierarchy.

To document how structure evolves from when galaxies first appeared
at redshift $z\sim5$ until the present epoch it is essential to understand
the evolution of visible baryons as density-biased tracers of the
underlying DM. Even studies that map DM more directly through gravitational
lensing must know the structure of the lens by its visible material.
Thus, all studies of structure evolution depend on a better understanding
of galaxies. 

Determination of our cosmological world model offers an exciting opportunity
for students of galaxy evolution. Prior interpretation of, for example,
number counts of galaxies as a function of redshift, depended on both
galaxy evolution and an uncertain world model. An accurate
model now decouples geometry from galaxy evolution observed
over a significant redshift, hence lookback time.

Advanced ground- and space-based telescope and detector technologies
from x-ray to millimetre wavelengths allow study of galaxy
evolution \textit{in situ} at high redshift. To mention only a few examples over the
past decade: the development of `optical' telescopes of up to 10.5-meter
aperture that are coupled to multi-slit/fibre spectrographs, advanced detectors
at infrared (IR) wavelengths, the advent of imaging at sub-millimetre
wavelengths, high-spatial resolution UV to near-IR images from
the Hubble Space Telescope (HST), detailed studies by
x-ray observatories of hot plasma in the Milky Way galaxy (MWg) interstellar
medium (ISM) and in the intracluster/group medium, and mid-IR views
from the Spitzer Space Telescope.

While there has thus been a stampede toward objects at high redshift,
to understand these challenging, photon-starved measurements require
multi-waveband constraints on comparable objects
nearby. We must assess how baryons have concentrated into galaxies
through, among other processes, star formation (SF) and evolution,
nucleosynthesis of heavy elements, and dynamical processing of stars
and dissipational ISM. Because galaxies are proximate enough
to interact strongly by tides, we must determine
how these processes are affected by environment (rich galaxy
cluster versus small group).

For those oriented toward cosmology, understanding galaxy evolution
is a necessary prerequisite to using galaxies in bulk to trace the
evolution of the dominant DM and to constrain the mysterious
dark energy. Others will research the diverse and fascinating physics
of galaxy evolution. Our review is aimed at those who are
preparing to follow either path. We assess what properties of galaxies
have been established recently in the \textit{nearby} universe
$(z\lesssim0.1,$ an insignificant lookback time compared to
the age of the universe). In number, 80\% are gas-rich spiral galaxies
(Sgs) and smaller dwarf irregular galaxies (dIrr's), the rest are
gas-poor elliptical (Eg) and S0 (called collectively, early-type)
galaxies and dwarf spheroidal galaxies (dSph's). We address questions
about galaxy structure and evolution using images and spectra of much
higher intrinsic spatial resolution and signal-to-noise ratio (SNR)
than can be obtained on high redshift counterparts. Our theme throughout
is the perhaps surprising degree of uncertainty that prevails over
most topics in galaxy structure and evolution.

Section \ref{sec:What-Can-We} introduces modern observations
of stars and DM. 
At the top level, we seek the main driver(s) of
galaxy evolution. Does a single process dominate? Our main observables
are the details within the bimodal separation of galaxies into luminous
`red and dead' early-type spheroids and blue star-forming discs.
How are these connected? Are Egs a single family differing only in
mass? Section \ref{sec:Spheroidal-Galaxies} examines their 3D structure
for universal characteristics. A consistent interpretation of their
surface photometry and kinematics requires the introduction of many dynamical
assumptions. Section \ref{sub:Dark-matter-content} discusses their
uncertain DM content. Section \ref{sec:Bulges-of-Spiral} examines
the apparently similar spheroidal bulges of Sgs, including those
of our MWg and our neighbour Sg Andromeda (M31). Are they primordial,
were they built from mergers,
or did they arise from secular (i.e., long-term) processes driven by instabilities?

In section \ref{sec:Thick-&-Thin} we study Sg discs of various thickness.
Are discs stable and axisymmetric? DM halos affect stability, so
we address secular effects in section \ref{sec:Matter-Transfer-Within}
and consider radial matter transfer. Are discs distinguished
only by their level of SF? We consider the multi-component ISM of
gas, dusty soot, magnetic field, and cosmic rays. At the bottom of
the potential well is often a supermassive black hole, best
studied by its induced stellar motions.

In section \ref{sec:Interactions-with-Other} we consider interactions
of stars with the hot gas and with other galaxies in rich clusters.
Do galaxies switch between spiral and ellipsoidal forms? How has environment
modified SF? We preview some new instrument capabilities in section 8, and
summarize in the last section. We assume that the
reader grasps the basics of the current cosmological world model 
\citep[for example the textbook by][]{Ryden02},
but we minimize astronomical jargon and suggest review articles as
broader topics emerge.
\ref{sec:What-(Astro-)Physical-Processes}
reviews key astrophysical processes to understand before considering
galaxy evolution; Table A.1 lists acronyms used throughout.

\section{\label{sec:What-Can-We}What do we learn from starlight?}

We first summarize observables from galaxy starlight.
Limiting uncertainties are \textit{systematic} (external), not statistical
(internal), in most astronomical datasets of relevance to galaxy
evolution because most dynamical parameters are not observed directly.
Instead, we shall see that they arise from correlations between, and
calibrations from, often heterogeneous photometric and kinematical
datasets whose inter-transformations introduce systematics. Uncertain
zero points and incompleteness biases that weigh points in a correlation
can be further confounded by the inadvertent inclusion into the study
sample of unrelated less luminous foreground or more luminous background
galaxies; the large range of galaxy luminosities and the complications
of patchy obscuration by dust mean that reliable distances are
a prerequisite to minimising such bias. Differential measurements
can minimize systematics. Computer simulations to quantify errors
are necessary in observational astrophysics, complicated by both the large and 
very small sizes of astronomical datasets.

\subsection{The Milky Way galaxy}

Observing stars over a wide range of masses and evolutionary states, 
one seeks to correlate their dynamical, SF, and chemical enrichment
histories with 3D space velocities, ages, and chemical compositions.
Stars live mostly in the relatively low luminosity phase of core hydrogen-to-helium
fusion. They spend the last $\sim$20\% of their lives in a series
of higher luminosity phases. High-mass stars have relatively brief lives
compared to low mass ones.

Only in the MWg can we obtain such extensive data on individual, especially
lower mass unevolved, stars. But only a minuscule fraction of its
$\sim10^{11}$ stars can be so detailed from our vantage point $8.0\pm0.4$
kpc \citep{Eisenhauer03} off-centre. 
Indeed, we can approximate
this program only within the $\sim50$ pc radius Solar Neighbourhood
where we can obtain accurate stellar distances by trigonometric
parallax from opposing points on the Earth's orbit. Otherwise, use
of secondary measures of distance built upon parallaxes amplifies
uncertainties. Over the last decade, parallaxes and photometry
from the Hipparcos satellite above atmospheric blur
has been combined with ground-based spectra and photometry
to calibrate the parameters of many nearby stars accurately \citep{Lebreton01}; future
astrometric space missions aim to expand this sphere a hundred-fold
in radius. Note that many stars with excellent parallax measurements
have relatively poor spectroscopy and even photometry; these deficiencies
are being rectified by ongoing surveys. 
Finally, our view of the MWg
in visible and UV light --- where the most powerful diagnostics
lie --- is obscured by interstellar dust. Only outside this waveband
does our view clarify along many sightlines.

Now consider velocity vectors. With parallax in hand and a radial
(line of sight, l.o.s.) velocity measured from the Doppler shift of
narrow absorption lines in the stellar photosphere, we must obtain
the star's tangential motion across the sky by measuring
at two widely spaced times
its sky position relative to more distant objects that define a fundamental
reference frame. As detailed
in \ref{sub:Gravitational-stellar-dynamics}, the orbit
inferred from the gravitational potential is the essential dynamical
input in the study of galaxy evolution. Key too are age
and chemical composition, which are extracted from spectral analysis
as summarized in \ref{sub:Stellar-Evolution}.

\subsection{Local Group galaxies}

The MWg helps to bind the Local Group of galaxies.
Across $\sim2.5$ million parsecs (Mpc), two large Sgs --- the MWg
and larger M31 --- dominate $\sim25$ satellites; a third smaller Sg
M33 ($\sim10\%$ of the mass of the MWg) is associated with M31. Together, the three Sgs emit 90\% of
the light of the Local Group. With M33, the other two large satellites of M31 are
the dwarf Eg M32 and dwarf S0 NGC 205. The largest satellites of the MWg are the barred
Large and irregular Small Magellanic Clouds (LMC and SMC). Local Group members
are still being discovered, especially near the dust obscured plane of the MWg.

Local Group members are close enough for some resolution into individual stars,
especially with HST. 
Luminous stars can be studied individually,
tending to be either young and massive or in
later evolutionary stages. 
The motions of the Magellanic Clouds across the sky
have been measured from stars \citep{Kallivayalil06a,Kallivayalil06b}.
The near side of the LMC is $10\%$
closer than the far. The resulting $20\%$ brightness difference of
stars of the same stellar type gives a rudimentary sense of the LMC's
3D structure. All other Local Group members are either too small or distant
to yield reliable information along the l.o.s. Hence the MWg, and
to a far lesser extent the LMC, are the only galaxies for which 3D
information is obtainable directly. 

High-resolution spectra, from which chemical abundances can be derived,
have been obtained for individual stars in some MWg satellites.
Abundances constrain the chemical history of a galaxy.
For example, in section \ref{sub:The-lowest-luminosity} we compare
abundances in the Sagittarius dIrr
to those in the thick disc and halo of the MWg, to constrain the assembly
of the MWg.

\subsection{Other {}`nearby' galaxies}

Individual stars outside the Local Group are virtually unmeasurable.
Instead, the 3D structure of a galaxy is extracted from
starlight projected onto the sky and integrated spatially across each
image pixel. 
One must
rely on the integrated l.o.s.\ velocity field (velocity function) that is broadened
by dispersed stellar velocities.
In addition to the pixel's spectral content and mean light flux,
variations in the surface brightness (SB) that arise
from Poisson fluctuations in the number of bright, evolved stars within
each pixel can be used to constrain the galaxy distance \citep{Tonry88}.

Most galaxies have only a cosmic redshift-implied distance; only 
$\sim10,000$ have distances determined independently with accuracy 
sufficient to separate a galaxy's l.o.s. \textit{peculiar velocity} 
due to gravity from the bulk cosmic expansion redshift.
In short, we know at best only three of the six spatio-kinematical
descriptors of galaxy structure. In situations of high
symmetry, we can invert to constrain the spatial variation throughout
some of the volume. As discussed in section \ref{sub:Internal-motions},
asymmetries in absorption line profiles can constrain the structure of
less symmetrical systems.

\subsection{\label{sub:State-of-the}State of the art}

At its average SB $(V=22.5)$, the moonless night sky over angular
area 1 arc-second$^{2}$ emits the radiant flux of a giant star at
0.2 Mpc or a Sun-like dwarf star at 32 kpc. Adaptive optics systems
on telescopes reduce the area of the effective
detection box, hence sky flux, at least another ten-fold, increasing
the detection distance of a single star at constant
SNR more than three-fold. From the ground, the SNR increases
at best as the square-root of the increased exposure time. HST
goes considerably deeper than ground-based facilities somewhat because of a
$\sim17\%$ on average reduced V-band background (which is still variable because of
unpredictable, cloud dependent Earthshine) but mostly because its diffraction
limited image tightens the detection box to $\sim0.02(\lambda/500\,\rm{nm})^{2}$
arc-second$^{2}$.

\begin{figure}
\begin{centering}\includegraphics[scale=0.37]{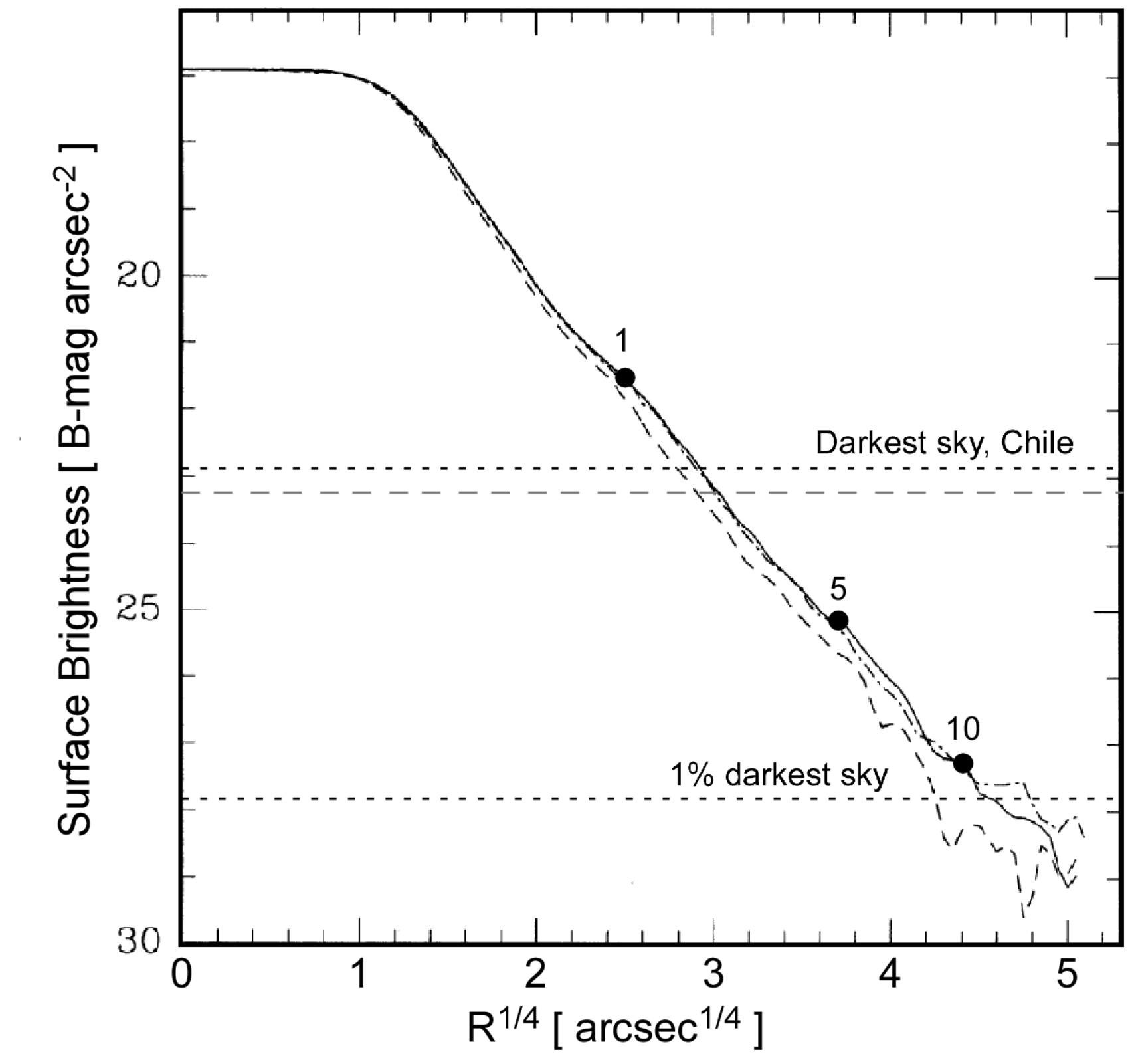}\par\end{centering}
\caption{\label{fig:SB-of-NGC}SB of the nearby Eg NGC 3379 (photometry from
\citealt{Capaccioli90}) with \fullcircle at radii $xR_{e}$
for $x=(1,5,10)$, the darkest night sky at a Chilean site (top \dotted)
and its 1\% level (lower \dotted), and the value at near-Earth orbit
(\broken). Because SB depends empirically on radius as the $1/4$ power,
this variable is plotted on the ordinate to highlight deviations; profiles
along both the apparent major and the minor axis are shown.
The galaxy profile within $\sim1.5$
arcsec radius is flattened by the blurring effects of the Earth's
atmosphere.}
\end{figure}

The radial light profiles of spheroidal galaxies
projected onto the plane of the sky
are characterized by an `effective radius' $R_{\rm{e}}$ within
which half of the galaxy luminosity lies. Figure \ref{fig:SB-of-NGC}
compares the SB of the moonless night-sky to those of spheroidal galaxies.
With diligent sky-subtraction, accurate photometry can be obtained
beyond $10R_{\rm{e}}$. Slit spectra can be obtained to $5R_{\rm{e}}$,
and micro-lens- or fibre-coupled spectrographs (field segmentation
techniques whose optics tend to scatter more light than open slits) can be
effective out to $1-2R_{\rm{e}}$. Until recently, selection effects
excluded galaxies of considerable angular extent but of constantly
low SB; section \ref{sub:Low-surface-brightness} discusses this population.

A complication, becoming historical, to transforming insights between
the local and distant universe is that we usually observe different
rest wavelengths at high redshift than we do nearby. Only recently
have UV-optimized spacecraft (for example, the Galaxy Evolution Explorer {[}\textit{GALEX}]
and Far Ultraviolet Spectral Explorer {[}\textit{FUSE}]) and near-IR ground-
and space-based spectrometers provided images and spectra with sufficient
wavelength coverage to transform accurately. Another complication
is the inadvertent inclusion of luminous background objects, often
star burst and/or active galactic nuclei (AGN), whose bright regions
are uncharacteristic of the underlying stellar mass. These ubiquitous contaminants
increase with survey volume and complicate the connection
to nearby objects.

\begin{figure}
\begin{centering}\includegraphics[scale=0.11]{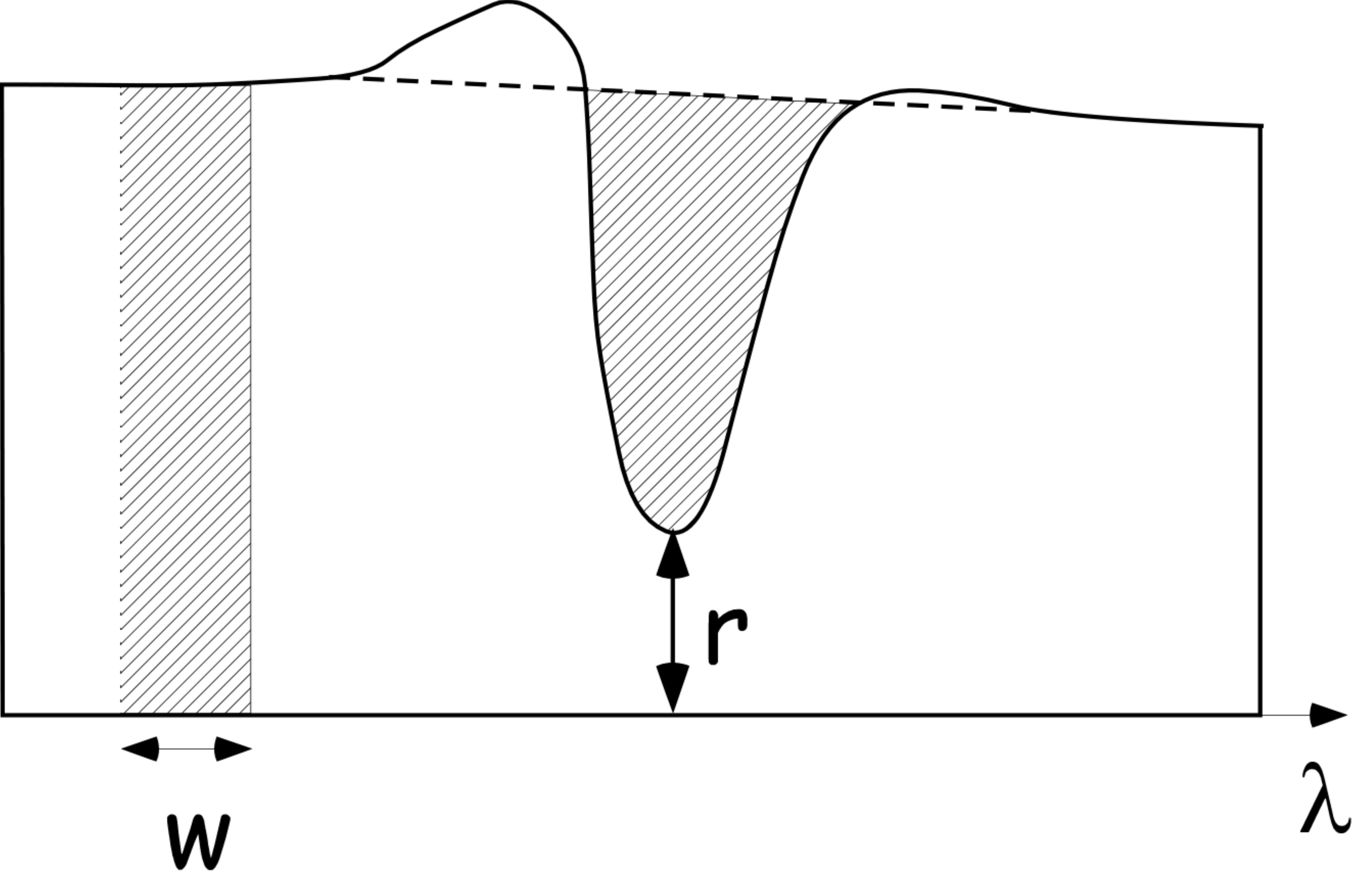}\par\end{centering}
\caption{\label{fig:equiv}Defining equivalent width $w$ and residual intensity $r$
of an absorption line spectrum. The vertical axis is radiant flux,
zero at bottom. A ratio of residual intensities defines
an index. To define $w$, the area notched from the slanted stellar
continuum is repeated at left (or right). As shown, equivalent width depends
even in data of infinite SNR on how one draws `the' continuum
across a line and where the two sidebands are defined; this is especially
true at blue wavelengths where lines crowd. For an index based on
$r$, the continuum level is irrelevant but scattered light must be controlled.
$\sim45\%$ of Egs show
weak emission lines, which must be excised for a pure absorption
spectrum.}
\end{figure}

Signatures of chemical abundances concentrate at wavelengths
shorter than $\lambda$550 nm \citep[see Figure 2 of][for example]{Bland-Hawthorn04},
the U, B, and V filter bands. One needs $\sim\lambda0.01$ nm resolution
$(R=\lambda/\Delta\lambda\sim40,000$) to measure accurately many elements mostly by minimising
systematic errors from otherwise blended lines. Stellar chemical abundances
are determined by high-dispersion spectroscopy of many, sometimes
weak absorption lines from key elements at various stages of ionization.
This program is feasible on even halo MWg stars with the largest
telescopes and most efficient spectrographs; an error of $\lambda10^{-4}$
nm in the equivalent width (defined in Figure \ref{fig:equiv})
of an absorption line yields an abundance error of 0.02 dex. F-K stars
emit most of their energy in the U- to L- ($\lambda$5000 nm) bands
that are all accessible from the ground. A SNR $>10$ measures {[}Fe/H]
with errors of $\pm0.1$ dex and errors in radial velocities of $\pm2$ \kms.
\footnote{A chemical abundance in brackets is measured in units of dex,
i.e. $[A/X]\equiv\log_{10}(N_A/N_X)-\log_{10}(N_A/N_X)_{\odot}$,
where $N_A$ and $N_X$ are the number density abundances
of elements $A$ and $X$. Positive dex means more metals than in the Sun.}
A six hour exposure with the latest spectrometers on the Keck telescopes measures the
intensities of H$\beta$ absorption lines in even dwarf Virgo cluster galaxies ($\sim16$ Mpc)
to $\sim\lambda0.02$ nm equivalent width
accuracy. In section \ref{sub:Evolutionary-Synthesis-Models} we show that
this translates into an age uncertainty of $\sim4$ Gyr for a 12 Gyr-old
star cluster; uncertainties are exponentially smaller for younger clusters.

Over the past decade, several wide-field surveys were made of the MWg
and other galaxies. Notable in the MWg are microlensing
experiments (section \ref{sub:Evidence-in-the}) and the Wisconsin H$\alpha$
Mapper \citep[WHAM,][]{Hafner03} survey of the ionized ISM.
Extragalactic surveys include the 
the Sloan Digital Sky Survey \citep[SDSS,][]{York00}, the Two-Degree
Field Galaxy Survey \citep{Colless02}, and the
near-IR 2MASS \citep{Strutskie06} and
Deep Near Infrared of the Southern Sky surveys \citep[DENIS,][]{Epchtein94}
that dip closer to the plane of the MWg to find galaxies
that are mostly dust obscured at shorter wavelengths.

\section{\label{sec:Spheroidal-Galaxies}Ellipsoidal Systems}

Egs (and the bulges of disc galaxies), by their simple SB profiles and
lack of an extensive ISM and ongoing SF, would seem to be the most straightforward
stellar systems hence the natural starting point to study
galaxy evolution. To focus discussion, we first frame basic
questions within the context of the evolution of structure
in the universe:

\begin{enumerate}
\item Can the true 3D shapes of Egs and disc galaxy bulges be recovered
uniquely from their 2D SB? Are they primarily oblate, prolate, or
perhaps more complex (triaxial) spheroids?
\item Do Egs form a single homologous family (hence formation mode)
or fundamentally different families (formation modes)
over their entire million-fold mass range?
\item Did SF/chemical evolution complete in only
a few dynamical times ($<1$ Gyr), or is there evidence for extended
(perhaps discontinuous) SF?
\item Are Egs dominated by DM, and how is DM spread compared to luminous matter?
\end{enumerate}
We summarize work that has sought to understand SF
and chemical evolution histories in diverse environments and as functions of galaxy
luminosity (and mass) to test homology,
to establish the distribution of their intrinsic shapes, to learn
the dynamical r\^ole of a central supermassive black-hole, and to constrain their DM
content.

\subsection{Current observational capabilities}

Before addressing the observed nature of Egs, it is useful to establish
the capabilities of current large (6.5-meter+ aperture) ground-based
telescopes with efficient visible-band/near-IR detector mosaics and multi-aperture
spectrographs, and of rapidly expanding spectroscopic surveys. What
are the relative merits of large angular-area, high-SNR, spatially-resolved
data of a single (or a few) nearby object(s), as opposed to more modest
data on huge numbers of galaxies in surveys such as the SDSS? The
datasets deliver complementary insights, but it is instructive to
examine for specific questions whether exhaustive data on one object
are more illuminating than less extensive data on many. One can often
recover the distribution in properties of a large sample from statistical
analyses of the distribution of key observational parameters by imposing
Bayesian prior information. The biggest challenge for the large sample
approach are selection biases that occur when working with flux-limited
samples of spatially extended but diffuse objects. The problem is
more severe when comparing distant to nearby samples, where it is
generally uncertain how the luminous objects detected at high redshift
relate to the bulk of galaxies at the present epoch.

During the 1980's and 1990's, long-slit spectrographs on 4-meter aperture
telescopes measured Eg kinematics, statistics improving with increased
CCD sensitivities and pixels. As noted in section \ref{sub:State-of-the},
by $R_{\rm{e}}$ in a massive Eg, the SB of B-bandpass starlight
has dropped typically to that of the moonless night sky. By $2R_{\rm{e}}$,
one is working $\sim6$ times fainter than the sky SB. With a 4-meter
aperture telescope, one obtains spectra typically out to $\sim1.5R_{\rm{e}}$
in reasonable exposures. To constrain the internal structure of the
galaxy requires spectra at multiple position angles (PA's) on the
sky, at least along the photometric major and minor axes. In contrast, photometric
profiles for Egs could be traced down to 1\% of the sky SB with $\pm30\%$
systematic uncertainty from the removal of its time-variable OH emission-line
glow. Another limiting factor is often the care taken in acquiring
`flat field' images to calibrate spatial variations in photometric
sensitivity across the detector and through the optics. Particularly damaging is scattered
light within the instrument+telescope.

Three major advances have increased sensitivity substantially.

\begin{enumerate}
\item Spectroscopic studies have progressed from sparse, long-slits along
select PA's reaching $\sim2R_{\rm{e}}$ \citep[for example][]{Bak00}
to the full spatial coverage over $\sim R_{\rm{e}}$ of the SAURON
integral-field spectrograph (IFS) \citep{deZeeuw02} on $>70$ early-type galaxies in both clusters
and the field. SAURON constrains within $R_{\rm{e}}$, a complicated
region containing counter-rotating stars, warps, and other transient
structures. Its optics deliver $>1500$ contiguous spectra per exposure
in either a lower spatial resolution mode used to maximize spectral
SNR at large radii, or a higher resolution mode used on brighter
regions near the galaxy core. While mapping kinematics to 6 \kms\
accuracy, its spectra also yield fluxes of some spectral indices to
$\lambda$0.01 nm accuracy in equivalent width. We introduce SAURON results below,
as appropriate. Other smaller-field IFS's (some using an adaptive optics system to sharpen
ground-based images by nulling part of the effects of atmospheric
turbulence) have detailed stellar cores.
\item Development of the `nod-and-shuffle' technique of coordinated
telescope and detector shifts \citep{Glazebrook01} to subtract night-sky
OH emission spectral bands to Poisson statistical limits. Implemented
on 8-meter+ aperture telescopes with efficient spectrographs, this
technique cancels many systematic errors to obtain spectra at much
fainter light levels than previously. Note that long-slit spectroscopy
is an inefficient data collection tactic, because it covers outer
radii only sparsely where large areas are needed to boost SNR. With
an IFS, coverage is continuous over the field of view, with outer
regions receiving the higher weight necessary to boost SNR. Nevertheless,
because of the purity of sky subtraction through an open long-slit
plus nod-and-shuffle, the sparse coverage at very low SB of long-slit
spectra still complements the full coverage at higher SB of an IFS
whose field-segmenting optics scatter more light.
\item The advent of wide-angle imaging and spectroscopic surveys.
Specifically, the 2dF survey on the 3.9-meter Anglo-Australian
Telescope produced $\sim$250,000 galaxy redshifts, and the SDSS on
a 2.5-meter telescope produced multi-bandpass photometry of $>10^{7}$
galaxies and multi-fibre spectra of $\sim10^{6}$ galaxies. With proliferating
multiple-CCD cameras and dedicated robotic, wide-field telescopes
for executing all-hemisphere imaging and spectroscopy, the future
of large scale surveys is extremely promising.
\end{enumerate}

\subsection{\label{sub:Intrinsic-shape}Intrinsic 3D shape}

We begin discussion\footnote{Readers who are unfamiliar with gravitational
stellar dynamics should first read the material in Appendix A.2.}
of Egs with a deceptively simple question:
what is their intrinsic shape? \ref{sub:Building-a-mass} noted
that obtaining the 3D distribution from the Abel inversion of
the 2D SB distribution is ill-conditioned, being highly unstable to details of noise and
data artifacts; the inversion can only be achieved with an assumed
3D symmetry. Thus, even very high quality SB data on the nearby, nearly
circularly symmetric `standard' Eg NGC 3379, yield no clear
answer on whether its slight ellipticity (Figure \ref{fig:SB-of-NGC})
implies oblateness, prolateness, or even triaxiality. It may well
harbour a weak, nearly face-on disc \citep{Capaccioli91}.

Eg shapes were therefore first constrained in galaxy samples by trending
the central SB against their degree of ellipticity.
galaxies of different ellipticity.
to map central SB versus degree of ellipticity. The seemingly well-posed
hypothesis that prolate galaxies viewed edge-on should have higher
SB than edge-on oblates has no clear-cut answer, although
the tendency has been to favour oblates \citep[for example][]{Marchant79}.
However, this statistical approach trades
the uncertainty of deprojecting a SB profile to 3D to that of extracting
the true parent distribution of 3D shapes from an observed distribution
of peak SB's and ellipticities, assuming of course that there \textit{is}
a well-defined parent distribution. In fact, \citet{Lambas92} conclude
that Egs are neither strictly oblate nor prolate.

Twisted isophotes whose ellipticity and PA of projected major axis change
with SB occur in many galaxies.
Section 4.2.3 of \citet{Binney98} shows that twisted isophotes
originate naturally when we view misaligned a triaxial system whose
principal axes are aligned and orthogonal at all intensities while
varying in axis ratio systematically with SB. 
But twists also result from strong tides between galaxies,
and indeed the most extreme ones are found in close galaxy pairs
\citep{diTullio79,Kormendy82}.
Add the aforementioned possible
disc lurking in some Egs, and 3D shapes cannot be determined.

In summary, a seemingly simple question about the supposedly
simplest galaxies leads to a surprisingly unsatisfactory result. To
progress, one must investigate kinematics.

\subsection{\label{sub:Internal-motions}Internal motions}

By the mid-1990's it was clear that most luminous Egs have isophotes
whose major axial PA's twist with radius, implying triaxial shapes.
Because triaxial shape implies a large fraction of stellar
orbits with additional isolating integral(s) (see Appendix A.2.3)
beyond $E$ and $L_{\rm{z}}$
\citep{Schwarzschild79}, the distribution function 
$f$ will have anisotropic velocity dispersions.
A signature of a rotating system with isophotes twisted by projected
triaxiality is rotation along \textit{both} projected axes, which
is seen in many Egs \citep[for example][]{Schechter79}.
Evidence for anisotropic velocity dispersions is more subtle. Any
such system will flatten more than could arise from the observed rotation.
This qualitative global statement lends itself naturally to quantification
by the tensor virial theorem (see Appendix A.2.5),
as developed by \citet{Binney78}, further discussed in
section 4.3 of BT, and reexamined by \citet{Binney05} in the light of
the new capabilities of IFS's.

The main indication of a triaxial shape is misaligned rotation, with
motion from tube orbits (Appendix A.2.4) streaming along the true minor-axis,
but observed along the \textit{projected} minor axis of an apparent
axisymmetric galaxy. 
Asymmetric absorption-line profiles are quantified with up to 6th-order Gauss-Hermite
velocity moments.
Combining absorption-line long-slit spectra to
map rotation speeds and surface photometry to map shapes, one interprets
with Jeans models (first two moments) from the tensor Virial Theorem or 
more general three-integral models (higher moments too)
to constrain the internal motions of spheroids. 

\begin{figure}[t]
\begin{centering}\includegraphics[scale=1.3]{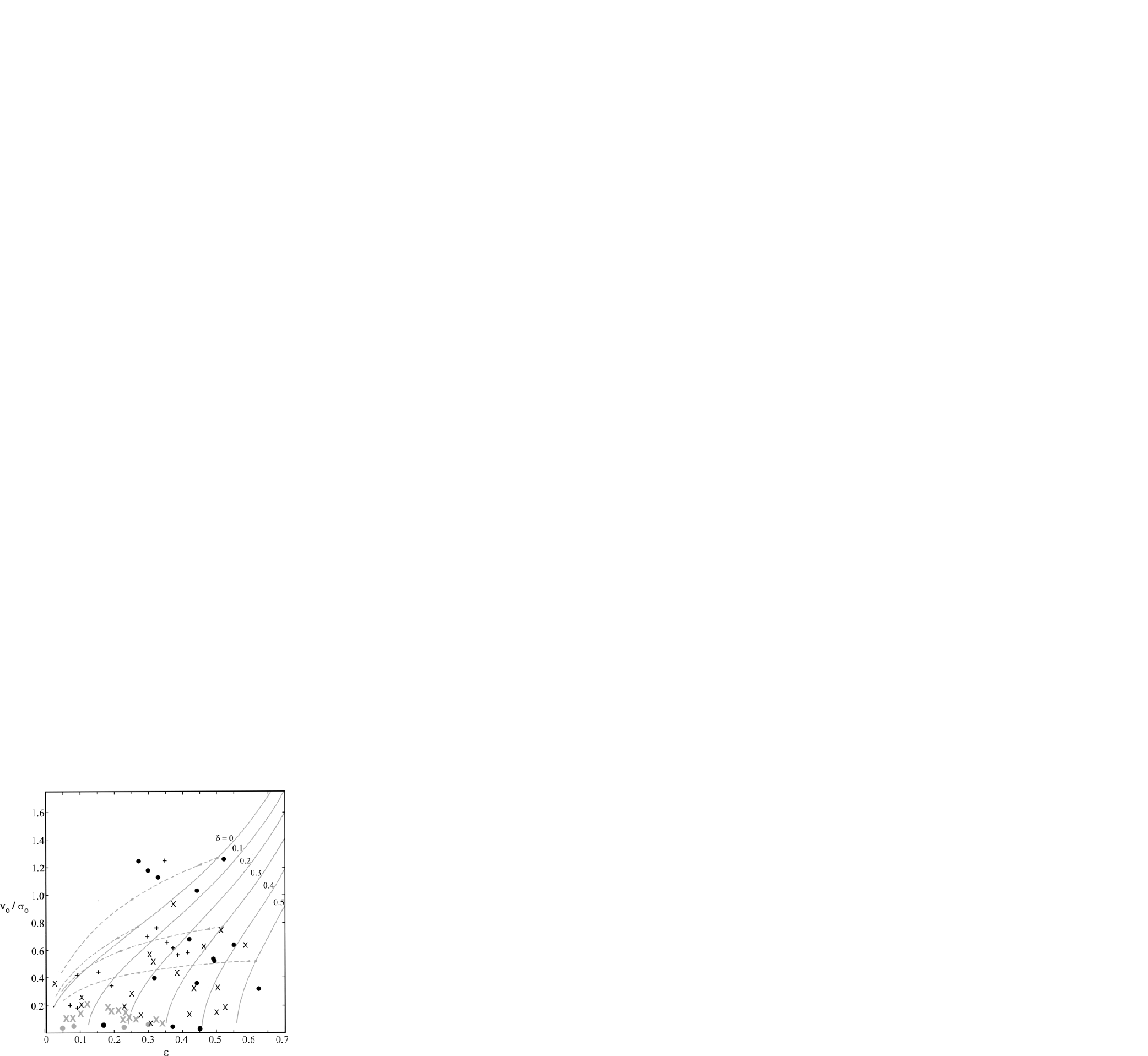}\par\end{centering}
\caption{\label{fig:kormendy}How observables (ratio of maximum rotational
to mean random motions $V_{0}/\sigma_{0}$ and photometric ellipticity $\epsilon=1-$axial
ratio) averaged over an oblate spheroidal Eg are related for various
values of the velocity anisotropy $\delta$ \full. Lines
(\broken) map to the left the theoretical $\delta$ curves drawn at right
as our viewing angle is varied from edge-on (right) to
face on (lower left). Plotted are light-weighted averages across individual
luminous Egs ($\times$ from \citealp{Illingworth77,Bertola75})
that mostly show no rotational support, and Virgo dwarf Egs (\fullcircle\
from \citealt{Geha02} and \citealt{vanZee04}) and lower luminosity
Egs ($+$ from \citealp{Davies83}) that mostly do; grey denotes
an upper limit of the measured $V_{0}/\sigma_{0}$.}
\end{figure}

In an axisymmetric galaxy viewed edge-on that rotates about the z-axis,
the potential energy tensor \begin{equation}
W_{xx}=W_{yy};\,\, W_{ij}=0\,\,(i\neq j)\end{equation}
and similarly for tensors
$\boldsymbol{\Pi}$ and $\boldsymbol{T}$ (random and ordered motion); in addition,
$T_{zz}=0$. The tensor Virial Theorem then gives\begin{equation}
\frac{2T_{xx}+\Pi_{xx}}{\Pi_{zz}}=\frac{W_{xx}}{W_{zz}}\end{equation}
If we define \begin{equation}
2T_{xx}=\frac{1}{2}\int\rho\overline{v_{\phi}}^{2}d^{3}x=\frac{1}{2}Mv_{0}^{2}\end{equation}
with $M$ the total mass, $v_{0}^{2}$ the mass-weighted mean-square
rotational part of the velocity function, and $\sigma_{0}^{2}$ the mass-weighted
mean square random part of the velocity function, we obtain\begin{equation}
\frac{v_{0}^{2}}{\sigma_{0}^{2}}=2(1-\delta)\frac{W_{xx}}{W_{zz}}-2\end{equation}
with $0\le\delta\le1$ measuring the anisotropy of the velocity dispersion
tensor (i.e., $\Pi_{zz}=(1-\delta)\Pi_{xx}$). If the isodensity surfaces
are ellipsoids, then $W_{xx}/W_{zz}$ depends only on their average
photometric ellipticity $\epsilon$ and not on their density profile
$\rho(r)$ along galaxy radii. If $v_{0}$ and $\sigma_{0}$ can also
be estimated by averaging spectra, then \full\ lines in Figure
\ref{fig:kormendy} plot the relation between the spatially averaged
$v_{0}/\sigma_{0}$, ellipticity $\epsilon$, and anisotropy
parameter $\delta$.

There are two complications. First, (3.4) assumed that we view
the galaxy edge-on to its true major axis. Otherwise the observed
eccentricity and velocity function alter. The dashed
lines in Figure \ref{fig:kormendy} show different $\delta$ hence
viewing angle $i$ that is usually unknown, so $\delta$ is degenerate.
However, different parts of the diagram indicate clear oblate rotation
or anisotropic flattening.
Second, although the tensor Virial Theorem is a global average of observables,
most of its results arise from a few long-slit spectra
through the nucleus along the major axis (and minor axis
and crossed at $45^{\circ}$ when more spectra are obtained).
Spectra normally do not extend beyond $R_{\rm{e}}$, so are not
global averages. An assumption must convert from observables
such as maximum rotational velocity and central velocity
dispersion to global values required by the tensor Virial Theorem. 

Despite these deficiencies, our understanding of Eg dynamics is based on
this Theorem.
Early work \citep{Illingworth77,Bertola75}
found that most luminous Egs flatten largely by velocity anisotropy,
not rotation. Subsequent work \citep{Davies83} led to the surprise
that lower luminosity dwarf Egs \textit{are} flattened mostly by
rotation. Recent attention has focused on Virgo cluster dwarfs
\citep{Geha02,vanZee04}. While most do not rotate,
some do so fast enough to flatten. Figure \ref{fig:kormendy}
summarizes these results.

\begin{figure}[H]
\begin{centering}\includegraphics[scale=3.5]{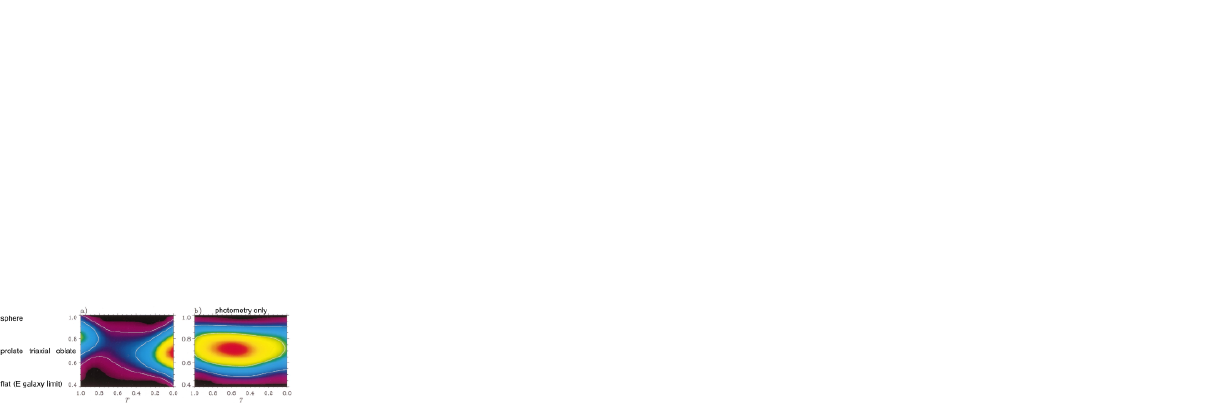}\par\end{centering}
\caption{\label{fig:Statler}\citep[From][used with permission.]{Bak00} The parent distribution
of intrinsic shapes of a sample of 13 Egs with high-quality, multiple position-angle
slit spectra, plotted in the space of triaxiality $T$ and flattening
$c_{L}$ of the light distribution. Contours enclose 68\% and 95\%
probabilities under the assumption that the parent population has
no preferred orientation to our l.o.s. a) Using photometric \textit{and}
kinematic data, a bimodal distribution is evident, with more probability
of oblate than prolate objects. The probability of them having very
triaxial shape is low. b) Photometric data alone constrain only the flattening.}
\end{figure}

The tensor Virial Theorem gives the global balance between the inferred rotation and
velocity dispersion anisotropy (hence further infers triaxiality and
the potential), at least when we have a reliable inclination.
Details on the internal structure of Egs can come
only by fitting to some dynamical model the spatially resolved spectra
of both streaming and random motions over an area of the galaxy. 
Statler and collaborators have fit spectra mostly from \citet{Davies88}
out to $\sim$1.4 $R_{\rm{e}}$ using Stäckel potential models with minimal
prior assumptions about dynamics. The resulting bivariate
probability distribution of flattening $c_{L}$ and degree of triaxiality
$T$ often yields an oblate best fit for an individual galaxy, but
the result is usually not well constrained. Instead, \citet{Bak00}
fit to the parent shape distribution of \citet{Davies88}'s 13 Egs,
using the same Bayesian models. They seek the shape distribution
of the entire family, which can be estimated more robustly than the
shape of any single galaxy. Figure \ref{fig:Statler} contours the intrinsic
shape of their sample as a bivariate likelihood distribution in $T$
and $c_{L}$. At left is the distribution from spectra and
surface photometry, with both prolates and oblates present but with
oblates more likely and strong triaxiality rare. In contrast, the
right panel shows the parent distribution if only photometry is used.
While flattening is still well constrained, triaxiality is not.

In summary, in proceeding from analyses of individual galaxies (e.g.,
NGC~3379) to a set drawn from a distribution, shapes are better constrained
but at the expense of added dynamical assumptions on the parent distribution.
\citet{Bak00} find that if rotation is assumed to be mainly `disc-like'
(decreasing rapidly from the major axis), then triaxiality is indeed
rare. However, if rotation is assumed to have `spheroid-like'
constant rotation speed at all latitudes, then triaxiality is common.
Results are still hostage to model assumptions.

The advent of IFS spectra that map the velocity function across the inner $\sim R_{\rm{e}}$
now allow application of the tensor Virial Theorem to a representative sample of Egs
with fewer assumptions \citep{Binney05}.
For example,
\citet{Verolme02} use SAURON IFS spectra to constrain
the shape of M32. Assuming axisymmetry, they obtain intrinsic flattening
$0.68\pm0.03$. \citet{Krajonovic05} is another illustration of modern orbit
fitting. \citet{Cappellari05}
is a preliminary analysis of part of the SAURON sample, in the spirit
of Figure \ref{fig:kormendy}. They show that those galaxies divide
into weakly triaxial and nearly isotropic slow-rotators, and nearly
oblate and anisotropic fast-rotators.

The reader may wonder if this discussion has led to a
result. Clearly, the CBE has not been solved uniquely.
That results (for example on whether a sample of Egs is
mainly prolate or oblate) are as varied as the underlying assumptions
shows that we are still far from uncovering shapes and establishing
the dynamics of Egs, especially using only photometry. Yet,
significant progress \textit{has} been made: the tensor Virial Theorem, supported by isophotal
twists evident in some Egs, shows conclusively that the more massive
Egs are flattened mainly by velocity dispersion anisotropy, while
intermediate mass Egs tend to be oblate rotators. This is a
key constraint on scenarios of galaxy evolution. Further progress
should come soon as IFS samples are analysed with statistical techniques
such as those of \citet{Bak00}; for example, see \citet{Statler04}.

\subsection{\label{sub:Global-scaling-parameters}Global scaling parameters}

Recall a question posed at the start of this Section: are Egs
a single homologous family over their full mass range?
While some investigators consider the range $\sim10^{6}$
to $\sim10^{12}$ \Ms, we adopt the conservative
approach that Eg luminosities range between $-18>M_{B}>-22$, dwarf Egs
range between $-13>M_{B}>-18$, and galaxies still fainter fall into
the dSph and/or globular cluster category. While the connection of dSph's
(studied extensively in the Local Group) to the main family of d/Egs is unclear,
we now ask: have all d/Egs had similar formation histories?
Does mass to light ratio ($\Upsilon$, in solar units) vary with luminosity?

\subsubsection{Photometric properties.}

Consider first the azimuthally averaged radial SB profiles of Egs.
While the azimuthal average may blur important differences between
objects (e.g., the possibility that different kinds of Egs tend to
show `discy' versus `boxy' isophotes), the radial SB profile
is a natural starting point. Given the degree of uncertainty that
arose from questioning the 3D structure of Egs in section \ref{sec:Spheroidal-Galaxies},
one should not be surprised that there is controversy over 
whether Egs form a single homologous
family in their radial SB profiles $I(R)$.
Much early work emphasized the difference between Eg and dwarf Eg profiles.
Specifically, luminous (hence, massive) Egs are well fit by the \citet{deVau48}
$r^{1/4}$ law\begin{equation}
\ln(I(R)/I_{\rm{e}})=-7.67[(R/R_{\rm{e}})^{1/4}-1]\end{equation}
with $I_{\rm{e}}\equiv I(R_{\rm{e}})$
and factor 7.67 to ensure that half of the total intensity comes from
$R<R_{\rm{e}.}.$ On the other hand, dwarf Egs (and Sg discs) are fit better by the
exponential profile\begin{equation}
\ln(I(R)/I_{\rm{e}})=-R/R_{\rm{e}}\end{equation}
A successful exponential fit does not imply that all Egs are
actually disc dominated, although it does raise the question of whether
some dwarf Egs masquerade as S0 galaxies. However, a different radial
SB profile does indicate that d/Egs may differ fundamentally. Further
evidence comes by plotting $\mu_{0}\equiv I(0)$ the central SB against
$M_{\rm{B}}$, the logarithm of the blue luminosity of the galaxy.
Starting from low luminosities, $\mu_{0}$ increases monotonically
with luminosity; at the higher luminosity end the trend reverses:
$\mu_{0}$ decreases as luminosity increases \citep[for example][]{Kormendy85}.
When few Egs with $M_{\rm{B}}\sim-18$ had accurate surface photometry,
it was generally thought that both the SB profiles and $\mu_{0}$'s
change character in a sharp transition near luminosity $M_{\rm{B}}\sim-18$.

\begin{figure}
\begin{centering}\includegraphics[scale=0.3]{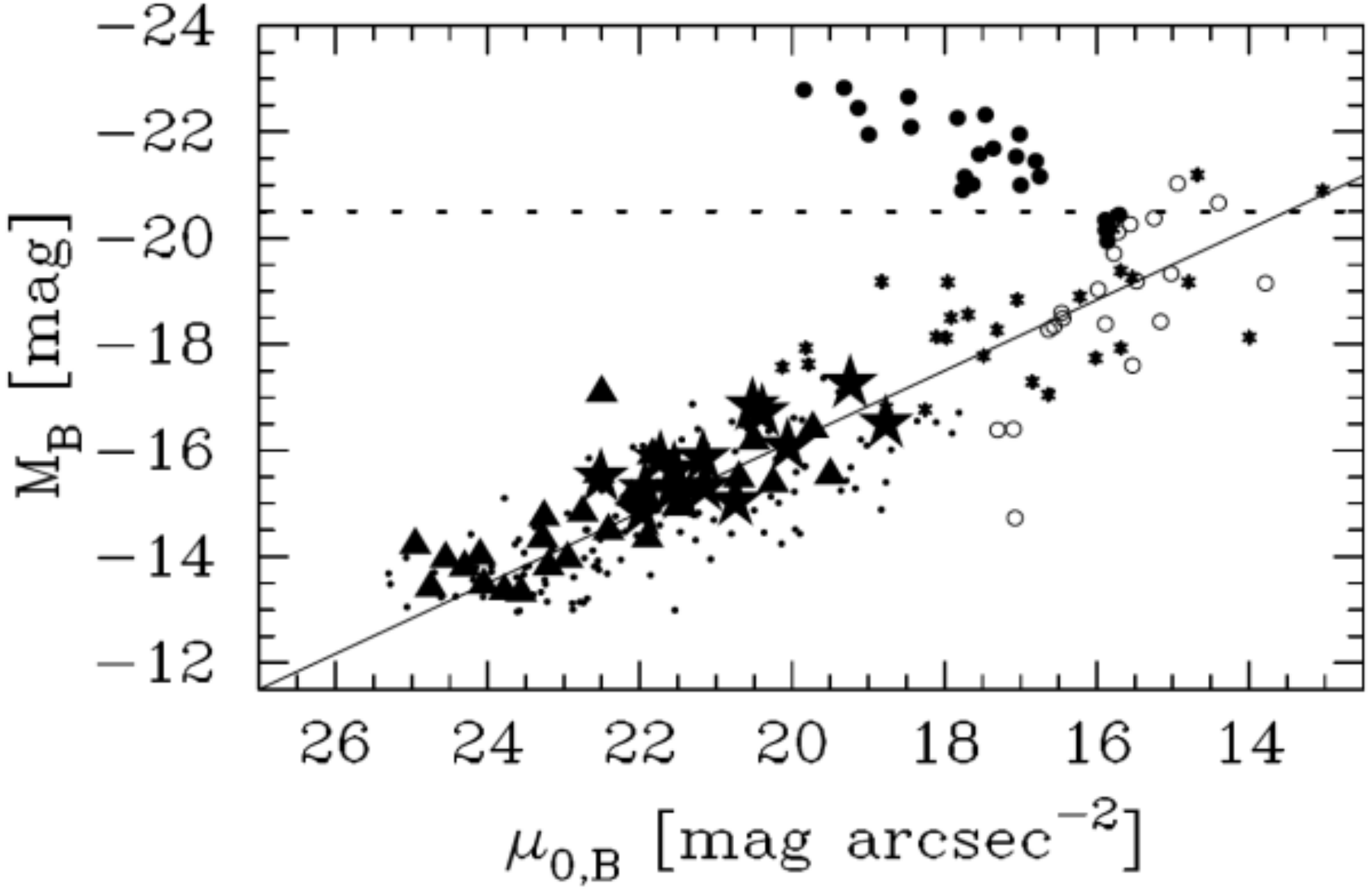}\par\end{centering}
\caption{\label{fig:Graham-plot.}From Figure 2 of \citet[used with permission]{Graham05},
showing the reversing trend in central blue SB at the horizontal \dotted\
for Egs more luminous than log galaxy
luminosity $M_{B}\sim-20.5$ in blue-magnitude units. The diagonal solid line (\full)
shows the correlation established from different data sets (different symbols, see paper) 
at low luminosity.}
\end{figure}

However, \citet{Caldwell83} showed that Egs form a continuous colour-magnitude
sequence, without breaking at $M_{\rm{B}}=-18$. In fact, recent
papers \citep[for example][]{Trujillo01,Graham03} have advocated
that Egs of all luminosities are fit by the more general \citet{Sersic68}
profile\begin{equation}
\ln(I(R)/I_{\rm{e}})=-b_{n}[(R/R_{\rm{e}})^{1/n}-1]\label{eq:Sersic}\end{equation}
with $b_{n}$ a function of the shape parameter $n$ set so that half
of the intensity is within $R<R_{\rm{e}}$ , and parameter $n$
varies gradually from 1 at the low luminosity end to $\ge4$ (the
de~Vaucoulers profile) at the high end. \citet{Graham03} find continuity
in the plot of $\mu_{0}$ (or $\mu_{\rm{e}}$) against $M_{\rm{B}}$:
Figure \ref{fig:Graham-plot.} shows that $\mu_{0}$ first
increases from low luminosity, then transitions smoothly near $M_{\rm{B}}\sim-20.5$
to decrease with further increasing luminosity. They show that a
S\'ersic profile fits very well throughout the galaxy for all lower
luminosity Egs; \citet{Graham05} presents the case that Egs form
a single family in their photometric profiles. Only at higher masses
must it be modified in the inner regions to avoid over-predicting
SB, and \citet{Graham03} argue that more massive Egs harbour a central
supermassive black-hole, which \citet{Valluri98} among others have shown may
make orbits near the centre chaotic, thereby lower the central
stellar mass density. However, \citet{Statler04} analyze one Eg studied
by SAURON and show this does not appear to have happened.

What to make of these results? On the one hand, enough free parameters
can always fit all galaxy profiles onto a single profile. On the other,
the evidence of Graham and collaborators does alleviate the
previously claimed discontinuity in Eg profiles from exponential to
$r^{1/4}$, and also explains plausibly the transition in central
SB as a function of galaxy luminosity. We therefore feel that, excluding
dSph's, there is no longer a strong case to be made from photometric
profiles that Egs divide into two distinct families.

\subsubsection{\label{sub:Kinematics:-the-fundamental}Kinematics: the fundamental
plane.}

Moving from photometry to kinematics, the tensor Virial Theorem has been applied to
other properties of early-type galaxies that correlate tightly over
at least $R_{\rm{e}}$, if not globally. The space defined by SB
$I_{\rm{e}}$ and velocity dispersion $\sigma$ is the surface\begin{equation}
R_{\rm{e}}\propto\sigma^{1.24\pm0.07}/I_{\rm{e}}^{0.82\pm0.02}\label{eq:FPlane}\end{equation}
\citep[for example][]{Jorgensen96} called the `fundamental plane'
(FP). The FP shows that the final radius of an Eg inside its DM halo
is set by the collapsing mass to within a factor of 2 over a hundredfold
range of mass. The narrow range of possible final configurations of
Egs is set by dissipation. A homologous collapse that ends in virial
equilibrium with constant $\Upsilon$ would yield $R_{\rm{e}}\propto\sigma^{2}/I_{\rm{e}}$.
Systematics of the observed deviation from (\ref{eq:FPlane}) through
breakdown of homology --- the so-called `tilt of the FP'
--- show the dependence on galaxy mass or other global parameters
of stellar $\Upsilon$ (hence the stellar population via the initial mass function, IMF) and
stellar and DM phase space densities. 
If the dynamical basis of these systematics can be understood, then their
trends with lookback time can provide insights into the mass build up of galaxies
and can test models of galaxy formation and evolution.

How these dependencies vary among galaxies sets the
intrinsic, thin scatter of the FP. For example, \citet{Jorgensen96}
find that\begin{equation}
\Upsilon_{\rm{fp}}\propto\sigma^{0.86}\label{eq:ML}\end{equation}
with 25\% scatter. 
(Such small scatter yields fairly accurate distances, enabling
study of Eg peculiar velocities.)
From extensive photometry by combining ground-based
and HST I-band images of 25 galaxies and SAURON spectra within
$R_{\rm{e}}$, \citet{Cappellari06} find that $\Upsilon_{\rm{fp}}$
can also be written within experimental uncertainties as a simple power law
of either luminosity or mass alone.  They derive mass densities
from a Jeans model and distribution function that depends only on $E$ and $L_{\rm{z}}$,
deprojecting their photometry and stellar velocities under an assumed
axisymmetry and fixed $\Upsilon$. 
They find only weak dependence of $\Upsilon$ on the
assumed inclination angle $i$ (good, because \citet{Krajonovic05} show that $i$ is
constrained only weakly by IFS spectra).
Next, fixing $i$ from the Jeans model, they use Schwarzschild's orbit
summing technique (\ref{sub:Stellar-orbits-and}) to construct
more realistic three-integral but still axisymmetric models that use more 
moments of the velocity function, not just the first and second used in the
Jeans model. Fits to spectra yield the same constant $\Upsilon$ as
from the Jeans model within scatter, demonstrating that the derived
$\Upsilon$ is independent of the Jeans assumption of a homologous
distribution function. The noisier $\Upsilon$ derived from their three-integral fits
to the velocity function yields the tight correlation \begin{equation}
\Upsilon(<R_{\rm{e}})=(3.8\pm0.14)(\frac{\sigma_{\rm{e}}}{200\,\rm{km s}^{-1}})^{0.84\pm0.07}\label{eq:MLsigma}\end{equation}
$\Upsilon$ is found to depend mainly on galaxy mass and then, at
fixed mass, on $\sigma_{\rm{e}}$, a trend related to galaxy compactness.
The scatter around (\ref{eq:MLsigma}) is small enough to exclude
significant triaxiality for at least the fast rotators in their sample, supporting
the assumptions that led to $\Upsilon$. Examining the scatter as
a function of luminosity, they find that the slow rotators, hence generally
higher mass Egs of triaxial shape (section \ref{sub:Internal-motions}),
have larger $\Upsilon$ than the fast rotators hence generally lower mass
oblate Egs. They show that non-homology is responsible for $\lesssim7\%$
of the variation in (\ref{eq:MLsigma}), most scatter coming instead
from intrinsic variations of $\Upsilon$ inside $R_{\rm{e}}$ around
typical stellar values that are consistent with simple stellar populations,
see section \ref{sub:Evolutionary-Synthesis-Models}) models.
They find that the `virial mass' defined as \begin{equation}
M{}_{\rm{virial}}=(5.0\pm0.1)R_{\rm{e}}\frac{\sigma_{\rm{e}}^{2}}{G}\end{equation}
\textit{is} a reliable, unbiased estimator of mass within $R_{\rm{e}}$.
This result confirms that
$R^{1/4}$ profiles and $\sigma_{\rm{e}}$ measured in a large aperture \textit{can}
measure mass build up even at high redshift.

\subsection{\label{sub:Dark-matter-content}Dark matter content}

\citet{Cappellari06} compare their dynamical estimate of $\Upsilon$
with that inferred from simple stellar population
synthesis to reproduce observed spectral
indices. They reject a \citet{Salpeter55} IMF because the \citet{Kroupa01}
form scatters less, and infer that the IMF barely varies among
galaxies. They conclude that $\sim30\%$ of the total mass
within the $R_{\rm{e}}$ of early-type galaxies is dark, consistent
with the few lensing cases \citep[for example][]{Ferreras05,Rusin05}.
Only a variable DM fraction within $R_{\rm{e}}$ can explain the
observed scatter.

\citet{Mateo98} reviews DM constraints from the internal kinematics
of Local Group dSph galaxies. Masses and central velocity dispersions are usually
derived using King models (Appendix A.2.2), which assume isotropic velocity dispersions
and that mass follows light. \citet{Kleyna01} obtain radial velocities
of many stars in the Draco dSph, find a flat/slowly rising velocity
dispersion at large radii, hence argue for an extended DM halo. As
a group, dSph's often have $\Upsilon\ge200$, making them nearly as
DM dominated as some low-SB galaxies (section \ref{sub:Low-surface-brightness}).
For example, \citet{Kleyna05} have studied 7 candidate stars in the
faint, 100 kpc distant, recently discovered \citep{Willman05} Ursa
Major dSph. They identify five likely members, assume an isotropic
velocity function with constant $\Upsilon$, and obtain a central
$\Upsilon=500$ solar from colours; they argue that this is a lower
limit to the global value because the galaxy is so dark. 
Is this system
a prototype of the `DM satellites' that linger around mature
galaxies at the end of $\Lambda$-CDM simulations \citep{Navarro97}?
\citet{Mateo98} obtained $\log\Upsilon=2.5+10^{7}/(L/\rm{L}_{\odot}),$
i.e., that each early-type dwarf has a luminous component with $\Upsilon=2.5$
solar that is embedded in a DM halo of $10^{7}$ \Ms. Central black-holes
more massive than $10^{4}$ \Ms\  are rejected in all systems except
Ursa Major \citep{Maccarone04} and Draco.

Further insights on DM have emerged from studies of planetary nebulae in Egs.
A planetary nebula converts up to 15\% of the white dwarf core's radiation to the $\lambda500$
nm {[}O~III] emission line, allowing velocities to be measured accurately
in early-type galaxies up to 30 Mpc distance on a 4-meter aperture
telescope. 
They are important probes in more massive spheroidal galaxies because their
emission lines are isolated easily from starlight with narrow-band
filters, and their kinematics can be mapped very efficiently using
a slitless spectrograph over a large field of view \citep{Douglas02}
and out to $5R_{\rm{e}}$ to make application of the tensor Virial Theorem more reliable.

However, they can be $<3$ Gyr old, a `young' population
in early-type galaxies. Thus their orbits have probably not relaxed
dynamically to the same state as their parent galaxy, for example
if their progenitor stars were formed in the late merger of a gas
rich system. \citet{Romanowsky03} have studied $\sim100$ planetary nebulae in each
of four lower-luminosity Egs. They found that velocities decline
by $\sim60\%$ in Keplerian fashion between one and $3R_{\rm{e}}$,
obviating need for DM. However, they note that such a decline could
also result from mostly radial stellar orbits, as might occur in an
Eg formed when two discs merge. They derive $\Upsilon=6.4\pm0.6$
solar that is consistent with their population synthesis, compared
to $20-40$ found in more luminous Egs. \citet{Napolitano05} obtain
similar results. These observations seem to leave little room for
DM within $2R_{\rm{e}}$ in lower luminosity Egs.

However, challenged by such compelling data, \citet{Dekel05} choose
more realistic stellar and DM density profiles, simulate disc galaxy
mergers, find that they can reproduce the observed planetary nebula velocity function with $\beta(r>R_{\rm{eff}})\equiv(1-\sigma_{\theta}^{2}/\sigma_{\rm{r}}^{2})\approx0.75$
and a significant DM halo, and establish that such values arise from
their simulations because radial orbits form preferentially in the
outer parts of the merger remnant whatever the gas content of the
progenitor galaxies. $\beta$ correlates with tidal strength, being
higher for more head-on collisions. It exceeds the $\le0.2$ predicted
for DM halos, suggesting that the planetary nebulae do not trace the DM. \citet{Merritt93}
reached a similar conclusion, estimating that hundreds of nebular velocities
would be needed to probe DM in the absence of other kinematical constraints.
In fact, \citet{Pierce06} find that the velocity dispersion of globular clusters
in NGC 3379 is instead constant with radius, arguing for a substantial
DM halo. Clusters on initial radial orbits will soon disrupt tidally,
reducing the orbital anisotropy of this population. 

These studies
emphasize the need for multiple DM tracers to provide consistency
and highlight the uncertainties from dynamically `young' populations.
Using globular clusters and planetary nebulae separately is perilous.

How is DM distributed spatially? Equipotential surfaces are always
more spherical than the matter distribution. It is therefore likely
that a DM halo around a spheroidal galaxy is close to spherical. It
may, however, be offset from the photometric centre of the galaxy,
and in simulations is also found at small radii to be `cuspy'
\citep[for example][]{Navarro97}. Kinematical evidence for a central
cusp of DM is inconclusive because of uncertain stellar disc and bulge
mass decompositions, the presence of distinct dynamical subsystems
in the core, and plausible variations in $\Upsilon$, dust and stellar populations
throughout the volume.

\subsection{Stellar Content}

Having discussed the structure and dynamics of Egs, we turn to questions
that bear on their formation and evolution.\footnote{A reader unfamiliar with the
basics of stellar evolution should first review Appendix A.1 where some standard
abbreviations are defined.} Unfortunately, despite
growing capabilities to detect and study the luminous content of galaxies
to lookback times of their formation, we cannot observe the assembly
of the dominant DM halos. Instead, we must rely on observations of
the SF and chemical evolution histories of the luminous baryons to infer how underlying
DM evolves. Given compelling evidence from both observations \citep[for example][]{Searle73,Larson80}
and numerical models \citep[for example][]{Mihos96} of galaxies undergoing
gravitational interactions/mergers, it is generally agreed that an
Eg that forms by merging anything other than two gas-free progenitors
will show enhanced SF and chemical evolution. So, a key question is: to 
what extent
can SF/chemical evolution histories of galaxies be extracted from their spatially
integrated starlight? In particular, how well can one determine the
mean age and chemical composition of stars in an Eg? And, if SF indeed
continued after a few dynamical times as the DM halo formed,
how reliably can one measure the SF/chemical evolution history? Naturally, to measure
discontinuities in this history would be particularly useful.

\subsubsection{\label{sub:Evolutionary-Synthesis-Models}Key issues for integrated
light spectroscopy.}

Before addressing results from integrated colours and spectra, we
highlight some important complications. 

\begin{itemize}
\item The evolution of a coeval stellar population is highly non-linear
in time. That is, the integrated spectrum first evolves very rapidly,
then slows as the MS turn-off (see Appendix A.1.5)
drops to lower masses. In fact, the models in Figure \ref{fig:Spectral-evolution-of}a
show that the population ages logarithmically.

\begin{figure}
\begin{centering}\includegraphics[scale=0.8]{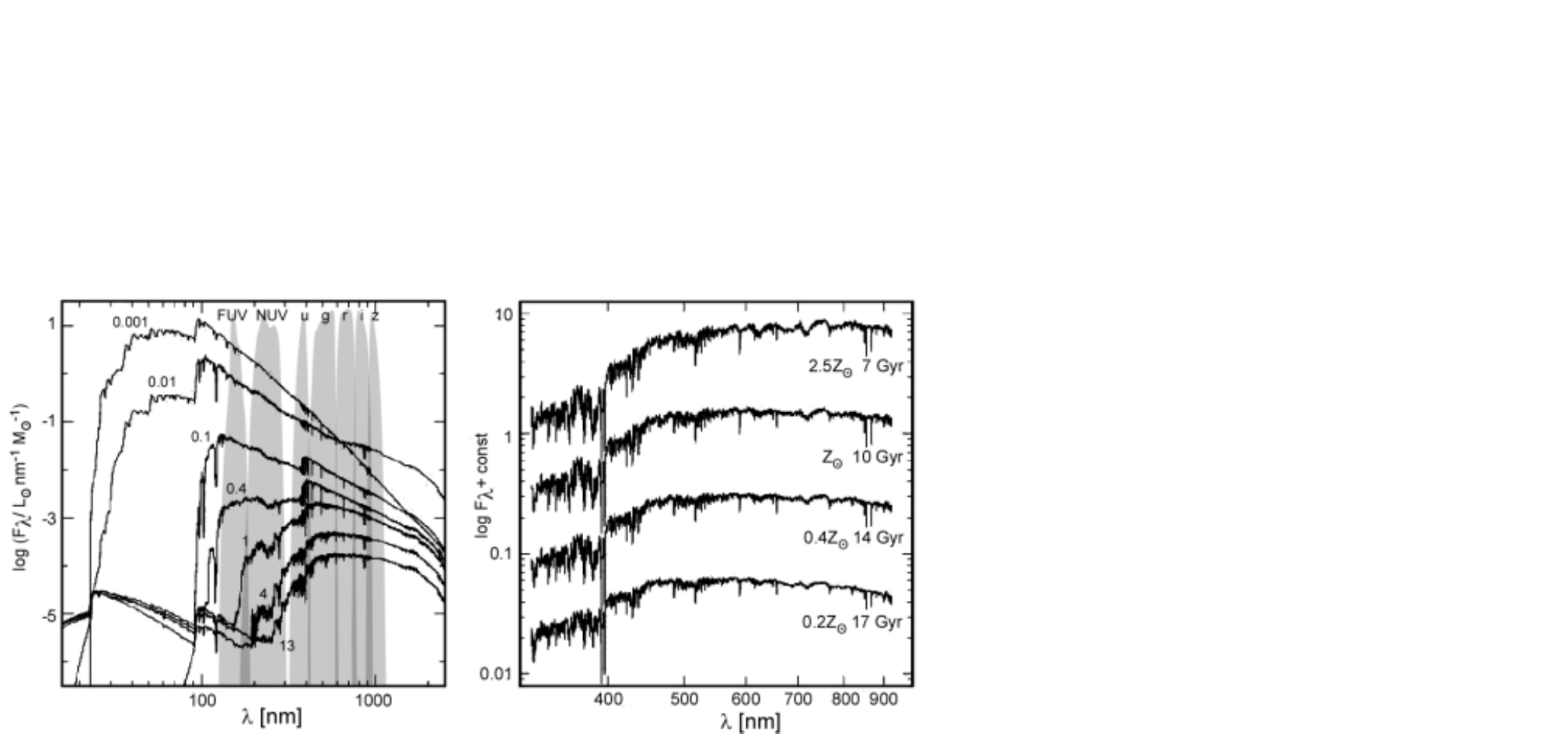}\par\end{centering}
\caption{\label{fig:Spectral-evolution-of}(a) Evolution of the standard
simple-population synthesis
model of \citep[][used with permission]{Bruzual03} for solar metallicity, 
using their GALAXEV models
(\url{http://www2.iap.fr/users/charlot/bc2003/index.html}). Times since
the end of the star burst are given in Gyr; note very small changes
$4-13$ Gyr after the burst. Between $0.1-1$ Gyr, all Balmer absorption
lines strengthen from late-B to early-F stars; this is the
standard diagnostic of a recent star burst. Over the same interval,
the `Balmer break' from the Balmer continuum limit evolves
into the `4000 \AA~ break' from numerous prominent
metal lines to the blue. In gray are the \textit{GALEX} FUV, \textit{GALEX}
NUV, and SDSS filter bandpasses. Sometimes, ages are determined crudely from the
relative intensities of light in adjacent bandpasses.
(b) A series of simple-population synthesis 
spectral models of increasing age and decreasing
metallicity (top to bottom) show how the two parameters are nearly degenerate.
However, close examination of the spectra reveals that the apparent degeneracy can
indeed be broken with spectra that resolve individual line strengths.}
\end{figure}

\item The integrated $\Upsilon$ of the evolving population also increases
dramatically and logarithmically from very small values for a young
population. Thus, even limited recent SF tends to overwhelm light
from the old population, skewing sharply the luminosity-weighted mean
age of a galaxy toward the age of the youngest stars. Such bias is,
of course, useful if one wishes to detect even small amounts of recent
SF from mergers.
\item Degeneracy between age and metallicity was the most serious complication to 
confront the first studies in integrated
light. Specifically, Figure \ref{fig:Theoretical-distribution-of}
shows that an older, metal-poorer population mimics a younger
metal-richer one. However, as detailed below, measuring mainly
age-sensitive hydrogen Balmer absorption lines coupled with metal features such
as Fe and/or Mg lifts this degeneracy \citep{Worthey94}. Age/metallicity
and dust reddening are a second degeneracy that alters photometric
colours and the overall spectral energy distribution. Dust reddens
starlight, making a galaxy look older and/or metal-richer. However,
individual spectral features are unaffected.
\item The most vexing hence interesting current issue for integrated light
studies is that element abundance ratios in Egs do not track those
in the Sun and Solar Neighbourhood stars. As we discuss below,
recent work shows that the $\alpha-$element/Fe ratio exceeds solar
in Egs, indicating more rapid chemical evolution than occurred locally.
While this result provides rich insights on galaxy evolution, it
unfortunately complicates models of integrated spectra.
\item Extracting a galaxy age from an integrated spectrum demands far more
of stellar evolution models than extracting the age of a resolved
star cluster from its CMD. For a CMD age, stellar modellers match
only the position and shape of the MSTO (Appendix A1.5). In contrast, an
integrated spectrum is a composite of light from many parts of the HR diagram. So, the
shape of isochrones must be correct across the HR diagram, not just
at the MSTO. Even more challenging, isochrones must have the correct
stellar \textit{number density} across the HR diagram. Equivalently,
the timescales for \textit{all} advanced phases of stellar evolution
must be correct. For example, \citet{Schiavon02a,Schiavon02b} show that discrepancies
between the age of the MWg globular cluster 47 Tuc derived from its integrated light
and from its CMD may arise largely because isochrones under-predict
stars along the upper part of the RGB.
\end{itemize}

\subsubsection{\label{sub:Results-for-the}Results for the dwarf elliptical galaxy
M32.}

Given these complications, the reader should question whether reliable
mean ages and chemical abundances of Egs can be extracted from integrated
spectra and photometric colours. In fact, this issue has been controversial
for decades. Pioneering models \citep{Spinrad71,Faber72} discovered
that Eg integrated spectra are dominated by the light of metal-rich, mostly old
stars. Moreover, there were strong indications that successful models
require stars with CN and Mg features that are only reproduced by
stars in the Solar Neighbourhood whose line strengths exceed greatly
those found in stars of solar abundance, a result that is now seen
in the context of enhanced-to-solar $\alpha$-element/Fe abundances.
\citet{Faber72} found that the mean metallicity of Egs increases
with mass. However, these studies could not constrain Eg ages beyond
showing that Eg spectra are clearly dominated by older stars.

\citet{Oconnell80}, in another landmark study of galaxy spectra,
analysed the integrated spectrum of the nucleus of the dwarf Eg M32 and
concluded that the Balmer absorption lines are too strong to be explained
just by an old, metal-rich population. Rather, M32 must contain MSTO
stars of intermediate age. Clearly, that Egs can contain such stars
has important implications on their formation and evolution, so O'Connell's
controversial assertion has been much scrutinized and has made M32
a `Rosetta Stone' for integrated light studies of galaxies.
In fact, its nucleus is bright enough to obtain a very high SNR spectrum
(SNR $>100$ per $\lambda$0.1 nm) in only a one-hour integration
on a small, 1.5-meter aperture telescope. Moreover, due to its low
($\sim$80 \kms) central velocity dispersion, spectral features
are better resolved than in the centres of massive Egs (central velocity
dispersion $\sim$300 \kms).

Several ways have been proposed to enhance metal and Balmer absorption
lines without resorting to a multi-age population \citep[see][for a summary]{Renzini86}.
Specifically, populations with a range in metal abundance can reproduce
the observed spectrum, i.e., an age-metallicity degeneracy argument.
Alternatively, a few hot ($\sim$10,000 K) stars could
strengthen Balmer absorption lines without resorting to a major intermediate
age population; these might be blue stragglers --- upper MS stars from
a very minor $\sim$1 Gyr old population --- or blue HB stars from a
metal-poor population. Finally, perhaps non-solar abundance ratios
have so contaminated the bandpasses of the indices used to measure
key spectral features that we are not modelling properly the integrated
spectrum.

Clarifying the situation of M32 and other Egs led to key
advances. Tinsley and collaborators \citep[for example][]{Gunn81}
introduced the now standard evolutionary population synthesis, i.e., comparing integrated
spectra of galaxies to a grid of model populations based on isochrones
and either empirical or synthetic spectra of stars.
(Synthesis replaced optimized fits to a galaxy spectrum from an observed
spectral library with few stellar astrophysics constraints.)
Another advance was the development of the Lick system of
spectral indices \citep[for example][]{Faber85}, a set of passbands
defined to isolate and quantify important spectral features. In a key
advance, \citet{Worthey94} modelled carefully
the age-metallicity degeneracy. He showed that plotting a Balmer line
index (e.g., strength of H$\beta$) against a metal-sensitive index
(e.g., strength of the Fe I $\lambda$5270 feature) decouples the
age-metallicity degeneracy. Finally, population synthesis models have
begun to incorporate non-solar abundance ratios (NSAR) to model spectra
and indices \citep{Maraston03}. This requires modelling the effects
of NSAR on the absorption features due to the abundance changes
on individual features \citep{Tripicco95}, and on the internal stellar
structure and evolution due to changes in, for example, the interior
opacity from changing proportions of major electron donors such C,
N, and O.

The present situation is encouraging. Simple stellar population models now
produce sensible ages and metallicities for well understood stellar
systems such as globular \citep{Schiavon02a,Schiavon02b} and open clusters
\citep{Schiavon04}. For M32, recent analyses
agree that an intermediate age population is unavoidable to reproduce
the integrated spectrum of its central $\sim$3 arc-seconds, both
\citet{Worthey04} and \citet{Rose05} finding that the light-weighted
mean age doubles from the central value ($\sim$3-4 Gyr) at
the nucleus to $R_{\rm{e}}$. Mean metallicity halves, so these
competing effects `conspire' to keep the integrated B-V colour
roughly constant. \citet{Rose05} demonstrate with a spectral indicator
sensitive to light from hot stats that the enhanced Balmer absorption
lines (over that expected for a uniformly old population) cannot be
attributed to a small population of hot stars. Hence, a more substantial
intermediate age population is indeed necessary to explain the strong
Balmer lines. Moreover, both \citet{Worthey04} and \citet{Rose05}
find only modest departures from solar abundance ratios in M32, simplifying
the interpretation of its integrated spectrum.

\subsubsection{Results for large samples of elliptical galaxies.}

We turn to results from large surveys of spatially integrated Eg spectra.
\citet{Trager00} studied 50 Egs in both the low-density field and
clusters, and found a significant range of light-weighted ages. The
range is widest for field Egs and for those with lower velocity dispersions
$\sigma$. As well as being older on average, larger $\sigma$ Egs
have larger NSAR, with {[}Mg/Fe] enhanced heavily in the most massive
ones. \citet{Caldwell03} obtain similar results in a study of nearly
200 Egs that cover a wide range in $\sigma$ (hence luminosity and
mass). Most striking is their result that low $\sigma$ galaxies scatter
in age far more than those with high $\sigma$, i.e., low-mass Egs
have had more prolonged SF histories. In contrast, in a study of early-type
galaxies in the Fornax cluster, \citet{Kuntschner00} found that all
Egs have similar old ages, but that metallicity increases with $\sigma$.
In a recent study, \citet{Sanchez06} analyse integrated spectra of
nearly 100 field and cluster Egs. They find that Egs in low-density
environments range in age more than their clustered counterparts,
with those in low-density environments tending to be more metal rich and younger
by $\sim$1.5 Gyr. For these galaxies,
an age-metallicity relation is found when metallicity is measured by Fe 
(but not by Mg). While a well-defined
mass-metallicity relation is found for clustered galaxies from both
Fe and Mg, for field galaxies a mass-metallicity relation is only
evident in Mg. They also find that lower luminosity galaxies spread
more in age than their higher luminosity counterparts. 

Remarkably, these studies have uncovered little evidence for a mass-metallicity
relation among field Egs. Most lower mass Egs have had prolonged
SF histories and slower chemical evolution than their higher mass counterparts. In
contrast, clustered Egs are systematically older than Egs in the
field, hence much more consistent with the hypothesis that SF occurred
rapidly and ended early to produce a well-defined mass-metallicity
relationship.

How do these results relate to predictions of the $\Lambda$-CDM hierarchy?
As mentioned, one cannot simply compare galaxy SF/chemical evolution histories
to the simulated build up of CDM halos. Rather, simulations must incorporate
parametrized simplifications of baryonic astrophysics to follow luminous particles.
For example, \citet{Somerville99}
have predicted SF histories of Egs, finding, unsurprisingly, that they follow
the assembly timescale of DM halos. Specifically, at all
masses (luminosities) one expects from the stochastic build-up of the hierarchy
to produce a range in light-weighted mean ages. Too, more massive
galaxies tend to have prolonged SF (hence younger light-weighted
mean ages) than low mass galaxies. Moreover, at given mass, SF is
prolonged for galaxies in low density field environments compared to
in rich clusters, as predicted by \citet{Kauffman98}. 
As \citet{Proctor02}
and \citet{Sanchez06} discuss, it is encouraging news for $\Lambda$-CDM models
that light-weighted
mean ages of Egs spread significantly, and also that Egs in clusters
are, in the mean, older than those elsewhere. However, that mean ages
of lower mass Egs are younger (and less enhanced $\alpha$-elements)
than for massive ones counters a straightforward
prediction of hierarchical models.

Inverted observation/model trends clearly require
a different model prescription for SF feedback, i.e.,
something that quenches SF in the deeper potentials of more massive
galaxies. It has been argued recently 
\citep[for example][]{Benson03,Bower06} that the solution to this problem
and to others in predicting the form of the galaxy luminosity function
at both low and high redshift, is connected with the formation of
supermassive black-holes in the centres of more massive galaxies and the subsequent
quenching of SF (see section 6).

Finally, one must recognize that 
existing data cover only the centres of Egs. Light-weighted
mean ages and metallicities are dominated by the nuclear region in
the spectrograph slit or few arc-second diameter fibre,
generally only a small fraction of $R_{\rm{e}}$. Long-slit studies
of M32 indicate that the intermediate age population concentrates
centrally \citep{Worthey04,Rose05}. Dwarf Egs in the Virgo cluster, which
typically have a younger light-weighted mean age \citep{Caldwell03}, 
are bluer in the nucleus
\citep{Vader88}, an `inverse' radial colour gradient. Together,
these results argue that the young and intermediate age populations
in Egs may be mostly nuclear. Unfortunately, little is known of population
gradients in Egs, due partly to centrally concentrated light that
makes extremely challenging the high SNR studies out to $R_{\rm{e}}$
necessary to disentangle age from metallicity. Radial colour gradients
traced to faint SB \citep[for example][]{Vader88,Peletier90,Wu05} reveal that
massive Egs have redder centres. The straightforward interpretation
is that their nuclei are more metal-rich. However, disentangling age
from metallicity with only visible-band colours is problematic. With
rapid progress in near-IR imaging, adding near-IR colours 
increases discrimination between age and metallicity thirtyfold
\citep{Cardiel03,MacArthur04}. (We
caution that modelling near-IR spectra requires understanding all the
issues of stellar mass loss and convection discussed in Appendix A.1.5;
see \citealt{Lee06}.) \citet{Kobayashi99}
summarized work on spectral index gradients in Egs from many sources.
For massive Egs at least, 
they concluded that universal NSAR's (as characterized
by {[}Mg/Fe] and significant metallicity gradients) exist.

\subsubsection{Stellar populations: the next steps.}

The bottom line from the previous section is that current data support
hierarchical formation in some ways, and challenge it in others. What
near-term, perhaps decisive, improvements in both observations and modelling of integrated
spectra can we anticipate? There are two data improvements.
First, results to date have solved the first-order
problem of the light-weighted mean ages and chemical compositions
of Egs. Now we must discriminate between multiple populations. Specifically,
to what extent is the $3-4$ Gyr mean age of M32's nucleus a composite
old and young population, or perhaps instead multi-episodic? Second, as mentioned
above, data on radial population gradients to constrain scenarios
of Eg assembly are sparse, especially for low mass Egs.

The first question of distinguishing mean ages and metallicities in
one or more populations is addressed by using multiple spectral features
that range widely in wavelength. A younger population, being bluer
than the older, will contribute proportionately more blue light than
red. Thus light-weighted mean ages from bluer spectral indices give
younger ages than those from redder ones \citep[for example][]{Sanchez06,Schiavon04}.
Such subtle distinctions require very high SNR and very reliable models
of different spectral indices. For example, clear problems in modelling
the H$\delta$ line as an age indicator come from `contamination'
by a nearby CN molecular band that can only be modelled confidently
by using the correct NSAR prescription \citep{Proc07}. So the problem
is not strictly one of better quality data; we also need better models.

Much modelling effort is underway to construct
fully consistent NSAR grids that incorporate not only the stellar
interior/evolution effects of NSAR, but also effects on the atmospheric
structure and its emergent spectrum. Ongoing work to synthesize
spectra for various NSAR prescriptions will eventually replace empirical
spectral libraries and their limitation to Solar Neighbourhood abundance
ratios \citep{Coelho05}.

Two tactics yield reliable radial age and abundance gradients
in galaxies. First, measure gradients out to interesting radii by going as
faint as is possible reliably.
This requires not only many
photons hence large-aperture telescopes and high-throughput
spectrographs and detectors, but also very accurate sky subtraction and
control of scattered light.
Second, cover an extended region with an IFS, do not view a narrow
slice of it through a slit.

The second advance is the advent of vast spectral surveys such
as the 2dF and the SDSS. Recent work on extracting the SF histories
of many galaxies in the SDSS have not targeted Egs \citep{Jimenez05}
specifically. In fact, fingering Egs for separate analysis is unwise
because morphology is ephemeral in the $\Lambda$-CDM
hierarchy. As discussed above, with exceptional SNR on a single galaxy
across many different radii,
one might distinguish multiple episodes of SF from a light-weighted
mean age. That would be a true advance.
A clever example of constraining SF by statistics
is \citet{Trager00},
who posit a `frosting' model for Egs to explain the distribution
of light-weighted mean ages in their sample. Frosting adds one episode
of young stars to an old population. Leverage increases
when galaxy ages and metallicities from intermediate and
high redshift surveys \citep[for example DEEP-2 and COMBO-17]{Faber05} are combined with
current epoch large samples like the SDSS. Clearly, constraints strengthen
when statistical properties of galaxies must be reproduced at
multiple epochs. It is beyond our scope to review the growing observations of
galaxy evolution at high redshift; we mention only
that recent data prefer `quenched' SF (SF that ends
abruptly rather than a single recent burst) over frosting
\citep{Faber05}.

\subsubsection{\label{sub:The-lowest-luminosity}The lowest luminosity (Local Group dwarf)
spheroidal galaxies: resolved stellar populations.}

In testing models to form and evolve Egs, much is gained by studying
the smallest and faintest galaxies because they are the most common.
Hundreds of dwarf Egs have been catalogued in the nearby Virgo \citep{Bingelli85}
and Fornax \citep{Ferguson89} clusters. To test the Eg model in the
extreme, one can study individual stars in Local Group dSph's. dSph's are very
low luminosity ($M_{\rm{V}}>-14$), low central SB ($\mu_{0,\rm{V}}>22$
mag arcsec$^{2}$), gas-poor galaxies that barely rotate. Their relation
to more massive Egs and to dwarf Egs (which are somewhat more luminous
and have brighter centres than dSph's) is unclear. \citet{Mateo98}
and \citet{Grebel05} review their properties. In galaxy groups, most
lie within $\sim$300 kpc of a massive galaxy, whereas dIrr's are
more dispersed \citep{Grebel05}.

\begin{figure}
\begin{centering}\includegraphics[scale=1.2]{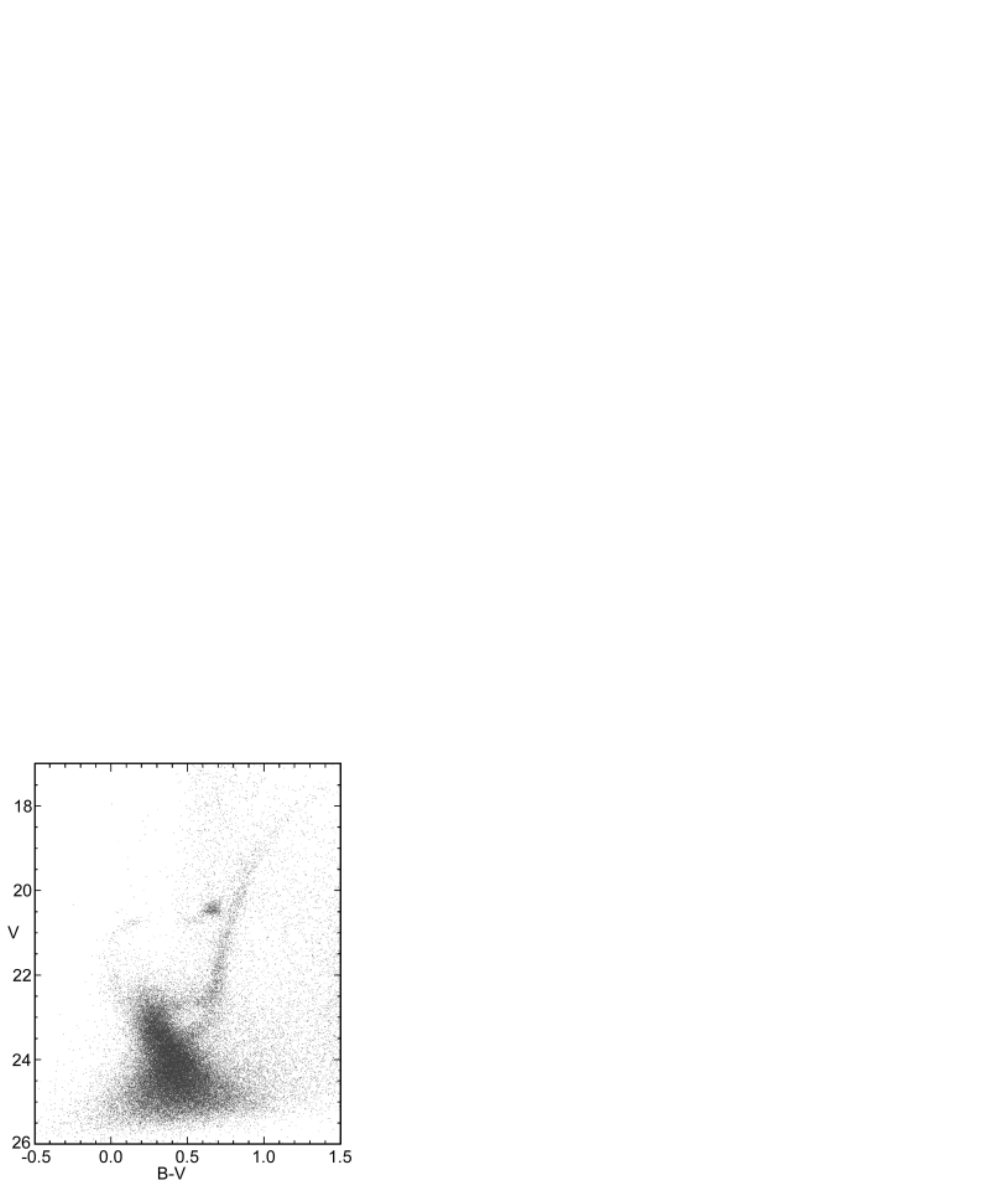}\par\end{centering}
\caption{\label{fig:CMD-of-Carina}CMD of the centre of the Carina dSph, a
MWg satellite, showing multiple sub-giant branches, and blue and red
HB stars \citep[][used with permission]{Monelli03}. The vertical axis
is related to log visible-light luminosity, and the horizontal
axis is colour with blue at left and red at right.
Isochrone fits yield distinct populations
aged 0.5, 5, and 11 Gyr from the present if helium and metal abundances are $Y=0.23$
and $Z=0.0004$, respectively.}
\end{figure}

SF and chemical evolution histories of Local Group dSph's can be detailed with ground-based
and HST multi-colour photometry to provide CMDs, and by using large
aperture telescopes to obtain intermediate-resolution spectra
of $\sim$100 stars simultaneously \citep[for example][]{Venn04}.
With metallicity from spectra, one can then fit isochrones to broad-band
photometry to pin down the SF history. Of course, a complex SF history
will superimpose multiple isochrones, increasing the likelihood
of a non-unique decomposition \citep{Gallart99}, for example Figure
\ref{fig:CMD-of-Carina}.

Key results come from CMD and metallicity studies. First, except for
the Ursa Minor dSph \citep{Carrerra02}, all Local Group dSph's show intermediate age stars
hence prolonged SF histories \citep{Koch06,Dolphin02,Saviane00,Gallart99}.
This is very surprising because energy
from the first $\sim1000$ SNe in the first star burst exceeds greatly the binding
energy in the shallow potential of the galaxy (total mass only $10^{7}$ \Ms),
implying the ejection of all ISM then. Consider the Carina dSph,
the only such system whose SF can be resolved:
Figure \ref{fig:CMD-of-Carina} shows that it
formed stars $<1$, $3-6$ ($\sim70\%$ of total), and $11-13$ Gyr
ago with few between.
Recent VLT spectra yield
mean metallicity $[Fe/H]=-1.7$, ranging from -3.0 to 0. Metal-rich
stars concentrate toward the centre. Carina has $\Upsilon\sim32$
with uncertain DM content and essentially no gas. Multiple star bursts
mean either that gas ejection was inefficient or that new gas was
accreted. Accretion or re-accretion would be difficult because of this dSph's
very low total mass, small cross section, and high orbital speed
in the MWg's outer halo ($\sim200$ \kms). So, merging gas
would have had to move in nearly the same orbit. Or, gas ejection
may not remove the densest giant molecular clouds (GMC) where SF
occurs. What causes the few Gyr latency between bursts is unknown.

A second indicator of a history of chemical evolution is the discovery that dSph's
like Carina range substantially in chemical composition. As expected,
mean metallicity increases with luminosity in dSph's. \citet{Grebel03}
among others have noted that the metallicity-luminosity relation in
dSph's offsets to higher metallicity at a given luminosity compared
to dIrr's (gas-rich low-luminosity galaxies, generally with substantial
rotation), thereby questioning if the latter can be precursors of
dSph's.

Third, is the question of whether the MWg halo, particularly
its component in retrograde motion, could have assembled from accreting
dSph's. \citet{Venn04} have analysed detailed abundance patterns
in the MWg halo and in dSph's. They find that dSph's have lower {[}$\alpha$/Fe]
at given {[}Fe/H] than halo stars. There is better consistency with
halo stars in extreme retrograde orbits, but other element abundance
ratios do not agree. Higher {[}Ba/Y] in dSph stars compared to that
in halo stars indicates slower chemical evolution in dSph's, because Ba forms in the
s-process. This result accords with the extended SF histories of dSph's.
So, while the halo is clearly not assembled from stars with the chemical evolution
of today's dSph's, it is unclear if earlier mergers are excluded entirely,
especially because the most metal-poor stars in dSph's appear
to have abundance patterns more consistent with the halo.

Finally, the distribution of Local Group satellites is interesting. \citet{Hartwick02}
finds evidence for dissipational collapse from the anisotropic velocity
ellipsoid of this population: it is flattened spatially along the
major axis, indicating a non-spherical potential. M31 and its satellite
retinue including M33 show a similar alignment \citep{Koch05}. Hartwick
argues that such is expected when a flattened matter filament
collapses rapidly to a plane as the globular cluster and dwarf galaxy population
formed. The collapse then progresses in the plane to form the giant
galaxy \citep{Larson69}. \citet{Azzaro06} examine
the alignment of satellite galaxies in the SDSS dataset, and find
that red ones are aligned along the major axis of red hosts
up to 0.5 Mpc away; the sample is too small to test for isotropy
around blue hosts.

Thus there is no consensus yet on the origin of
dSph's because of their surprisingly large spread in stellar age and
chemical composition. The nature of these extremely
low mass galaxies remains a particularly important challenge for any
scenario of galaxy evolution.

\subsubsection{\label{sub:Local Group-globular-clusters}Local Group globular clusters.}

The next rung down in mass and our main probe of baryonic halos beyond
the Local Group are globular clusters. There are $\sim500$ in M31, and 
$\sim150$ in the MWg \footnote{See the tabulation at
\url{http://physun.physics.mcmaster.ca/Globular.html}} with continuing
discoveries from near-IR surveys. Some Egs
have more than 10,000. HST or an adaptive optics system resolve all but the
most distant Local Group globular clusters to their core.
Not all are thought to be bound in galaxies, 
some may be `intergalactic tramps'. While
there are counter examples of globular clusters younger than 1 Gyr in the LMC and
especially in merging systems, the MWg, M31, LMC, Fornax dSph, and
Sagittarius dIrr's all have globular clusters of comparable ancient age within $\pm1$
Gyr. An exception is the massive cluster $\omega$ Cen, which has
multiple stellar populations and anomalous chemical abundances (especially
copper) compared to other globular clusters.

Globular clusters in most galaxies have a bi-modal colour distribution
\citep[for example][]{Ashman92}, usually interpreted
as two populations and the signature of a hierarchical
build-up. However, \citet{Yoon06} have challenged this interpretation
by noting the strong non-linear dependence of globular cluster colour on metallicity
through the dependence of HB morphology on metallicity. They argue
that the colour distribution is due instead to a continuous
range of metallicity in a single old population, thereby avoiding
multiple metallicity and/or age populations in globular cluster distributions.

\subsubsection{\label{sub:Chemo-dynamical-clues-in}Local Group archaeology: chemo-dynamical
clues in Local Group baryon halos.}

Local Group archaeology aims to sequence the events that formed
those galaxy discs and halos, and thereby to constrain models of galaxy formation.
Including recently identified objects,
the MWg baryon halo has $\sim3\times10^{9}$ \Ms\  within 50 kpc
radius ($<0.6\%$ of the total), density profile $\rho(r)\propto r^{-3.5}$
out to radius $\sim100$ kpc, and may have two components (inner Halo
I moderately flattened, outer Halo II more spherical according to
\citealt{Gould98}).
Analyses of both globular clusters \citep{Rodg84,Lee99}
and halo field RR Lyraes \citep{Bork03,Kin07}
and subdwarfs \citep{Maj96} identify two kinematic
components to the halo: one in mean retrograde motion, the other in slow prograde 
rotation \citep{Chiba01,Reid05}.
The former is believed to result
from tidal disruption of accreted satellites.  The key question of
whether most of the halo formed in a single monolithic collapse,
\citep{Eggen62}, or through the
more stochastic accretion of fragments \citep{Searle78}, is still
under debate \citep[for example][]{Chiba01,Gratton03}.

Phase space structure in the halo may be preserved, but the relative importance
of dissipation over accretion during halo formation is unknown. With enough
stellar kinematical data, \citet{Freeman02} argue
that one can identify the stream from each accreted satellite.
Few stars have been observed at sufficient spectral resolution for this, 
and in any event spectra can today be obtained only
of inner halo objects. In the Solar
Neighbourhood only 0.5\% of the stars are in this component, a few
thousand high-velocity or metal-poor stars \citep{Carney96}. Halo
I and II stars are $10-13$ Gyr old, but the bulge seems to be younger,
$8-11$ Gyr. There is a large range in {[}Fe/H] = -5 to +0.5, but
element abundances in Halo I differ clearly from those measured in Local Group dSph's.

Photometry is much more extensive, for example from the SDSS. At $V>18$,
density enhancements in Halo II are large \citep{Ibata03,Newberg02},
especially tidal debris from the Magellanic Stream plus 
Sagittarius dwarf galaxy
\citep[our nearest neighbour galaxy,][]{Majewski03,Majewski03a,Belokurov06} 
that together span the sky.
A few moving groups have been identified ($1^{\circ}\times10^{\circ}$
stream associated with globular cluster Pal 5, \citealt{Odenkirchen03}), and the
mean and dispersions of their metallicities and kinematics have been
measured; the Monoceros stream has been interpreted either as an accreted
galaxy at $\sim15$ kpc Galactic radius \citep[for example][]{Helmi03,Newberg02}
or as part of the MWg's warped disc \citep{Momany04}.

M31 is the other Local Group galaxy with a detectable halo (the halo of M33
is elusive, \citealp{McConnachie06}, but see \citealp{Ibata07}).
The $V-$ and i-band Isaac Newton Wide-Field
Camera Survey of M31 resolves stars 16 times less luminous than those
at the tip of the RGB and extends out to 55 kpc radii \citep{Ibata01}.
The Canada-France-Hawaii Telescope MegaCam survey goes deeper and covers a larger field.
In these images \citet{Ibata01} and \citet{Ibata07} find a half-dozen star streams
in that bulge/halo, including one at 70 kpc (the Giant Stream, comparable to the Magellanic
Stream) that extends for 100 kpc and whose large velocity deviation indicates no association
with either of M31's large satellites M32 and NGC 205 \citep{Chapman06}. Assuming
an orbit apocentre of 125 kpc, they find \citep{Ibata04} 
$7.5_{-1.3}^{+2.5}\times10^{11}$ \Ms\  halo mass within.
\citet{Guhathakurta06} study the halo at 10 \kms\ resolution and obtain metallicities
from the Ca~II triplet.
\citet{Chapman06} detect a distinct, non-rotating,
metal poor ($[Fe/H]\sim-1.4\pm0.2$) stellar
component from 10 to 70 kpc radius, of abundance similar to
Halo I of the MWg \citep{Chiba01}. Evidently, M31 has had a more exciting life than the MWg,
including a surprisingly recent merger (0.25 Gyr according to simulations by \citealt{Font06}).

In summary, the seeming simplicity of Egs once suggested that they
are homologous in mass, largely supported by random motions but
flattened by rotation, and assembled quickly
(a few collapse times) while forming stars efficiently. Improved observational tools
allowed study of `ellipsoids' of all 
masses, and we have seen in this section that none of these assumptions have turned out to be true.

\section{\label{sec:Bulges-of-Spiral}Bulges, pseudo bulges and bars in disc galaxies}

An Sg generally contains a spheroidal stellar component that is more
centrally concentrated than its flattened disc. This `bulge'
provides clues for galaxy evolution. As section 3 explained, the $\Lambda$-CDM
scenario assembles a bulge through a near-equal mass merger. The result
is an Eg until it reacquires a disc by cooling the
residual diffuse halo of ISM that was heated by the merger. Hence
in the $\Lambda$-CDM scenario, Sg bulges should resemble Egs in their
internal dynamics and stellar populations. So, is a massive bulge
also flattened mainly by an anisotropic velocity dispersion tensor?
What has the bulge of the MWg told us about other bulges?
Are bulge properties consistent with hierarchical mergers, or are
other formation processes relevant?

To answer, one must address two issues that regard the bulge as a
separate entity from the often dusty disc. First, we must separate
their light profiles. Large bulges show the centrally concentrated
SB profiles that characterize Egs, a S\'ersic profile with index
$n\sim4$. Disk light declines in an exponential \begin{equation}
I(R)=I_{0}\exp(-R/R_{\rm{d}})\end{equation}
of $I_{0}$ central SB and $R_{\rm{d}}$ scale length. Sometimes
the disc dominates at large radii so can constrain $R_{\rm{d}}$
while the bulge dominates at small. When separating disc and bulge
light, the outer fit is then extrapolated inward to constrain $I_{0}$
and is subtracted from the observed profile; the residual is fit with
a S\'ersic profile to define the bulge \citep[for example][]{Schombert87,MacArthur03}.
Thus, the bulge profile is only reliable for a pure exponential disc
profile. 2D surface photometry is required and, in early-type (i.e.,
bulge-dominant) Sgs, projection effects must be considered \citep{Noordermeer06}.
Dust grains attenuate bulge starlight. In the B-band, \citet{Driver07} find that
only 29\% of the bulge starlight escapes into the intergalactic medium.

\subsection{Pseudo bulges}

Second, it seems that not all bulges arise from mergers. Instead,
as reviewed by \citet{Kormendy04}, `pseudo bulges' evident
in disc-dominant (so-called late-type) Sgs are often built by secular 
processes; \citet{Noordermeer06} finds similar flattened bulges
even in some early-type Sgs. 
Specifically, non-axisymmetric disturbances (bars) can trigger gas inflow and subsequent star 
formation to build up a pseudo-bulge.
That its scale length
is well-correlated with that of the disc \citep{Courteau96,MacArthur03}
further suggests its secular origin.

Other structures that at low spatial resolution mimic `classical'
bulges (but resolve into more complex structures in HST and
adaptive optics system images) are inner discs, bars within bars, and box-shaped nuclear
bulges (perhaps nuclear bars seen edge-on). Near-IR surveys that are sensitive
to the stellar mass distribution have found many bars in visible-light
unbarred Sgs. For example, the bulge of M31 is barred \citep{Beaton06},
and is twice as massive as that in the MWg, perhaps \citep{Brown06}
having been augmented by a merger with a galaxy of MWg mass.
\citet{Kormendy04}
conclude that there are comparable numbers of early- and late-type
galaxies, classical bulges and pseudo bulges. However, pseudo bulges
in late-type Sgs have masses only $1-10\%$ of bulges in Egs.
\citet{Driver07} find from their Millenium Galaxy Catalog that 60\% of stellar baryons
lie in galaxy discs, and 27\% in classical bulges.

\citet{Corollo03} reviews photo-kinematical distinctions between
classical and pseudo bulges, finds that they are quite blurred, and
challenges their division into distinct components. To the extent
that this distinction is tenable, the consensus is that 
the central light excess over an inward-extrapolated exponential disc tends,
in late-type Sgs, to be a pseudo bulge from secular evolution, but
is a classical bulge in bulge dominant Sgs.

\subsection{Bars and other non-axisymmetric distortions}

As was touched on above, bars and other non-axisymmetric distortions
can play an important r\^ole in the secular evolution of pseudo bulges.
They are reviewed by \citet{Sellwood93}, \citet[focused primarily on
galactic rings, but discussing bar dynamics too]{Buta96},
\citet{Knapen99}, \citet{Knapen00}, \citet{Shlosman01},
and \citet{Kormendy04}.  
Specifically, bars drive gas inward to form stars, and they
scatter disc stars into the pseudo bulge.  
Here we address only
questions related to the issue of bulge versus disc development:
How are bars generated?
How long do they last?
How often are they found?

Bars form either through internal secular processes or through external
tidal triggering.  Under appropriate conditions set by the mass distribution
in the disc and halo, and the disc velocity dispersion, gas discs are unstable
to bar and spiral structure formation through the `swing amplification'
mechanism \citep[for example][]{Toomre81}.  In brief, low-amplitude leading spiral wave
patterns propagate through the galaxy centre and emerge as amplified trailing 
waves.  A bar develops by exchanging angular momentum with the outer stellar disc,
DM halo, and/or gas disc.  In hot
stellar systems (e.g., Egs and stellar interiors), any energy
exchange produces a bifurcated core-halo structure \citep{Lynden68} because of
the negative specific heat of a self-gravitating system.  Similarly,
its angular momentum components separate \citep{Lynden72,Kormendy05},
with resonance between the rotation frequency and the frequency of
radial oscillation delineating the in/outward flows of angular momentum.
Extensive explanations for bar formation based strictly on orbit theory are
provided in, for example \citet{Sellwood93,Buta96}.
Naturally, numerical simulations of spiral discs are a crucial reality check on
simplified Hamiltonians.  Early simulations were restricted to 2-D, and were
sufficiently grainy due to the small number of particles tracked that the secular bars
that formed could plausibly have been artifacts of numerical
instability.  However, with greatly increased sophisticated numerical
models now in 3-D, it is clear that real bar instabilities
occur and that their strength depends on the mass distribution and velocity dispersion of the disc
and halo \citep[for example][]{Knapen99,Shlosman01}.

While bar instabilities may form in isolated discs only through internal processes,
tidal interactions and minor mergers are also effective triggers to
drain angular momentum from the gas and inflow it
to perhaps form stars and build up a pseudo bulge.  Thus
it is of considerable interest whether most pseudo bulges have been created by
external (interaction/merger) processes, which would be more
consistent with the hierarchical merger picture \citep{Kannappan04},
or by internal processes, which would pose problems for that
scenario \citep{Kormendy05}.  In support of the external trigger, \citet{Kannappan04}
find a correlation between central blue colours in pseudo bulges
with morphological evidence of tidal encounters or merging.
On the other hand, \citet{Kormendy06} examined two
pseudo bulges in detail, and found no evidence for an external trigger.  

The longevity of a bar depends upon the importance of
gas in its evolution \citep[for example, the simulations of][]{Debatt06}.  
For purely stellar systems, bars
tend to be long-lived, in fact are only weakened by tides \citep{Athanassoula03b}.
This is because angular momentum flows
from the bar to either the outer disc or the DM halo, which slows the
pattern speed and increases the strength of the bar
\citep[for example][]{Athanassoula03a,Athanassoula03b}.
On the other hand, when gas is included, there
is general consensus that its interaction with the bar weakens
the bar, but little agreement on how this occurs.
\citet{Bournaud05} propose that the bar drains
angular momentum from the gas, which increases bar pattern speed and weakens the
bar.  \citet{Berentzen07} argue that the gas has little direct effect.
Instead, the bar is destroyed by the build up of the pseudo bulge
as gas inflows to form stars, a less direct r\^ole for gas than 
\citet{Bournaud05} propose.  The
timescale for bar destruction is similarly unclear, with some investigators
finding a quick $\sim$1-2 Gyr one \citep{Friedli94,Bournaud05},
and others much longer \citep[for example][]{Berentzen07}.  

Finally, observations in the near-IR H band, which better reveal the bulk of
the underlying stellar distribution than do optical passbands, find that
$\sim2/3$ of Sgs show a bar
\citep{Knapen99,Knapen00,Eskridge00,Whyte02,Marinova07}.  
In addition, bars are more common among galaxies with a central starburst 
\citep[for example][]{Hunt99}, but whether
they are more common in galaxies with AGN than those without remains controversial 
\citep[for example][]{Mulchaey97,Knapen00,Laurikainen04}.

\subsection{Constraints from stellar populations}

Stellar population studies consider mostly visible-light
and near-IR colours, not line strengths. As noted in section \ref{sub:Results-for-the},
combined visible-band and near-IR colours break the age-metallicity
degeneracy. However, dust in Sgs dust extinguishes starlight to introduce
a new degeneracy. Near-IR colours are relatively immune to dust, but
their combination with dust-sensitive visible-band colours
ameliorates the age-metallicity degeneracy. To avoid uncertain
reddening and contamination by young stars in a superimposed disc,
\citet{Peletier99} studied nearly edge-on Sgs near
$\sim R_{\rm{e}}$. They found that
the massive bulges of early-type Sgs resemble those of early-type
galaxies in the Coma cluster. Specifically, derived ages at $\sim R_{\rm{e}}$
scatter by $\lesssim2$ Gyr around $\sim$10 Gyr mean; the absolute
age is uncertain from uncertain modelling of near-IR colours. In contrast,
smaller bulges in late-type Sgs are bluer, hence younger. As for
Egs, where more massive galaxies are older in the mean, the trend
of older more massive bulges seems to contradict the prediction that
hierarchical mergers form smaller things first. However, one must
recall that smaller Sg bulges are probably dominated by secular
processes.

When examined spectroscopically, the centres of Sg bulges are found
by \citet{Proctor02} to be younger and sparser in $\alpha$-elements
than Egs. \citet{Moorthey05} find that large bulges in Sgs resemble
massive Egs, both being old and red, with enhanced $\alpha$-elements.
\citet{Norris06} obtained long-slit spectra to $\sim2R_{\rm{e}}$
in the edge-on S0 galaxy NGC 3115 and find an old, light-weighted
mean age for its bulge, while the disc is $\sim5-8$ Gyr old.
Given the confused disentanglement of bulge and disc light \citep{Corollo03},
detailed maps of spectral line indices over a wide range of `bulge'/disc
contributions are necessary to interpret indices obtained
from sparse slits on each galaxy.

Dust restricts our view of the MWg's inner bulge in visible-band light
to `Baade's Windows', a few sightlines of reduced extinction.
In the near-IR, extinction is hugely reduced to reveal a clear triaxial
bulge in for example the \textit{COBE} image \citep{Dwek95}. \citet{Babusiaux05}
summarize recent work that shows the triaxial bar ending in a possible
inner ring whose longest axis is inclined to our l.o.s.\ by $\sim22^{\circ}$.
This barred inner bulge, prominent in counts of RC stars, appears
to be secular in origin. So, it is surprising and perhaps problematic
for the secular evolution picture that stars in Baade's largest `Window'
(at Galactic latitude $-4^{\circ}$) are old \citep{Zoccali03,Kuijken02}
and have enhanced $\alpha$-element/Fe abundances \citep{McWilliam94,Barbuy99}.
near-IR imaging spectroscopy of this region will continue to be important
for understanding secular bulges in general and the stellar populations
of the MWg bulge in particular.

\section{\label{sec:Thick-&-Thin}Disk Systems }

\subsection{Luminous Structures}

Aside from the aforementioned (pseudo-)bulge population, and a small
halo population (as evidenced by metal-poor stars and globular clusters in the MWg
and by globular clusters in other galaxies), Sgs are characterized by a prominent
disc. Disks are highly dissipated, kinematically cold (rotating) structures,
conventionally assumed to have forgotten their progenitors, despite having made only a
few tens of rotations since their formation. 

The azimuthally
averaged SB profiles of discs are exponential in both radial and vertical
directions.
Many discs are dusty;
\citet{Driver07} find that 63\% of B-band photons released in discs actually escape from
these galaxies.
As was discussed in section 4,
discs suffer non-axisymmetric instabilities that can develop into distortions, 
predominantly two-armed spirals. The nature of spiral structure
is a major subject discussed by, for example \citet{Bertin96}. 
Here we concentrate just on issues that relate directly to
the formation and evolution of whole galaxies.
Specifically, we address

\begin{itemize}
\item The existence of thin and thick discs, and the relationship between
these apparently distinct components.
\item That both ionized and neutral gas can trace Sg discs out to large
radii, enabling a more definitive measure of the DM in Sgs than is
possible in Egs.
\item That Sg discs harbour most of the ongoing SF in the universe, making
them important laboratories for studying SF and feedback processes.
\end{itemize}

\subsubsection{Thick disks.}

\noindent \citet{Burstein79} and \citet{Tsikoudi79} discovered that
the vertical distribution of light in some edge-on galaxies has two
exponential components; that with larger scale height is the thick
disc. Later, \citet{Gilmore83} established a thick disc in the MWg
by counting stars by apparent brightness perpendicular to the Galactic
plane. Such counts are the convolution of the vertical density function
with the very broad stellar luminosity function, so are hard to interpret
uniquely because any feature in the luminosity function can be misinterpreted
as a feature in the density function. The stellar luminosity function
is broad both because of the mass-luminosity relation for MS stars
and because evolved stars range widely in luminosity. The thin disc
has $280-350$ pc scale height at $R_{0}$ and contains most of the
baryonic angular momentum. The thick disc has $0.9-1.2$ kpc scale
height and $\rho_{\rm{thick}}/\rho_{\rm{thin}}=0.085_{-0.065}^{+0.045}$
locally \citep{Siegel02}. The two discs overlap maximally at $1-1.5$
kpc at $R_{0}$; sparse, faint stars complicate their separation and
normalization. Adding colours improves deconvolution prospects.

It is now customary to separate the discs kinematically \citep[for example][]{Carney96},
a statistical separation because assigning any star to one disc is
uncertain. While, on the whole, stellar metallicity declines from
the thin disc to thick to halo (the thin disc has {[}Fe/H] = -0.7
to +0.3 and the thick has {[}Fe/H] = -2.2 to -0.1), there is enough
overlap that again only statistical assignment is possible. \citet{Reddy06}
have observed a sample of $\sim100$ thick disc stars, plus fewer
stars in the thin disc and halo. It is clear that thick and thin disc
stars have strikingly different abundance patterns with little overlap.
Specifically, $\alpha$-elements are relatively enhanced in thick
disc stars. In addition, these stars are on average 5 Gyr older.

Suggested origins of the thick disc include a population of dissolved
super-star clusters of $\sim10^{6}$ \Ms\ \citep{Kroupa02},
or the infall of satellites \citep{Bekki00}. \citet{Reid05} argues
from the white dwarf luminosity function for a continuous sequence of formation
from halo ($1\%$ of white dwarfs in the Solar Neighbourhood) to thick ($20\%$)
and ultimately thin discs ($\sim80\%$). The low-luminosity cutoff
of the white dwarf cooling curve tells us that the thick disc is $\sim9.5\pm1$
Gyr old, with most of its stars having formed over $\sim1$ Gyr; it
has {[}Mg/Fe] $>$ 0.2 (average in one sample {[}Fe/H] = -0.49). Its
abundances overlap with those of metal-poor stars in the thin disc. Cooling
curves also show that the thin disc began to form many stars $7-8$
Gyr ago. An age-metallicity relation is evident in the thick disc:
{[}Fe/H] increases by $\sim0.5$ as the age decreases by $\sim5$
Gyr. \citet{Reddy06} note that chemical abundance patterns in thick
disc stars are distinct from those of stars in dSph's and in the Magellanic
Clouds \citep{Shetrone03,Venn04}. Thus, the thick disc cannot have
been built from mergers of satellites that resemble our surviving
current dSph population. Moreover, the distinct abundance patterns
of thick and thin discs exclude the possibility that the thick disc
could be a dynamically heated thin disc \citep[for example][]{Feltzing04}.

\citet{Bland-Hawthorn04} propose new instruments on 8-meter aperture
telescopes to `chemically tag' stars, to learn the dissipation
history of the MWg's disc. Many elements cannot form in normal stellar
evolution, so reflect the chemistry of the progenitor ISM.  Stars
of expected similar age in the Solar Neighbourhood travel in distinct
kinematical `moving groups', suggesting diffusion from common
birth sites. Tagging aims to use detailed abundance patterns that
are correlated with kinematical streams to track stars with identical
abundances back to specific formation sites. In the first test of
homogeneity, \citet{deSilva05} find negligible intrinsic scatter
in heavy element abundances (Zr, Ba, La, Ce, Nd) among Hyades cluster
members. The key uncertainty to resolve by study of other clusters
is if there is sufficient
chemical diversity across the disc to tag all sites uniquely.

Thick discs are ubiquitous.
While some early studies found them in early-type Sgs but not in types
Sc and later, \citet{Yoachim06} have found them in virtually all Sgs sampled.
They argue that thick discs result from early mergers of satellites.

\subsubsection{Warps and the outer edges of discs.}

Outer discs are seldom flat. The warp of the MWg was well described by 
\citet{Oort61} and extragalactic warps were mapped in HI with early radio interferometers
\citep[for example, M33, by][]{Rogstad76}.
As sensitivity improved, photometry of edge-on \citep[for example][]{Garcia02}
Sgs and the rotation curves of highly inclined Sgs \citep[for example][]{Noordermeer07}
often show warps in the HI disc beyond the starlight. For example,
the HI layer of the MWg thickens monotonically from $R_{0}$, and
beyond $1.5R_{0}$ warps up to $\sim6$ kpc ($\pm25^{\circ}$) from
the disc plane on one side at 30 kpc radius (the other side is much
flatter) \citep{Levine06}. Stars follow the warp.  Adding a tidal
wake in DM to the direct tides from the Magellanic Clouds \citep{Hunter69},
\citet{Weinberg06} explain most of these aspects by tides that
resonate in the outer gas disc of the MWg; the $m=2$ mode is an important
constraint that bears on dynamical alternatives to DM and on the shape
of the DM halo. The warp is very dynamic, `like a flag flapping
in the breeze'. The two other Local Group Sgs have prominent, asymmetric
warps that may arise from mutual tidal interaction, triaxial DM halos,
or accretion of cold gas.

While some discs truncate at $\sim4$ exponential
scale lengths \citep{deGrijs01}, others do not \citep[for example][]{Erwin05}.
Deep images of NGC 300 \citep{BH05}
show stars out to 10 radial scale lengths. The discs of a few
low-SB galaxies (section \ref{sub:Low-surface-brightness}) are seen
beyond 30 scale lengths, including the prototype of this class, Malin 1, which was
found by \citet{Barth07} to have a normal disc component
embedded within the 100 kpc extended component.
At such depths, modelling the contamination
of starlight by the noise of compact background galaxies is critical;
assumptions have been validated against deep galaxy counts
from HST.

\subsection{Global scaling relations: Mass-Velocity correlation in Sgs}

Section 3 showed that constraining Eg internal dynamics is frustratingly
difficult. The tensor Virial Theorem shows that massive Egs are flattened by velocity
anisotropy not rotation, but we have much to learn about their internals.
Considering the homology of global properties
of Egs leads to the Fundamental Plane, a global spatio-kinematical
correlation. 

However, Sgs provide a more favourable situation
for deciphering internal dynamics and mass distributions because
kinematical tracers in the cold, rotating disc can be measured out to
many radial scale lengths. We first consider global scaling relations
for Sgs, in analogy with study of Egs that uncovered the FP. In
section \ref{sub:Dark-matter-content1}, we consider what can be learned
about the distributions of light and dark matter from fits to
the measured radially variable rotation speed of the disc (the
rotation curve, RC). We consider what can be inferred on internal
properties from an optimum RC for a single galaxy as opposed to statistical
analysis of a sample.

Sg RC's are measured with emission lines from regions of
ionized hydrogen, and with the $\lambda21$ cm hyperfine spin-flip
transition in the ground-state of hydrogen (HI). The latter line
is unobscured by dust, and \citet{Tully76} measured its full width integrated
across an Sg disc to establish a correlation between the rotation
speeds of discs and their B-band luminosities. This Tully-Fisher relation
(TFR) is the Sg analogy and precedent of the FP. Its small scatter
spurred work on the extragalactic distance scale, galaxy peculiar
velocities, and 3D cartography of the local universe. Here, we consider
what can be inferred about homology of the global properties of Sgs
from the existence of, and small scatter in, the TFR.

The TFR establishes that the disc rotation speed (section \ref{sub:Evidence-from-other})
--- defined optimally \citep{Verheijen01,Noordermeer06} by the `flat
part' of the RC (although RC's are not particularly flat, section 5)
--- correlates with total baryonic mass --- as measured
originally from the B-band luminosity and now including ISM \citep{McGaugh01}
to account for all baryons.
The tightest correlation
uses K-band luminosity and HI to map baryonic mass while minimising
corrections for dust extinction and variable SF to find \begin{equation}
V_{\rm{flat}}\propto L_{\rm{K}^{\prime}}^{x}(\rm{baryonic}),\end{equation}
with \citep{Verheijen01,Noordermeer06} $x=4$. One converts to mass
with an assumed $\Upsilon_{\rm{d}}\equiv[M/L]_{\rm{d}}$. $\Upsilon_{\rm{d}}$
is constrained by galaxy disc colours (from, for example \citealt{Bell01}'s
grid of stellar population models, to yield $x\sim3.5$) and by a
universal mass-discrepancy relation \citep{McGaugh04} that is consistent
with both colours and standard IMF's. $\Upsilon_{\rm{d}}$ seems
to vary little among different Sgs \citep{Bell01,McGaugh04}.

This tight correlation, coupled with its independence of parameters
such as disc size and mean SB \citep[for example][]{Courteau99},
shows that the internal dynamics of Sgs do not depend on how baryons
are distributed within the potential. Indeed, \citet{McGaugh05a} finds
the same behaviour in low-SB galaxies (LSBg's) --- those whose average
SB is fainter than 40\% of the moonless night sky \citep{Bothun97,Impey97}.
We adopt the conventional explanation that the fundamental homologies
of Sgs arise from ubiquitous DM halos, with baryons simply `garnishing'
the result. An alternative interpretation, MOdified Newtonian Dynamics
\citep[reviewed by][]{Sanders02}, proposes to alter Newtonian acceleration
at the small value reached toward the visible-band edges of galaxy
discs where RC's tend to flatten. Whether or not MOND is real physics, \citet{McGaugh04}
finds that its dependence on acceleration gives the tightest possible
description of the discrepancy between baryons and Sg RC's, with scatter arising
only from observational uncertainties.

\begin{figure}
\begin{centering}\includegraphics[scale=2]{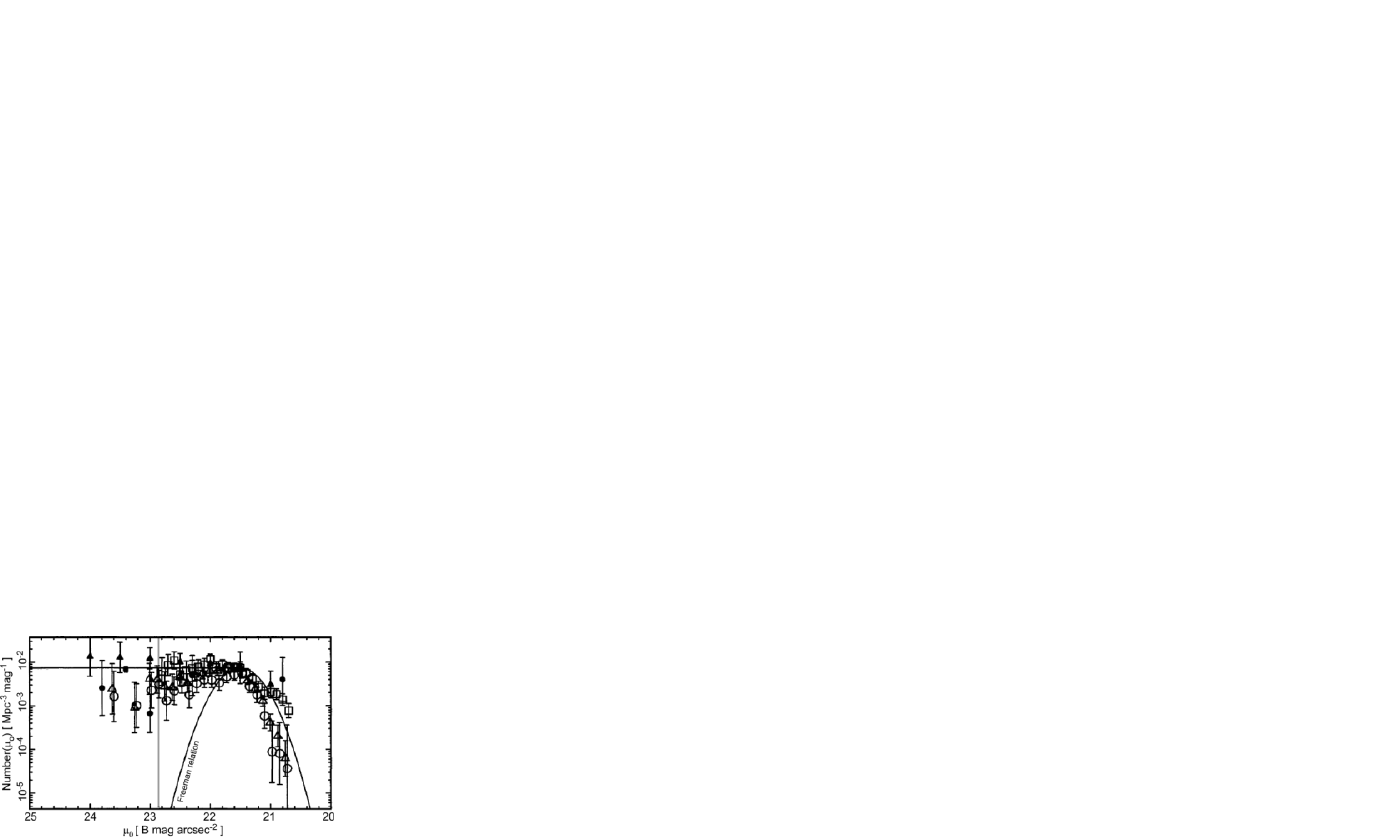}\par\end{centering}
\caption{\label{fig:Space-density-of}Number density of galaxies versus
their central SB $\mu_{0}$. LSBg's start
slightly to the right of the vertical grey line (the SB of the darkest
night sky), and extend to the left where their outer regions can currently be detected down
to $\sim10^{-3}$ of this limit. They can have the same blue luminosity
as high SB galaxies, but span a much larger area on the sky. Freeman's selection
for high SB galaxies is at right.
This figure
originally appeared in the Publications of the Astronomical Society of the
Pacific \citep[PASP, 109, 745]{Bothun97}.  Copyright 1997,
Astronomical Society of the Pacific; reproduced with permission of the
Editors.}
\end{figure}

LSBg's are the strongest test of the homology of Sgs through
the TFR; they make up to 30\% of the gaseous mass of the universe,
a negligible fraction of its stellar mass \citep{Driver07},
and at least half of the total galaxy population (Figure \ref{fig:Space-density-of}).
While mostly detected as gas-rich Sgs in the field (a few being
AGN hosts) and in poor, spiral-rich clusters, a few dwarf Egs of low SB
have been found. LSBg's challenge $\Lambda$-CDM by testing the proposition
that baryonic luminosity functions are biased toward high-SB galaxies.
To focus on this problem, we now consider Sg RC's and DM in
more detail.

\subsection{\label{sub:Dark-matter-content1}Dark matter content}

\subsubsection{\label{sub:Evidence-from-other}Evidence from galaxy rotation curves.}

Contrasting the limited work on DM in Egs, the RC's of gas in the
discs of spiral and S0 galaxies provide compelling and direct evidence of
DM \citep[for example][]{Broeils92,Cote00} because, at sufficient
radius, the DM halo mass is responsible for most of the observed rotation.
While evidence for important DM originated from RC's of ionized gas
in regions of active SF \citep{Rubin78} (HII regions in astronomical
jargon), ambiguity lingers \citep[for example][]{Kent86} because
this gas is generally too faint to see at radii large enough to pin
down the DM halo. Specifically, one cannot distinguish between a maximal-mass
stellar disc (i.e., the largest $\Upsilon_{\rm{d}}$ whose rotational
contribution does not exceed the full RC; \citealp{Sackett97,Palunas00,Weiner01,vanAlbada86})
and a sub-maximal disc of lower $\Upsilon_{\rm{d}}$ ($\sim6.3$
according to \citealt{Bottema93}'s study of stellar velocity dispersions)
plus a more massive DM halo \citep[for example][]{Courteau99,Pizagno05}.
This stellar disc --- DM halo mass degeneracy exists from S0 to Sd
galaxy types.

While the radial SB profile tracks disc baryons,
normalization by $\Upsilon_{\rm{d}}$ depends on vagaries of
stellar population models and, in discs, on the inevitable presence of dust.
Light comes mainly from the upper IMF,
in discs from young hot stars that lie near dusty regions. But mass comes from numerous low mass
stars. Uncertainty in the universality of an IMF established from
the Solar Neighbourhood makes $\Upsilon_{\rm{d}}$ hence disc mass
uncertain, thence the rest of the RC that is ascribed to the DM halo.

Evidence for DM and its isolation from baryons is compelling in galaxies
with the most extended HI, where most maps from radio interferometers
such as the \textit{VLA} and Westerbork Synthesis Radio Telescope show
a flat or increasing RC at the largest measurable radius
\citep{Rogstad78,Bosma81,Noordermeer07}.
If the RC spans sufficient radii, one can see variations in the
DM/baryon ratio.
To establish their individual contributions
one can then assume constant $\Upsilon_{\rm{d}}$ (questionable)
for baryons, and a simple radial distribution of DM, usually an
isothermal sphere or the cuspy \citet[NFW]{Navarro97} halo profile 
motivated by simulations (section 5.3.1) but with little observational
support.
In analogy to the relative clarity of bulge-disc decompositions when
the observed SB profiles span
both bulge- (inner) and disc-dominated (outer) regions, RC's that
span radii from where bulge+disc dominate (inner) to where DM halo
dominates (outer) allow robust decomposition of baryonic (bulge+disc+gas)
and DM components. 

\begin{figure}
\begin{centering}\includegraphics[scale=0.41]{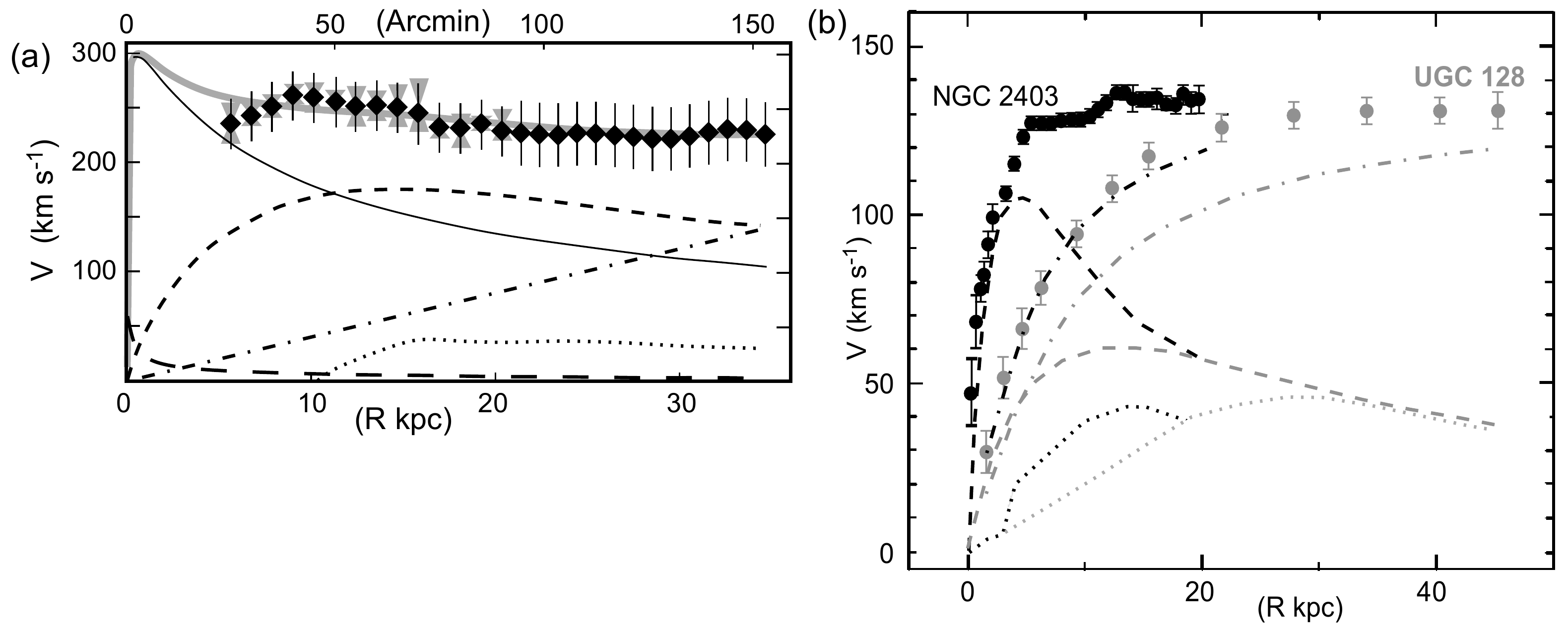}\par\end{centering}
\caption{\label{fig:The-rotation-curve}RC decompositions of (a)
M31 (grey line and \fullcircle\ \citealt{Carignan06}) with a central 
$10^{8}$ \Ms\ supermassive black-hole (\longbroken) \citep{Bender05} and bulge stars (\full); 
(b) a high SB galaxy (NGC 2403, black lines and \fullcircle) and a LSBg
(UGC 128, grey lines and \fullcircle, adapted from
\citealt{deBlok96}). Shown in both panels are DM halo (\chain), stars in the
disc (\broken) and gas (\dotted).}
\end{figure}

RC's from long-slit spectra often jump to 200-300 \kms\
within a few hundred pc radii.
The rise to and location
of peak rotation correlates with starlight concentration at the
nucleus, arguing that bulge mass dominates dynamics \citep{Noordermeer07}.
RC's often decline 10-20\% from the peak before flattening,
falling fastest in luminous galaxies. For example, Figure \ref{fig:The-rotation-curve}a
shows \citet{Carignan06}'s data and maximal-disc decomposition of M31 within
35 kpc radius, which yields $M_{\rm{stellar}}=2.3\times10^{11}$ \Ms, 
$M_{\rm{HI}}=5.0\times10^{9}$ \Ms, and $M_{\rm{dm}}=1.1\times10^{11}$ \Ms; 
this galaxy has negligible HI inside 5 kpc radius. DM dominates
only at $\gtrsim30$ kpc. 

Indeed, the trend is for high luminosity discs like that of M31 to
be maximal in stars, whereas low luminosity galaxies and LSBg's are
dominated by DM \citep[for example][]{deBlok97,Kranz03}. 
Figure \ref{fig:The-rotation-curve}b compares
mass decompositions of high- and low-SB galaxies.
The DM halos (\chain) are very similar despite completely different baryon
curves, and DM dominates throughout the LSBg. \citet{Oneil04}
map in HI nearly half of 81 LSBg's that have HI mass $>10^{10}$ \Ms; 
all RC's rise out to the last point measured. LSBg's have the
largest $\Upsilon_{d}$ measured. Their RC's \citep{McGaugh01} are
shallower, baryon surface densities are smaller at small radii
than those of high SB galaxies, and are best fit with minimal discs.

Flat RC's imply that the DM density must decline no faster than $1/r^{2}$.
Although no erstwhile `flat' curve has yet been seen to decline
at largest radius, the DM profile must eventually steepen to yield
finite mass. Halo extents probed with weak gravitational lensing \citep{Hoekstra04}
and faint H$\alpha$ emission \citep{Bland-Hawthorn97} both suggest
steep DM profiles beyond radii traced by HI. \citet{Persic96}
argued that all (mostly visible-light from ionized gas) RC's measured
by \citet{Mathewson92} outside the bulges of Sb-Sd galaxies
can be explained by a Universal Rotation Curve (URC) set only by galaxy
luminosity. Once a maximal stellar bulge contribution is removed,
\citet{Noordermeer07} find that agreement with the URC is often,
but not always, very good. At larger radii, all curves flatten but
are too complex to be described by a URC that depends only on galaxy
luminosity.

A recent estimate of mass distributions in Sg discs and DM halos
from the best RC's is \citet{Noordermeer06}, who analysed HII and
HI (out to 20 scale lengths) curves and multicolour surface photometry of 19 S0-Sab
galaxies. 
Rotational velocity totals 
\begin{equation}
V_{\rm{circ}} = 
\sqrt{\Upsilon_{\rm{b}}V(r)+\Upsilon_{\rm{d}}V_{\rm{d}}^2(r)+\eta V_{\rm{HI}}^2(r)+V_{\rm{DM}}^2(r)+V_{\rm{pm}}^2(r)}
\ \ \label{eq:rotation}\end{equation}
\begin{figure}
\begin{centering}\includegraphics[scale=5.1]{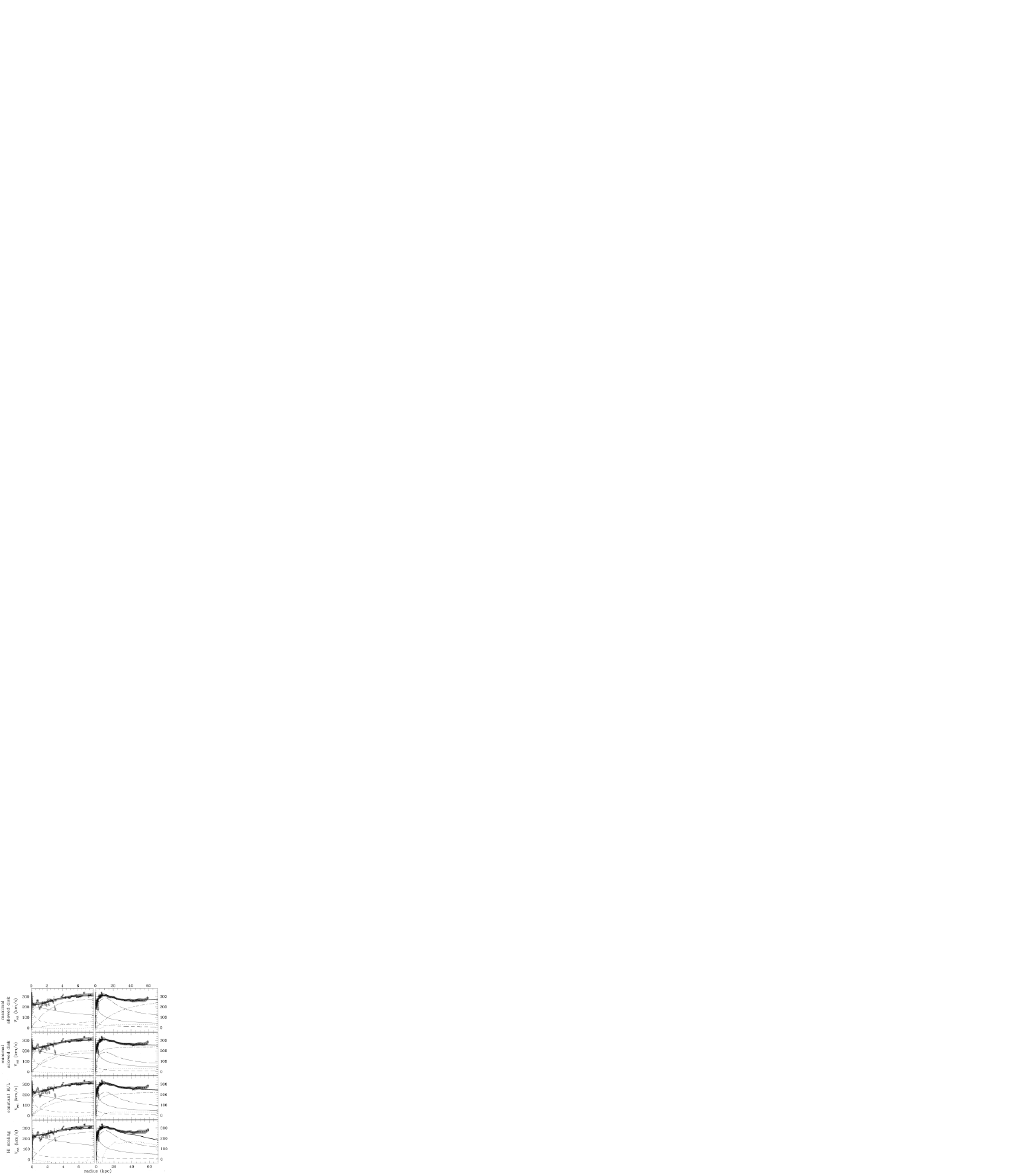}\par\end{centering}
\caption{\label{fig:rotcurvedecomp}\citet[][used with permission]{Noordermeer06}'s 
mass decompositions
of SA galaxy IC 356, using an isothermal DM halo. (Similar results
are obtained using an NFW DM profile.) The top two
rows assume maximal and minimal mass stellar discs, respectively; the DM
contribution is small for the former and large for the latter.
the third assumes constant $\Upsilon$, and the bottom assumes a scaled
HI disc of (baryonic) DM and \textbf{no} CDM halo. The left column
shows the central region with \opensquare showing kinematical
long-slit measurements from the ionized gas, the right shows the full
RC from HI spectra; the curve width in grey shows the range spanned
by the azimuthal average. Bumps and wiggles arise from non-circular
motions associated with spiral arms and their resonances. Decompositions are plotted with
the following lines: stellar bulge (\full) and
disc (\longbroken), gas disc (\dotted),
central supermassive black-hole (\broken), and DM halo (\chain).
Bold (\full) is the combined model RC (\ref{eq:rotation}).}
\end{figure}
with $\eta=1.4$ accounting for the mass of hydrogen and other abundant
elements (gas accounts for only 4\% of the total mass), $V_{\rm{pm}}(r)$
the contribution of a central point mass (perhaps a supermassive black-hole), and the
bulge $(M/L)_{\rm{b}}\equiv\Upsilon_{\rm{b}}.$ Both $\Upsilon_{\rm{d}}$
and $\Upsilon_{\rm{b}}$ are held constant, despite bluer discs
at larger radii that may result from different age/metallicity.
\citeauthor{Noordermeer07} derived mass models for 17 galaxies; Figure
\ref{fig:rotcurvedecomp} shows decompositions for
IC~356. A maximal stellar bulge contributes little farther out; the
main fitting parameters are the disc scale length and an uncertain
$\Upsilon_{\rm{d}}$ that yields an uncertain DM halo profile. Nevertheless,
the outermost 20 kpc of this RC (and others in his dataset) show clear
need for a DM halo. The bottom panel is the best fit without a halo,
boosting disc mass density $10-50$ fold \citep[using an additional baryonic disc component,][]{Pfenninger05}.
While residuals are within uncertainties and the fit at $\lesssim30$
kpc reproduces RC bumps and wiggles of unknown dynamical significance,
it is clearly untenable beyond. (He notes that ionization by the metagalactic
radiation field may decouple the DM from gas, \citealt{Bland-Hawthorn97}.)

A satisfactory fit uses a DM halo with isothermal profile and a maximal
stellar disc (top panel) of larger $\Upsilon_{\rm{d}}$
than expected from visible-band colours. An alternative, excellent
fit (second panel from top) uses a \textit{minimal} stellar disc with
reasonable $\Upsilon_{\rm{d}}$ and an NFW DM profile. Hence,
definitive assessment of the DM halo profile, and its contribution
relative to the disc at maximum rotation speed, is possible only if
$\Upsilon_{\rm{d}}$ can be found independently. As Noordermeer
notes, the most direct constraint on disc mass would come from the
velocity dispersion of stars in the outer disc. One would need the
very largest telescopes to measure the $\sim10$ \kms\
velocity dispersion of stars at that very low SB. His minimal
discs are consistent with simulated DM halos. But, his data
show a larger range of central concentrations and his fitted DM halo
is therefore more concentrated than simulations find. He finds that core radius
increases with scale length, independent of the choice
of minimal/maximal discs. 

\citeauthor{Noordermeer06}'s study shows the advantages of pursing
RC's to the largest possible radius. A complementary approach is to
constrain DM and baryonic contributions through statistical analysis
of many Sg RC's that range over galaxy mass, SB, and visible-band
diameter. Differing contributions of disc or DM halo to the maximum
rotation velocity give different patterns of correlated residuals
around the mean TFR. For example, \citet{Courteau99} studied a sample
of luminous, non-barred galaxies. They found little correlation between
residuals from their version of the TFR, specifically $\Delta\log V_{2.2}$
versus $\Delta\log R_{\rm{exp}}$, with $R_{\rm{exp}}$ the scale
length of the exponential disc and $V_{2.2}$ the velocity at $2.15R_{\rm{exp}}$
where rotation peaks in a pure exponential disc. If discs are maximal
with no DM halo inside $2.15R_{\rm{exp}}$, residuals should correlate
with $\Delta\log V_{2.2}/\Delta\log R_{\rm{e}}=-0.5$. From this
discrepancy and the behaviour of other residuals, they conclude that
the DM halo contributes 60\% of the total mass within $2.15R_{\rm{exp}}$,
so even high SB galaxies are not maximal discs at that radius. This largely
model independent result has been further reinforced in barred Sgs
\citep{Courteau03}.

\subsubsection{\label{sub:Evidence-in-the}Evidence in the MWg.}

We are supposedly surrounded by DM, but tracers through the halo are
hard to identify. \citet{Ollig01} note another frustration:
the local RC of the MWg $\Omega(R_{0})$ is uncertain. Errors on
`consensus' values are overwhelmingly systematic, so their propagation
is complicated. Thankfully, parallaxes of stars orbiting close to
the supermassive black-hole at the centre of the MWg will soon measure $R_{0}$ to dispense
with some of these irritations.

\paragraph{Microlensing surveys}

Light variability curves of background stars being microlensed by
MWg halo objects have been obtained by the MACHO \citep{Alcock00}, OGLE
\citep{Udalski94}, and EROS \citep{Aubourg93} projects
as they stare at the LMC (MACHO) or at the MWg bulge (others). Microlensing
measures directly the mass distribution along the l.o.s. As the Earth's
orbital motion causes a parallax shift in the observed light curve, the duration
of the microlensing event (the time to cross the Einstein ring of
angular width $\propto\sqrt{m}$) is set by three parameters that
are degenerate observationally: the transverse velocity of the lens
to us, the ratio of lens to source distances, and the lens mass $m$.

Surveys of microlensing toward the LMC have produced somewhat inconsistent
results and contradictory interpretations. MACHO toward the LMC finds
fewer events than expected if the total MWg halo mass arose solely from
compact objects of mass $m$. Lensing optical depths $\tau\sim3\times10^{-7}$
based on $\sim13-17$ events suggest that 20\% of the MWg halo mass
is in $0.5_{-0.3}^{+0.2}$ \Ms\  objects \citep{Alcock00,Bennett05}.
However, \citet{Novati06} proposes that many MACHO events are actually
self-lensing by the LMC halo, thereby lowering the estimated DM halo
fraction. Moreover, the EROS-2 collaboration finds a microlensing
$\tau$ only 10\% that of MACHO \citep{Jetzer05}. Proposed new microlensing
surveys will eventually resolve these controversies.

The OGLE/EROS dataset of more than 100 events provides the average
$\tau$ along several sightlines toward the MWg bulge and the distribution
of event durations. Several groups \citep[for example][]{Bissantz02}
have used these data with far-IR $\lambda240\,\mu$m maps of the MWg
bulge along sightlines through the stellar bar to argue for a near
maximal stellar disc, as found in other Sgs (see section \ref{sub:Evidence-from-other}).
For example, \citet{Bissantz02} reproduce the MWg RC out to 5 kpc
radius without a DM halo, unsurprising given the small radius probed,
and use a near-maximal stellar $\Upsilon_{\rm{d}}$. Streaming motions
associated with a triaxial bar account naturally for the $>40\%$
fraction of lensing events that exceed 50 days \citep{Evans02}. In
short, while microlensing surveys of the bulge have not located a
DM halo, they have helped to constrain the inner structure of the
MWg.

\paragraph{Stellar densities: disc and halo populations}

The local mass density can be estimated from the Jeans equation by
assuming that the disc potential is steady state, and is separable
into $r$ and $z$ motions because vertical variations in mass exceed
greatly radial ones. One then obtains\begin{equation}
\frac{\partial}{\partial z}\left[\frac{1}{\nu}\frac{\partial(\nu\overline{v_{\rm{z}}^{2}})}{\partial z}\right]=-4\pi G\rho(R_{0},z)\end{equation}
with $R_{0}=8.0\pm0.4$ kpc the solar radius. Measuring the \textit{volume}
variations of star density $\nu(z)$ and $\overline{v_{z}^{2}}(z)$
above the MWg plane, yields $\rho(R_{0},z)$ \citep{Eisenhauer03}.
Note that disc star counts must be differentiated three times, once
to obtain $\nu(z)$ and twice in (5.3), strongly amplifying observational
errors. \citet{Oort32} obtained $\rho(R_{0},z=0)\approx0.15$ \Ms\ 
pc$^{-3}$. Hipparcos 65 years later has halved this
to $0.076\pm0.015$ \Ms\  pc$^{-3}$ \citep{Creze98}.

Most halo stellar remnants are old white dwarfs that have cooled below
4000 K.
Counts within 1 kpc height yield a halo stellar mass 
density $\rho_{*}\sim0.044$ \Ms\ pc$^{-3}$. 
Deeper surveys are underway, using
kinematic cuts to isolate halo and disc white dwarfs, and selecting for
bluer colours caused by molecular hydrogen (H$_{2}$) absorption in
the red; cool white dwarfs have no other absorption lines. 
From 34
with halo kinematics, \citet{Oppenheimer01} conclude that such
remnants comprise 2\% of the halo mass locally.
However, \citet{Reid01} argue
that this sample mixes at least two kinematical populations, with
20\% of `halo' white dwarfs actually in the thick disc (section \ref{sub:Chemo-dynamical-clues-in}).
Because their disc density declines off the plane, the 
true white dwarf halo population
would be too small to account for much DM. This and the MACHO results
exclude a maximal baryon thick disc at least locally.

The disc stellar \textit{column} density within a few scale heights
of the plane is better determined because it requires \textit{only}
a double differentiation of the star counts\begin{equation}
\Sigma(<z\,\rm{kpc})=-\frac{1}{2\pi G\nu}\frac{\partial(\nu\overline{v_{\rm{z}}^{2}})}{\partial z};\end{equation}
\citet{Oort32} found $\sigma(0.7)\sim90\,\rm{M}_{\odot}$ pc$^{-2}$; \citet{Creze98}
review the methodology and efforts to refine his limit. \citet{Gould97}
use deep HST images to count common disc M-dwarf stars far
above the disc plane, normalize these to the locally determined density,
and obtain $\sigma(1)=26\pm4$ \Ms\  pc$^{-2}$. \citet{Ollig01}
obtain $\sigma(1.1)=35\pm10$ \Ms\  pc$^{-2}$ and show that the
main uncertainty comes from the observationally restricted height
into the halo; reliable tracers are sparse, blended with the populations
of the discs, and cannot reach the several kpc height required to
constrain tightly the DM halo.

Only beyond 30 kpc do we recover DM tracers, now dwarf satellites,
GCs, and a smoothly distributed population of evolved RR Lyrae stars
beyond 60 kpc \citep{Ivezic04}, their kinematics all implying DM
\citep{Mateo98}. Examining these tracers, \citet{Ollig01} use mass
models to conclude that the density toward the centre of the MWg's
DM halo is uncertain by a factor of 1000. The relative contributions
of stars and DM to the force above the disc plane cannot be separated
easily, a `vertical disc-halo conspiracy'.

\paragraph{Gas surface density}

The MWg RC $\Theta(r)$ is constrained by HI data to $\sim10$ \kms\
internal accuracy. However, \citet{Lockman02} uncovers contradictions
in the patchwork of datasets that comprise this curve in textbooks.
He cautions us not to take it too seriously, especially between $1-2R_{0}$
where its very uncertain gradient yields a 100\% uncertain HI surface
density. The curve yields $v_{\rm{c}}^{2}/2\pi GR_{0}\sim226$ \Ms\ 
pc$^{-2}$, $\sim2.5$ times that derived from luminous material within
700 pc of the disc \citep{Ollig98}. Most DM must be beyond this radius,
in the halo. The IAU `consensus' rotational speed
at the solar radius, $\Theta_{0}\equiv\Theta(R_{0}=8.0\pm0.4\,\rm{kpc})=220$
\kms, exceeds slightly \citet{Ollig01}'s best estimate $200\pm10$ \kms.

The observed flaring of the HI layer beyond $2R_{0}$ is a constraint
with different dependence on $R_{0}$ and $\Theta_{0}$. If the ISM 
is pressurized only by thermally induced turbulent motions in the steady state,
hydrostatic equilibrium applies, so \begin{equation}
\frac{d\sigma_{\rm{g}}^{2}\rho_{\rm{g}}(z)}{dz}=\rho_{\rm{g}}(z)K_{\rm{z}}(z)\end{equation}
with gas velocity dispersion $\sigma_{\rm{g}}=9-10$ km~s$^{-1}$
and mass density $\rho_{\rm{g}}(0)\sim0.042$ \Ms\  pc$^{-3}$
\citep{Malhotra95}, and $K_{\rm{z}}(z)$ the vertical force per
unit mass. \citeauthor{Ollig01} estimate that neglecting pressures
in the local ISM from the magnetic field (which is frozen to the ionized
ISM) and cosmic rays (which are intense only near SF regions) makes
less than 10\% error in the thickness of the gas layer at $R_{0}$
and less at $2R_{0}$ where there is little SF.

\begin{figure}[H]
\begin{centering}\includegraphics[scale=2]{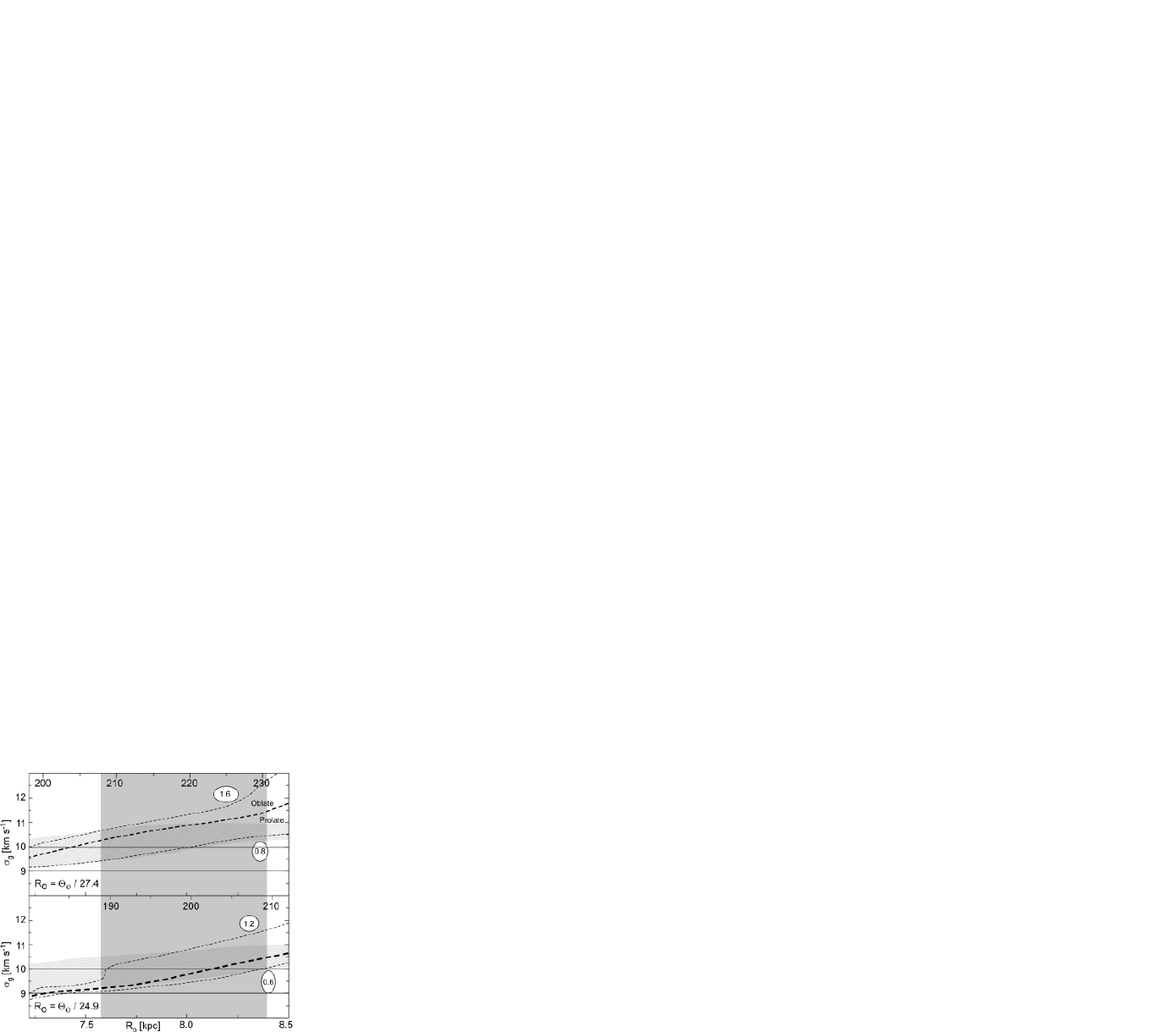}\par\end{centering}
\caption{\label{fig:ollingDMfig}(Simplified from Figure 3 of \citealt{Ollig01}.)
How constraints on MWg scaling constants relate to the gaseous velocity
dispersion $\sigma_{g}$ and DM halo flattening $q$. Each panel chooses
a $\Theta_{0}/R_{0}$, together spanning their uncertainties;
numbers along the top of each panel give the local rotation speed $\Theta_{0}(R_{o})$
in \kms. Gray shading spans current estimates of $R_{0}$ as obtained
from parallax measurements of the Galactic Centre \citep{Eisenhauer03}.
The vertical axis shows a large range over $\sigma_{g}$, with the
two solid (\full) horizontal lines at \citet{Malhotra95}'s
bounds. Hatched shading spans uncertainties in the column density
of the stellar disc. The dark dashed (\broken) curve shows the
spherical boundary between oblate and prolate shapes of the DM halo
and the lighter dashed (\dashed) curves show the range. }
\end{figure}

A plausible DM density distribution is an isothermal spheroid\begin{equation}
\rho_{\rm{dm}}(r,z;q)=\rho_{0}(q)(\frac{R_{\rm{c}}^{2}(q)}{R_{\rm{c}}^{2}(q)+r^{2}+(z/q)^{2}})\end{equation}
with $R_{\rm{c}}$ the dark halo's core radius, $q$ the flattening
from the $q=1$ sphere, and central density $\rho_{0}$ absorbs all
dependence on $q$. With $R_{0}$ in kpc and $\Theta_{0}$ in \kms.
\citet{Ollig01} find that\begin{equation}
\frac{\rho_{\rm{dm}}(R_{0},\Theta_{0})}{10^{-3}\,\rm{M}_{\odot}\rm{pc}^{-3}}=\frac{11.5+3.8\times(R_{0}-7.8)\pm2}{q(26.7R_{0}/\Theta_{0})^{2}}\end{equation}
The non-linear 
coupling between these parameters is illustrated in Figure \ref{fig:ollingDMfig}.

\paragraph{Implications for the DM halo of the MWg}

Motions of the Sagittarius dwarf galaxy permit either a prolate or spherical 
DM halo for the MWg \citep{Ibata01a}. However, a `Field of Streams' discovered
in the SDSS by \citet{Belokurov06} shows that the star stream
from this galaxy wraps around the MWg two or three times. \citet{Fellhauer06}
argue that this coherence requires a close to spherical inner DM halo.
(Section \ref{sub:Chemo-dynamical-clues-in} discusses MWg star streams.)
\citet{Ollig01} use \citet{Malhotra95}'s data on the vertical velocity
dispersion of HI in the Solar Neighbourhood to constrain flattening
of the DM halo; for allowable limits on $R_{0}$ and on $\Omega(R_{0})$,
Figure \ref{fig:ollingDMfig} shows that the MWg halo is spherical
or barely oblate and has density in the Solar Neighbourhood $(11\pm5)\times10^{-3}$ \Ms\  
pc$^{-3}$ (0.42 GeV/c$^{2}$ cm$^{-3}$). Note that this is
within the $\pm0.015$ \Ms\  pc$^{-3}$ uncertainty of the stellar
density from Hipparcos as mentioned above \citep{Creze98}, avoiding
substantial DM in the local disc.

\subsection{Modes of ongoing SF}

One cannot hope to model the evolution of galaxy baryons without understanding
SF quantitatively. Unfortunately, dust obscures most SF at visible-light wavelengths.
Stars form within
an active, multi-phase ISM whose dynamics are only starting to become
tractable as long-IR and sub-mm detector arrays have grown in number of
pixels, and as
3D hydrodynamical simulations have become more realistic. It
now appears that stars tend to form today in the
coldest, molecular phase in galaxy discs. SF is stimulated above average
levels by any non-axisymmetric instability that sweeps through the
disc, usually triggered by close tidal passage or merger of a
neighbouring galaxy. This has been quantified observationally \citep[for example][]{Larson78},
and simulated in detail \citep{Mihos96}. Examining SF
is beyond our scope, see \citet{Evans99} for example.
Here we concentrate on its global aspects that are crucial
to modelling galaxy evolution.

\subsubsection{\label{sub:Quantification-of-ongoing}Quantification of ongoing SF.}

How can SF rates be measured accurately in different types of galaxies and environments?
SF manifests mainly through the high luminosities of the most massive
MS stars formed. So, its unique signature is brief
(few Myr), intense UV ionizing radiation from hot stars that is
absorbed by gas in the star forming complex. Such HII
regions emit hydrogen recombination lines in the visible-band
(Balmer series), UV (Lyman), and IR (Paschen and Brackett), forbidden (i.e., electrical non-dipole
transition) spectral lines of partially ionized oxygen and other elements,
strong UV fluxes, free-free
radiation in the radio continuum, and synchrotron radiation from energetic
particles produced in SNe \citep[for example][]{Dopita05}. 
In fact, because stars form in dusty GMCs, 
graphitic and silicaceous grains quench UV and visible radiation, and
recycles much radiation into IR and mm wavebands. 
The signatures of SF at shorter wavelengths
are easily extinguished. Thus IR photometry from space of heated dust
is most effective at measuring the true SF rate. Abundances in the gas-phase
are set both by astrophysical processes within stars and by chemistry
on the surfaces of grains and gas molecules. 

The SF rate is set by the energetic photons
(and particles) emitted by the most massive stars.
Determining the total mass of a star burst therefore
requires very uncertain extrapolation of the IMF to the lower masses
that constitute most of the event. We encountered this problem in
section \ref{sub:Evidence-from-other} when constraining stellar $\Upsilon_{d}$.
What is observed as ongoing SF is the integrated result of
many HII regions of diverse age, mass, size, pressure, and chemical abundance.
K-band luminosity measures the integrated light of
older stars  to establish 
the efficiency of SF across a galaxy\begin{equation}
b=\frac{\rm{SF~rate}_{\rm{present}}}{\left\langle \rm{SF~rate}_{\rm{past}}\right\rangle }\end{equation}
(for example $\sim5\%$ in isolated dwarf galaxies), with
average past SF rate $\left\langle \rm{SF~rate}_{\rm{past}}\right\rangle $.
H$\alpha$ flux can be converted into the total flux of ionizing photons
using recombination line transition probabilities, determining the
rate of forming massive stars; by assuming a form of the IMF, $\rm{SF~rate}_{\rm{present}}$
can be inferred \citep[for example][]{Kennicutt94}. More readily
observable {[}O~II]$\lambda\lambda$3726,3729 is often a surrogate
for H$\alpha$ in studies of high-redshift galaxies, but is less certain
because of the strong temperature dependence of collisionally excited
forbidden lines. \citet{Kennicutt98} reviews determinations of SF rate;
he shows that the present SF rate characterized by parameter $b$ varies
strongly with Hubble type: $b$ is often $<0.07$ for Sa discs, 0.3
for Sb discs, and 1 for Sc discs, all with large dispersions. If the
SF rate is taken to decline exponentially, then the e-fold timescale
is only $\sim2$ Gyr for Sa's, rising to $\sim5$ Gyr for Sb's.

Can global SF rates be parametrized for semi-analytic models (and for
mixed N-body and gas dynamical models) of galaxy evolution?
Are such prescriptions sensitive to environment? For example, tides
between galaxies in rich clusters occur more often than in low-density
environments. But such encounters at the large velocity dispersion
of the cluster will be `impulsive', with relatively small
effect compared to the slow, nearly parabolic encounters of galaxies
in groups. Thus, there is a complex trade-off between destabilization
from frequent impulses (`harassment', Moore
\etal 1996), and the effects of slow encounters, as modelled for
example by \citet{Mihos96}.

Galaxies undergo major bursts of SF,
often triggered by tides from a passing companion. A galaxy is
star bursting mildly if $b=2-3$ and strongly if $b>10$.
Bursts trigger on a dynamical collapse time: a 10 kpc radius with
$10^{10}$ \Ms\  collapses in 200 Myr; mergers can speed this up.
$b$ is measured most effectively by combining SDSS visible-band and
\textit{GALEX} space-UV photometry; thus \citet{Salim05} find that
$\sim20\%$ of galaxies with $\sim10^{8}$ \Ms\  to $\sim5\%$ of
those with $\sim10^{11}$ \Ms\  have had a significant star burst
within the last Gyr. Similarly, \citet{Heckman05} have studied 74
nearby galaxies with high far-UV luminosities ($3-30$ \Ms\  yr$^{-1}$
converting into stars) and find that galaxy mass and UV SB correlate
inversely. Using the SDSS, \citet{Brinchmann04} find that in the
local universe most stars form in Sgs that exceed $10^{10}$ \Ms\ 
($\sim15\%$ being AGN) in regions $100-500$ pc across, with $\sim20\%$
in mild star bursts and 3\% in intense ones $(b>10)$.

`Passive' SF in high SB galaxies occurs within GMC cores at densities
$10^{2}-10^{3}$ cm$^{-3}$; section \ref{sub:Low-surface-brightness}
discusses SF in LSBg's. Young stars are dust enshrouded for their
first 1-2 Myr, and radiate most ionizing UV within the first $\sim6$
Myr, thereafter photo-dissociating their molecular envelopes over
$\sim10$ Myr. For the next $50-100$ Myr, burst luminosity
increases by only 10\% \citep{Leitherer99} as GMC destruction
ends SF.

Broad-band colours of visible light age SF crudely \citep{Whitmore95}.
Sensitivity to SF within the last 100 Myr is much improved by adding
the \textit{GALEX} NUV and FUV filter bands (Figure \ref{fig:Spectral-evolution-of}a).
More accurate timing of an ongoing or recently ended burst follows
from details of the gaseous emission-line flux ratios. \citet{Gonzalez99}
show that the He~I/H$\beta$ ratio varies with the age of the
burst: $0.10-0.12$ until 4 Myr, then declines steeply below $0.05$.
The {[}O~III]$\lambda$5007/H$\beta$ ratio starts near 4.5, declines
to $\sim1.2$ by 2.5 Myr, and reaches 0.5 by 3 Myr. Near-IR images of
hydrogen and helium emission can map SF in dust shrouded discs. In
the MWg, emission from transitions between barely bound atomic levels
of hydrogen and helium are mapped with radio interferometer arrays
across very extinguished parts of GMCs \citep[for example][]{Gordon03}.

\subsubsection{Sites of SF and dependence on gas density.}

Where do stars form? How does gas surface density
set the disc SF rate \citep{Schmidt65}? 
How the rate of `passive' SF depends on the surface density
of hydrogen $\sigma_{\rm{gas }}$, and the threshold above which
stars can form, was estimated empirically by \citet{Kennicutt98}
from the discs of small-bulge Sgs as\begin{equation}
\sigma_{\rm{SF}}\propto\sigma_{\rm{gas}}^{1.4}\end{equation}
His value agrees with the theoretical `disc instability' criterion
for cloud collisions \citep{Toomre84}, although both \citet{Schaye04}
and \citet{Noordermeer06} have subsequently mapped ongoing SF in
large bulge high SB galaxies below half this value. \citet{Silk77} gives another
estimate\begin{equation}
\sigma_{\rm{SF}}\propto\sigma_{\rm{gas}}\Omega_{\rm{gas}}\end{equation}
with $\Omega_{\rm{gas}}$ the orbital speed of the gas clouds. \citet{Thilker05}
have used \textit{GALEX} to find that stars are forming in some galaxies
at $2-4$ times the radius of the visible-band disc with $\sigma_{\rm{gas}}$
below Kennicutt's thresholds as established at smaller radii. $\rm{SF~rate}_{\rm{present}}$
depends on the mass fraction in dense gas, whose value is set by turbulence
\citep{Elmegreen02}. \citet{Kennicutt98} finds\begin{equation}
\rm{SF~rate}_{\rm{present}}\,\rm{(M}_{\odot}\rm{\, yr}^{-1})=\frac{L(H\alpha)}{1.26\times10^{34}\rm{\, W}}\end{equation}
although this misses the, perhaps majority, SF that is extinguished
by dust.

Until SF ends with the conversion of all ISM into a warm phase, it can propagate over
a $\lesssim1$ kpc scale \citep{Zhang01};
this behaviour is evident in mergers and also as a clear sequence
of `super-shells' in the LMC where up to half of the stars form
adjacent to such high-pressure HII regions (true also in the MWg).

\subsubsection{\label{sub:Raw-material----}Raw material --- effects of the state
of the ISM.}

A major uncertainty in the physical state of the ISM and
also the spectrum from hot stars that bathes HII regions is
the depletion of elements from the gas phase onto
dust grains and polycyclic aromatic hydrocarbon (PAH, a planar-hexagonal
carbon lattice built from benzene rings) molecules formed in dusty
bubbles around AGB and RGB carbon-rich stars; the size distribution
of grains (power-law up to large sizes, declining exponential thereafter);
grain evaporation by thermal and perhaps chemical sputtering, grain
shattering, and shock heating from SN; the photo-dissociation of H$_2$ into HII;
and PAH photo-dissociation
in HII regions. Chemical abundances derived from emission lines
show how metals still in the gas phase vary across unobscured HII
regions. Most otherwise abundant elements, originally as gas, have
depleted onto dust grains; sulphur is one exception, but its emission
intensity depends on its temperature, abundance, and level of ionization.

The ISM of the MWg has been reviewed lucidly by \citet{Ferriere01}.
Despite $\sim1000$ generations of SF in the MWg disc each of $\sim10$
Myr duration \citep{Bland-Hawthorn04}, $\sim10^{7}$ \Ms\  of dust
and $4\times10^{9}$ \Ms\  of gas remain to extend SF to 30 kpc radius.
Gas subdivides into a cool layer of $\lesssim0.3$ kpc scale height,
and a warm neutral (Lockman) and ionized (Reynolds)
layer to 0.6 kpc. Half of its mass is molecular hydrogen
(H$_{2}$) within cores of GMCs of mass $(0.1-2)\times10^{6}$ \Ms. 
Velocity crowding of multiple clouds blurs our view. Hence,
over the last decade, less edge-on Local Group galaxies mapped in surveys with
the BIMA array (N. hemisphere) and 4-meter aperture NANTEN dish (S.
hemisphere) have provided our best views. \citet{Blitz06} review what
these surveys have revealed about the molecular
ISM, using emission from the CO molecule as a tracer for the molecular
gas. This gas is correlated with HI in that every GMC is found on
an HI filament but there are many filaments without CO. CO concentrates
toward galaxy centres, with exponential scale lengths comparable to
that of the stellar disc. To form stars, HI must clump to become
a filamentary GMC, increasing its surface density up to twenty-fold;
this conversion is almost complete in high-pressure, HI deficient
regions such as galactic nuclei.

HII regions \citep{Oey97} and wind bubbles from massive stars break
out of the GMC along its density gradient, expanding as the star burst
progresses until the gas pressure balances with the lower-density,
diffuse ISM. In balance, bubbles then coast in a momentum-conserving
state (in which they pass most of their time) that is set by the density
of the ISM phase with the largest volume filling factor, until the
swept-up shell of ISM fragments by turbulence. In this way and with
SN, the GMC core dissolves $\sim1$ Myr after SF is triggered \citep{Dopita06}.
A sequence of cluster ages and correlated CO emission in the LMC shows
that a cloud dissolves in $\sim20-25$ Myr \citep[and references therein]{Blitz06};
HII regions then pass $\sim\frac{1}{4}$ of their lives in a
quiescent phase.

The other half of the ISM mass in the MWg is HI in a complex \citep{Cox05},
multiphase topology that is too uncertain to model
evolving SF \citep[although, see the daring simulations of][]{deAvillez00}.
Instead, SF is usually modelled in a uniform pressure and density
ISM \citep[for example][and references therein]{Dopita05}. In the
disc, HI is often optically thick and the uncertain transverse component of its
non-circular motions makes uncertain
$\sigma_{\rm{gas}}$ across much of the MWg (section \ref{sub:Evidence-in-the}).
Cosmic rays, heavy
nuclei accelerated at the shock fronts of expanding supernova bubbles \citep{Ferriere01},
deposit a third of the energy into the ISM.

Once more than $\sim30$
successive SNe have detonated, their individual `chimneys' (in
simulations like \citealt{deAvillez00} rising 500 pc above the disc)
merge into a `superbubble' \citep[see the review of][]{Veilleux05}.
These x-ray emitters (thermal spectrum, typically $\sim10^{5-6}$ \Ms)
are surrounded by H$\alpha$-emitting shells up to $\sim1.5$
kpc diameter; the LMC has particularly clear examples. If either structure reaches high enough,
its thermalized gas redevelops momentum
down the pressure gradient, and the flow can entrain and `break out'
of the disc as a `galactic fountain': once above $\sim$250
pc altitude, it becomes Rayleigh-Taylor unstable from declining ambient
densities and eventually drips as descending
sheets that pump entrained ISM (including
metals, for example \citealt{Kenney99}) across a galaxy to
smear out primordial abundance gradients. 
Indeed, at $<1$ kpc \citet{Lockman03} finds dense HI clouds
that are too low mass to self-bind gravitationally; they have disc kinematics
so were presumably uplifted by correlated SNe and the resulting `fountain'.
\citet{Pidopryhora06} map a superb example $\sim7$ kpc distant
that reaches $\sim1$ kpc into the MWg halo, contains $\sim10^{6}$ \Ms\  
of hydrogen, and has total energy $\sim10^{53}$ erg.

ISM porosity to supernova blast waves sets the lateral extent
of a star burst, the chemical yield, and how efficiently gas is ejected
from a potential (see section \ref{sub:Feedback}).
The other important but poorly observed parameter \citep[for example][]{Veilleux05}
is the thermalization efficiency $\epsilon$ --- the fraction of mechanical
energy from all SNe not yet radiated before their blast waves merge;
it is degraded into turbulence thence entropic heating. $\epsilon$
depends on the pressures and scale heights of
the various ISM phases, and on the clustering of SNe. Estimated values
in star bursts vary widely because of uncertain ISM properties, from
$\epsilon=0.005$ (appropriate for a cloudy ISM with many GMCs and
a mix of SNe Ia and II) to $\epsilon=1$ (appropriate for a galaxy
bulge with only SNe Ia). ISM has been incorporated into semi-analytic
models of SF by scaling parameters from a three-phase description
\citep[see the review by][]{Cox05}. In this picture, a poorly constrained
topology of static cool clouds, warm interface, and pervasive hot
phase from intersecting supernova remnants forms and evolves toward $\epsilon=1.$
Blast waves evaporate clouds, and the resulting warm ablata balance
cooling from denser regions.

\subsubsection{\label{sub:Low-surface-brightness}SF in low-SB galaxies.}

\begin{figure}
\begin{centering}\includegraphics[scale=1.55]{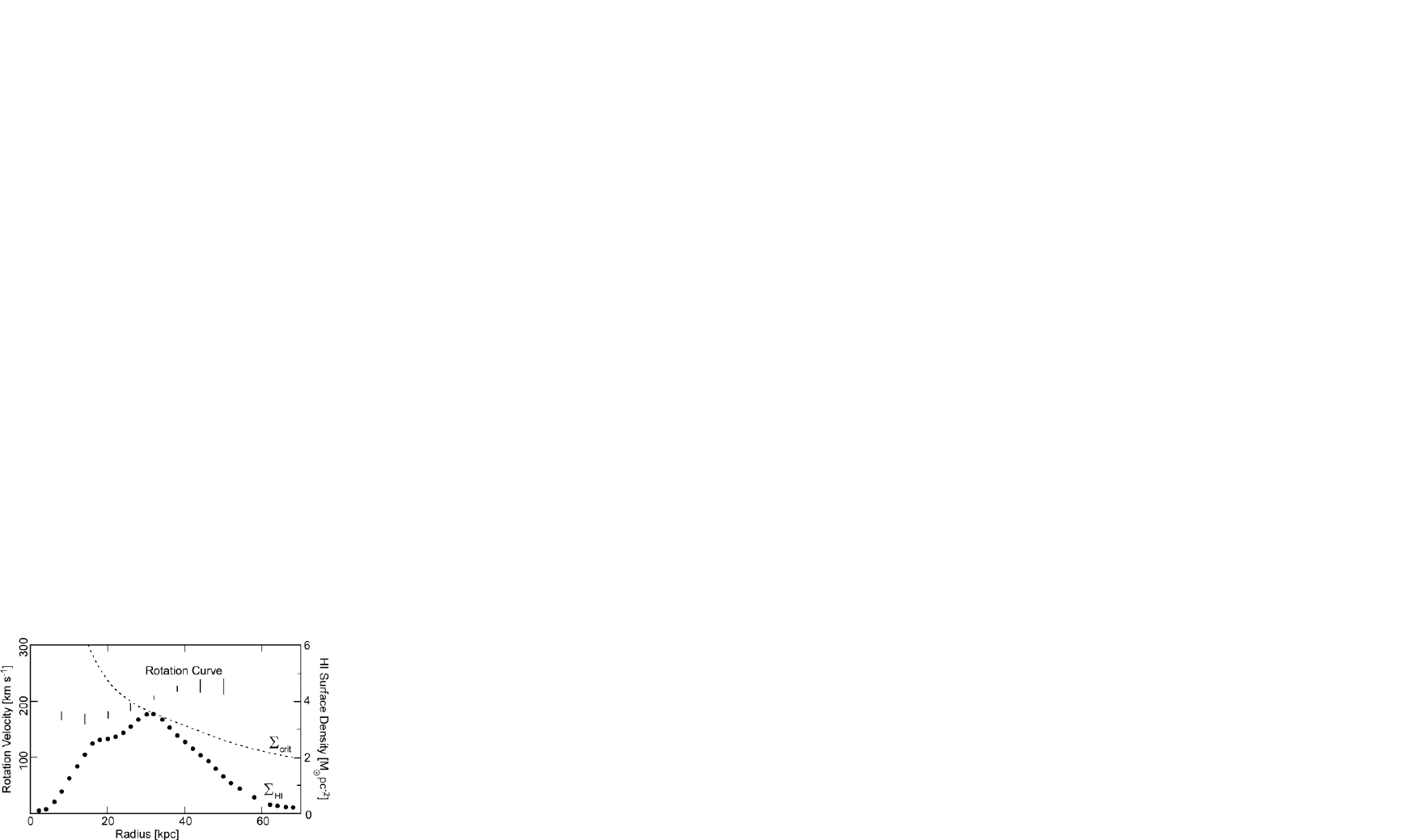}\par\end{centering}
\caption{\label{fig:Example-LSBg-rotation}RC (vertical lines)
and HI surface density (\fullcircle) of LSBg
UGC 6614. Only at 30 kpc does the surface density rise to meet
the critical value of \citet{Kennicutt98} (\dotted) to form stars.
This figure
originally appeared in the Publications of the Astronomical Society of the
Pacific \citep[PASP, 109, 745]{Bothun97}.  Copyright 1997,
Astronomical Society of the Pacific; reproduced with permission of the
Editors.}
\end{figure}

Selections inhibit discovery of LSBg's, so the correlations discussed
previously are only now being extended as targeted searches and surveys
like the SDSS uncover examples over the full range of galaxy mass,
in voids, and on the rims of matter filaments. Most LSBg's have
no bulge or bar and little dust.
Their blue colours (young stars hence slow, continuous SF), 
metallicities ($0.05-0.5$ solar, a few $>1$), and
gas content are less evolved than in high SB galaxies; they are not faded.
They are often gas rich yet have HI
densities below \citet{Kennicutt98}'s threshold for SF (for example
Figure \ref{fig:Example-LSBg-rotation}), and sparse CO suggesting
no GMCs. \citet{Mihos99} model their ISM and conclude that H$_{2}$
can be substantial within a few kpc radius but its warmth
$\sim30-50$ K inhibits SF. 
They have extreme HI masses for their blue luminosity, up to 10 times
those of high SB galaxies. 

\subsubsection{\label{sub:Jet-and-radio-lobe}Jet and radio-lobe induced SF.}

Jets can trigger some SF by crushing ISM through direct impact or thermal shocks
driven by the expansion of the cocoon/radio lobe \citep[and references therein]{Bicknell00,Fragile04,Odea04},
to explain aligned radio continuum jets and SF
(for example Minkowski's Object near NGC 541, \citealt{Croft06}).
However, in section \ref{sub:Feedback} we discuss suppression of SF (`feedback') by jets.
\\

In summary, disc baryons provide our best opportunity to trace motions
in DM dominated regions. 
We have shown how to measure the SF rate and how it depends on the
densities of the diffuse, multi-phase ISM. 
The diversity of SF complicates interpretation:
uncertainties about dust content and IMF make $\Upsilon$ uncertain, complicating isolation of the
DM component.

\section{\label{sec:Matter-Transfer-Within}Feedback}

To parametrize
models of the global evolution of star-forming galaxies, the key question
is to what extent does energy injected into the surrounding
ISM (through UV radiation and winds/SNe from massive stars, production
of heavy elements that alter the gas cooling rate, etc) either promote
or discourage continuing SF, and redistribute the remaining ISM among
its phases? The astrophysical processes responsible are denoted collectively
\citep{White78} \textbf{feedback}. Because it acts within the underlying
galaxy DM potential well, feedback alters the dependences of the SF and chemical evolution rates
on galaxy mass.

\begin{figure}
\begin{centering}\includegraphics[scale=0.28]{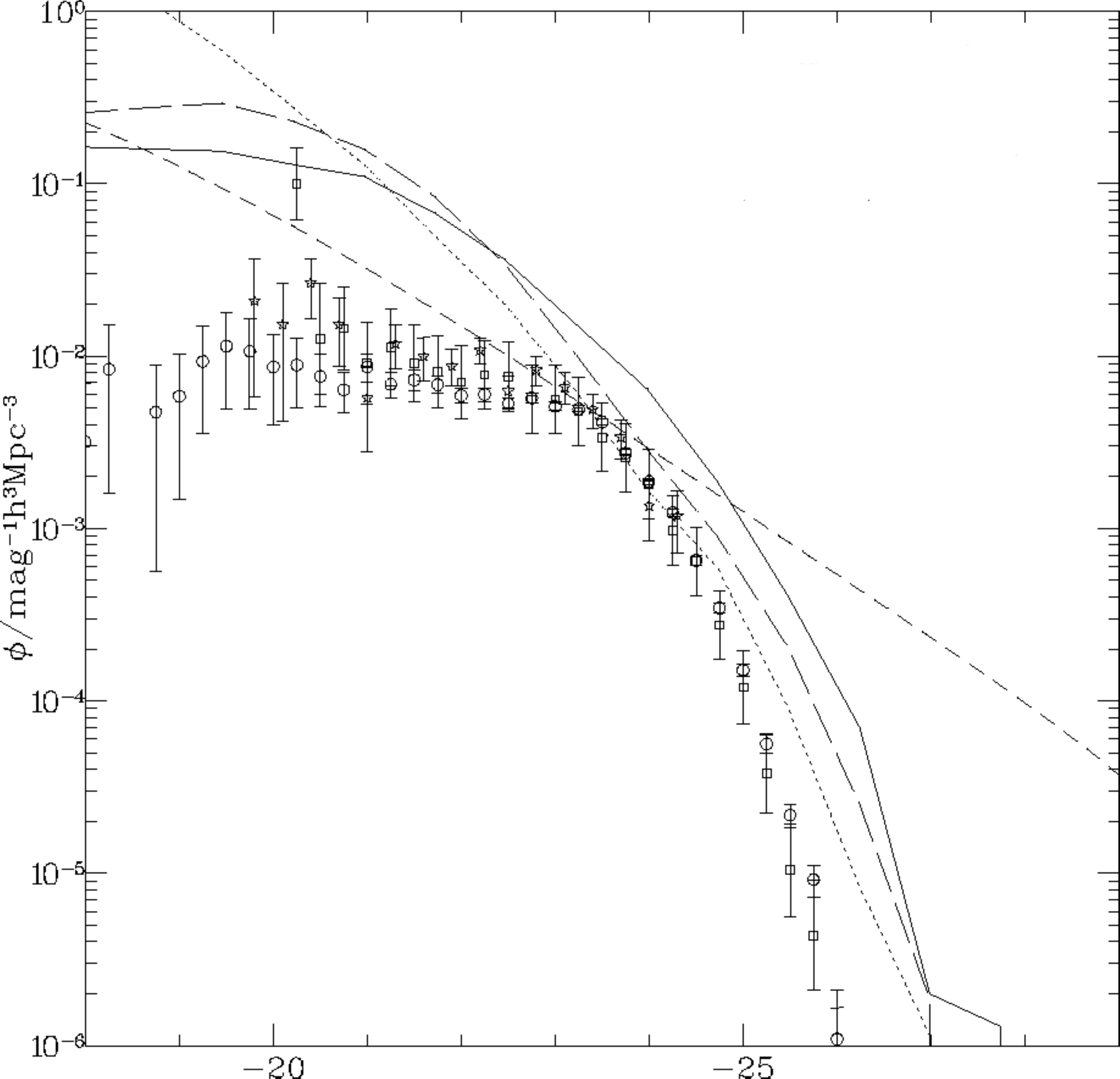}\par\end{centering}
\caption{\label{fig:Data-for-the}\citep[From][used with permission]{Benson03}. The
galaxy luminosity function $\Phi$:
the number of galaxies per million pc$^3$ volume and per
magnitude in apparent brightness versus the $-\log$ luminosity of a galaxy (in
units of its absolute magnitude, the MWg being $-20.7\pm0.2$).  Data
are \opencircle, \opensquare,
and stars with error bars. Shown are models of the DM halo-mass function
without feedback, (diagonal \broken), and with various feedbacks
mentioned in the text as curves (cooling (\dotted); photoionization
(\longbroken); merging (\full)).}
\end{figure}

The crucial r\^ole of feedback in galaxy evolution appears when
modelling the observed luminosity function (LF) of galaxies. In the
$\Lambda$-CDM scenario, the redshift dependence of the hierarchical assembly
of DM halos can be modelled straightforwardly because only gravity operates.
The DM halo mass distribution is predicted either from
N-body simulations \citep[for example][]{Springel05} or through the
analytical framework of the extended Press-Schechter theory \citep{Press74,Somerville99,Cole00}.
Because DM halos build by amplifying and merging an initial Gaussian
distribution of density fluctuations with a roughly scale-invariant
power spectrum that has grown into the non-linear regime, the resulting mass
distribution has an exponential high-end
and a roughly power-law low-end, as observed \citep{Schechter74}.
However, if one assumes that CDM is modified
by a single $\Upsilon$ to yield the baryonic mass distribution (but
recall in section \ref{sub:Kinematics:-the-fundamental} the tight $\Upsilon-\sigma_{\rm{e}}$
correlation in Egs), one finds that model and observed LF's disagree.
This model LF (\broken\ in Figure \ref{fig:Data-for-the})
has two problems: its low-luminosity
slope over-predicts faint galaxies, and it declines steeply at much
higher luminosity (mass) than observed. Adding a simple prescription
for gas cooling (\dotted) further over-predicts the faint end
because lower mass halos are cooler virially. However, cooling helps
at the bright end because gas cools slowly in the
massive, virially hot DM halos. Low masses are further suppressed
by re-ionization after the first stars and AGN form
(\longbroken). Adding mergers of small galaxies into larger ones (\full) further boosts
the luminous end.
Conversely,
this end can be reduced by suppressing cooling in massive Egs with AGN via reheating shocks
from their jet-emitting supermassive black holes \citep{Croton06,Bower06}, see Appendix B.
 
\begin{figure}
\begin{centering}\includegraphics[scale=0.4]{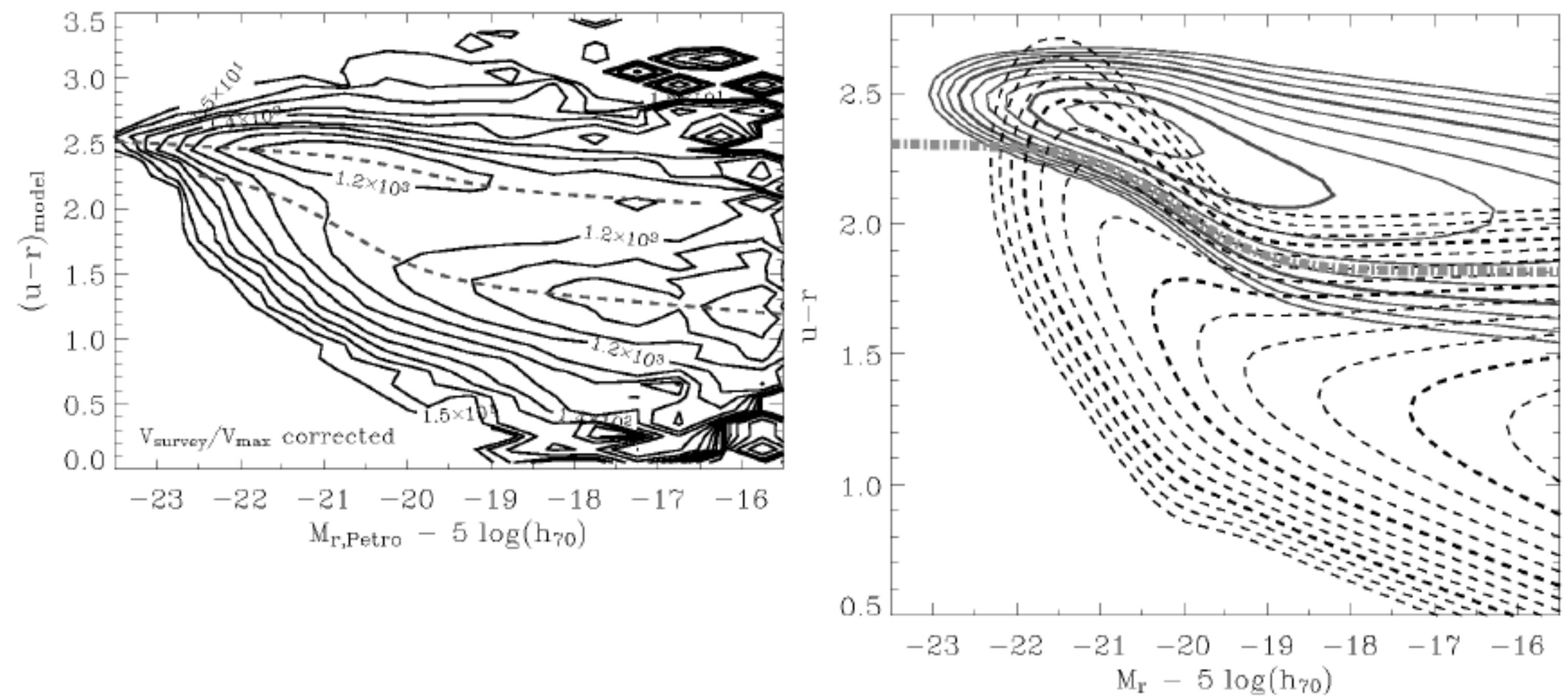}\par\end{centering}
\caption{\label{fig:Galaxy-color-distribution}Galaxy colour distribution
from SDSS \citep[used with permission]{Baldry04}. 
Left: data corrected for incompleteness, with contours showing the number
of galaxies. Right: deconvolution into
`red and dead' (top ($\full$)) and star forming (bottom ($\broken$))
components.
Separation is even cleaner when \textit{GALEX} colours are used too.}
\end{figure}

While matching the observed galaxy LF, a successful model of galaxy
evolution must also reproduce the observed K-band TFR. The
zeropoint and slope of the TFR depends on what fraction of baryons
cool to form K-luminous stars, while their scale length in
the DM halo modifies the RC. Moreover, the $\Lambda$-CDM scenario
yields a continuum of SF rates in galaxies because it does not
inhibit accretion of cold gas that would equip all galaxies with a
reservoir from which to form stars today. Yet, the SDSS has quantified
(Figure \ref{fig:Galaxy-color-distribution}) earlier impressions
\citep{Kauffman03} that galaxies divide bimodally
by colour and mass into a blue sequence of active, SF  galaxies and a sequence of 
`red and dead' early-types that appeared after $z=1$ (for example, \citealp{Wake06})
with AGN-like emission spectra \citep{Cooper06}.

\begin{figure}
\begin{centering}\includegraphics[scale=0.57]{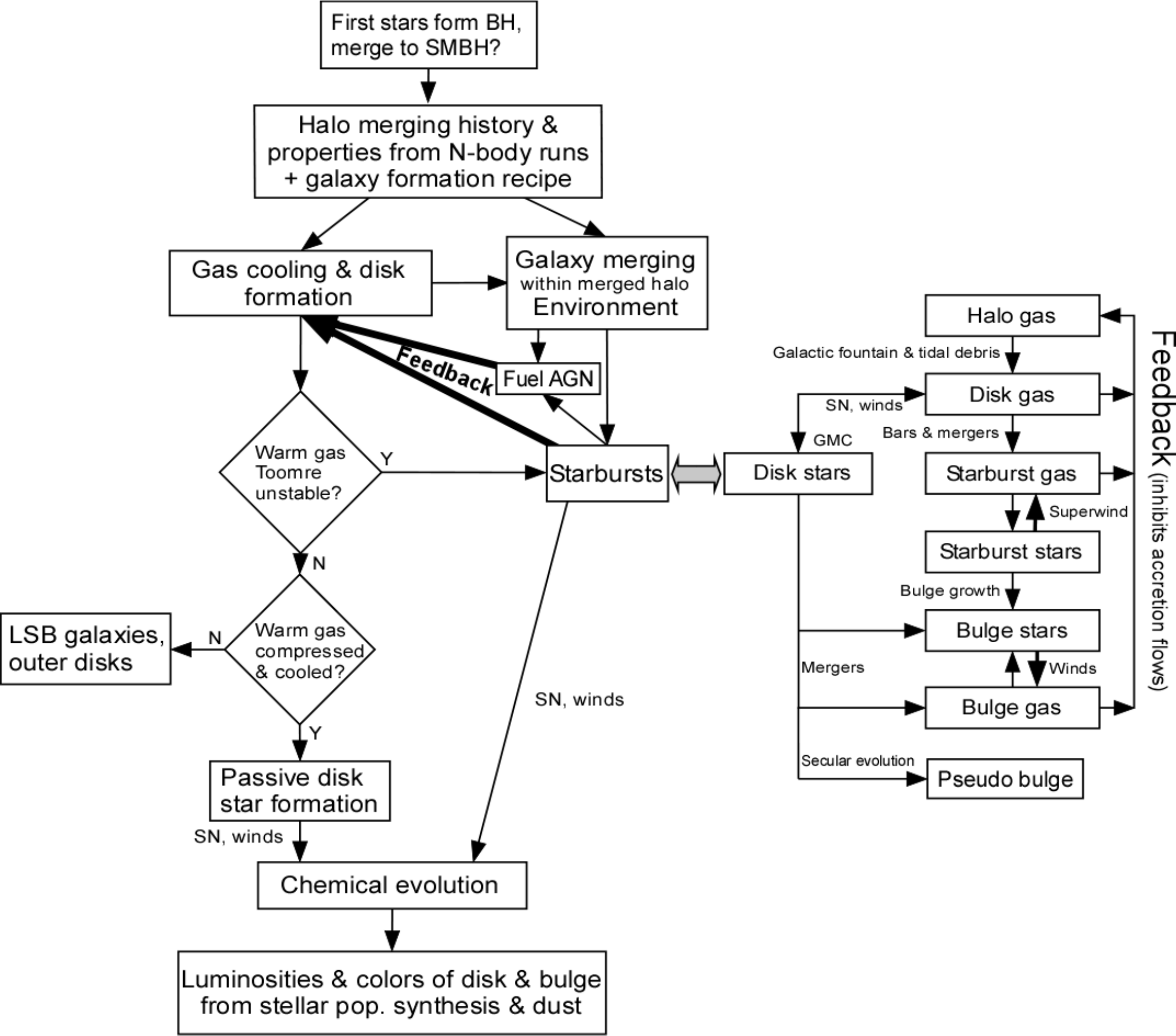}\par\end{centering}
\caption{\label{fig:Feedback-flow-chart.}The baryon flows that operate in
stellar and AGN feedbacks, and their outcomes. 
The second box from the top is treated currently with numerical
simulations because observational constraints are rudimentary. The other steps are discussed
in the text.
The flows operating in disc starbursts are highlighted at right.}
\end{figure}

In summary, feedback and other baryonic gas dynamics alter galaxy
properties drastically; they `quench' subsequent SF
and build up mass by inhibiting flows of energy and angular momentum.
Multiple routes in Figure \ref{fig:Feedback-flow-chart.} 
build the red galaxy sequence from the blue;
the effectiveness of some depend strongly on environment so are detailed
in section \ref{sec:Interactions-with-Other}. \citet{Baugh06} and \citet{Cole00}
review comprehensively the parameters of galaxy evolution models.
In the following, we merely outline the most important processes.

\subsection{Hot and cold modes of gas accretion}

Most accretion occured at high redshift, in two modes: 1) `hot'
gas falls into the DM halo at its virial temperature, heating
faster than the cooling time in an almost spherical shock.
2) `cold' merging gas cools faster
than it is shock heated so collapses first into and then perhaps along
a filament \citep{Keres05}. In the present epoch, the cold mode is
located by simulations \citep{Keres05} at $<1.8\times10^{11}$ \Ms\ 
because of lower infall velocities. 
In contrast, the simulations by \citet{Cox04} highlight how the final merger
can shift gas from a cold, SF phase to a hot, non-SF one whose ionization
depends on the metagalactic UV radiation field, and the uncertain leakage
of ionizing photons from the disc and from halo planetary nebulae.

Unclear is how a cold accretor sheds its gravitational energy to bind
to a galaxy: through a strong shock near the disc, or gradually enough
for pressure gradients to settle gas gently into disc rotation? 
\citet{Fraternali06}
consider the extended HI gaseous halos of disc galaxies \citep[for example][]{Swaters97}, and 
conclude from the failure of velocity models to avoid an HI outflow
that the galaxy must be accreting from the intergalactic medium.
A massive yet
isolated HI cloud observed by \citet{Minchin05} cannot be tidal
debris \citep{Bekki05}. Will it become a cold accretion? Metallicity
measurements can discriminate provided that ionized gas is present
or if there is a background quasar spectrum to absorb its light with
neutral gas. Other clues come from the orbits of Local Group metal-poor globular clusters
and other satellites of the MWg and M31 \citep[including M33,][]{Koch05},
which are flattened along the major axis of their velocity ellipsoids
to indicate non-spherical potentials. No comparable constraints exist
for hot accretion at high redshifts, probably because its spherical
distribution would be too diffuse to detect with present day x-ray
telescopes.

\subsection{\label{sub:Feedback}Mechanisms and scales of feedback}

`Classical' feedback in a forming, early-type galaxy arises from a large-scale
wind established in the ISM by overlapping hot, blast-wave bubbles
of supernova explosions (each $\sim10^{51}$ erg) and stellar winds (each $10\%$
as strong). Recently identified as a major feedback are the
supermassive black-holes in AGN that drive particle jets 
(section \ref{sub:Jet-and-radio-lobe}), their
shocked cocoons and associated thermal winds over multi-kpc scales
and each potentially $\sim10^{47}$ erg~yr$^{-1}$ for up to $0.1$ Gyr. 
\citet{Veilleux05} discuss these outflows. An outflow
is prominent where directed motion thermalizes, and usually has peak
contrast against unrelated emission at a characteristic energy. For
example, gas decelerating from $v_{\rm{s}}$ \kms\ clumps
and attains post-shock temperature $T_{\rm{ps}}=0.11v_{\rm{s}}/(100$
\kms) $10^{6}$ K that radiates x-rays strongly and is extinguished by
photoelectric absorption, not directly by dust.
The shocked gas may be further ionized by the AGN's `ionization cone', 
although the AGN's radiation field dilutes rapidly with distance.
Note that outflows
are not the only way to sweep gas from a galaxy; in section \ref{sec:Interactions-with-Other}
we discuss `ram pressure' stripping as a galaxy passes near
the core of a cluster.

The rudimentary treatment of ISM in most semi-analytic models propagates
blasts too rapidly through the hot phase, so clouds are
crushed quickly to overproduce stars. An ISM more in tune with modern
understanding (section \ref{sub:Raw-material----}) reduces porosity and
confines blast bubbles adiabatically to retain most of the supernova energy.
For example, \citet{Springel03} use a two-phase ISM whose components
exchange mass by condensing into cool clouds, by evaporating in blast
waves, by supernova heating, and are replenished by enriching, energetic wind
outflows from massive stars. With this ISM they do predict a quiescent cosmic 
history of SF that is consistent with
red massive, early-type galaxies (Figure \ref{fig:Galaxy-color-distribution})
and blue, lower mass galaxies (below $10^{12}$ \Ms;
\citealt{Cattaneo06}) that cool and form stars.

Gas is also reheated and ejected by AGN. In simulations, an AGN is
fuelled by mergers that usually stir star orbits into a bar or
other transient structure, which then torque and shock gas into infall.
The efficiency of feedback from a supermassive black-hole is set by the nature of the
shock propagating medium, and by the efficiency of gravitational energy
release $\epsilon_{0.1}$ (units of 10\%)
such that $E=2\times10^{61}M_{\rm{BH},8}\epsilon_{0.1}$ erg 
(black-hole mass in $10^{8}$ \Ms). To remove $M_{\rm{g},11}$
(units $10^{11}$ \Ms) from a region of velocity dispersion
$\sigma_{200}$ (units 200 \kms) requires $E=4\times10^{58}M_{\rm{g},11}\sigma_{200}^{2}$
erg. The main accelerator at small radii is radiation pressure $P_{\rm{rad}}/k\sim2\times10^{7}L_{46}r_{\rm{kpc}}^{-2}$
for ionizing luminosity in units of $10^{46}$ erg and radius in kpc, respectively;
the process is only effective within $\sim10$ pc from AGN found locally
and $\sim1$ kpc in quasars. Coupling AGN ionizing photons to gas is greatly
enhanced in dusty clouds \citep{Dopita02}.

\citet{Springel05} find that a rapid star burst or AGN outflow can
end SF abruptly. \citet{Cattaneo07} find that intermittent jets are
most effective because they do not burrow beyond small radii, hence
deposit their energy in the densest gas. Such jets reproduce quantitatively
the colour-luminosity,
-environment, and -morphology relations of the SDSS dataset,
and form quickly the red and dead population of
massive early-type galaxies. AGN feedback therefore
establishes successfully the bimodality of galaxy types in massive
galaxies whose potential wells are too deep for supernova feedback to eject
gas, see Figure \ref{fig:Galaxy-color-distribution}. If accreted
gas settled in while a luminous AGN was powered down, the subsequent
effects on it after power up could be strong. Dust grains would charge
electrically and be expelled, the cold phase removed, the density
of the diffuse phase reduced, and the porosity of the ISM to supernova increased
from $\sim10^{-4}$ to 1 over $\sim8$ Myr, all by a galaxy-wide ionization
front.

\subsubsection{\label{sub:Galaxy-scale-winds}Galaxy-scale winds.}

A galaxy-scale wind regulates SF by heating and entraining ISM into
the halo or beyond. Cooling $10^{6-6.5}$ K gas is imaged by the Chandra
x-ray Observatory (\textit{CXO}), the kinematics of gas at $\sim10^{5.5}$
K are constrained by \textit{FUSE} spectra of OVI$\lambda\lambda1032+1038$
emission at the interface between the cooler bulk cloud and the enveloping
ISM \citep[see the simulations of][]{Marcolini05}, post-shock gas
at $10^{4}$ K is mapped across outflows in multiple visible-band
and near-IR emission lines that can diagnose gas conditions, mid-IR emission
from hot dust is mapped by the Spitzer Space Telescope, 
and HI and molecular lines are mapped with radio interferometers. 

M82 and NGC~3079 \citep{Cecil01}
provide particularly clear, multi-frequency views of SN-driven, galaxy-winds,
while \citet{Strickland04} use \textit{CXO} to survey diffuse soft
x-ray emission in a sample of edge-on Sgs that spans the full range
of SF activity. In thin, star bursting discs, superbubbles blow out
vertically, quickly energizing at least the galaxy halo and perhaps
the IGM (although, $\Lambda$-CDM says that the virial radius of the
dark halo of a massive galaxy is $\sim250$ kpc and existing x-ray
telescopes are too insensitive to track gas to such low SB). In
the sharp images of \textit{CXO}, \citeauthor{Strickland04} isolate
diffuse emission from point sources (mostly low-mass x-ray binaries) and
find that its SB in edge-on Sgs
correlates with the mean SF rate per unit area. Most of the
clumped emission in `windy' galaxies turns out to be ambient ISM clouds that have been
crushed into pancakes and accelerated by the wind. Depending on how
well it resists Kelvin-Helmholtz and Rayleigh-Taylor fluid instabilities,
the shredding cloud may approach the wind terminal velocity \citep{Marcolini05}.

\subsubsection{\label{sub:Ubiquitous-AGN-and}Ubiquitous AGN and supermassive black-holes.}\

\textit{SWIFT} satellite hard x-ray images and HST/ground-based
long-slit spectra have uncovered supermassive black-holes
in essentially all nearby galaxies with bulges \citep[for example][]{Kormendy95,Ho97},
including some starbursts. The region of gravitational influence
is small (at 4 Mpc distance, 1 arcsec for $2.5\times10^{8}$ \Ms)
and in only three cases can one distinguish unambiguously between
a dense star cluster and a supermassive black-hole (the MWg, M31 and NGC~4258, see the review
of \citealt{Merritt06}). A few masses can be obtained from the circular
Keplerian motion of compact masers within GMCs that orbit at the outer
edge of a nuclear disc (NGC~4258, NGC~4945, and Circinus and NGC~1068
with the complication of severely warped circumnuclear discs), see
the review of \citet{Lo05}. More model dependent techniques to bound
masses in more distant galaxies are reverberation maps from correlated
broad line-AGN continuum flux variations \citep{Peterson04}, or when
the AGN gravitationally lenses a background galaxy into multiple images
\citep{Rusin05a}. 

Our sharpest view of an AGN comes from studies of the supermassive black-hole in the
MWg, Sag.~A{*}, which flares in the near-IR/x-ray periodically as it
is fuelled from a relativistic orbit
\citep{Schodel03}. From precise orbits of stars
S02 ($15.3\pm0.34$ yr period) and S0-16 (comes within 600 Schwarzschild
radii, 45 AU), \citet{Mouawad05} and \citet{Ghez04} derive mass
$3.5\pm0.3\times10^{6}$ \Ms, and indeed will soon measure relativity
effects such as perigalacticon advance. 

Masses of central black holes correlate with host galaxy properties \citep{Marconi03,Haring04,Ferrarese00,Gebhardt00}
\begin{equation}
M_{\rm{BH},8}=1.66\pm0.24(\sigma/200)^{4.86\pm0.43},\end{equation}
(units of $10^{8}$ \Ms) implying that the black hole has $\sim0.3\%$ of a galaxy's
baryon mass. There is some scatter, for example the $\sigma$'s
of the MWg \citep{Tremaine02} and M87 \citep{Cappellari06} are too
small for their correlation masses, whereas the supermassive black-hole masses in M31
($1.1\times10^{8}$ \Ms\  \citealt{Bender05}),
M32 ($(2.5\pm0.5)10^{6}$ \Ms\  \citealt{Verolme02})
and Cen~A \citep{Krajonovic06} are twice those predicted, so perhaps are powered
down.
Indeed,
\citet{Revnivtsev04} suggest that the x-ray bright reflector at Sag. B2 $\sim100$
pc away indicates that the MWg's 
supermassive black-hole emitted $L\sim1.5\times10^{39}$ erg~s$^{-1}$
(i.e., $10^{6}$ brighter than today but still only 1\% that of a
typical AGN) in the band 2-200 keV for at least a decade 300-400 years
ago.

\citet{Wyithe03} fit simultaneously
the $M_{\rm{bh}}-\sigma_{\rm{gal}}$ relation and the evolution,
shape, and zero-points of the quasar visible-band and x-ray luminosity
functions out to $z=5.5$ in a model wherein a black hole powers up whenever
it feasts on gas delivered by a major merger 
\citep[for example the DM simulations of][]{Volonteri03}.
They predict $M_{\rm{bh}}\propto V_{c}^{5}$, independent of redshift.
 \citet{Rafferty06} find cavities in the intracluster medium that
were inflated by large radio jets. AGN are able energetically to balance
cooling in more than half of the 33 clusters studied. The most powerful
example of this heat source was found by \citet{McNamara05} in a
cluster at $z=0.22$: a radio source $\sim550$ kpc long of $\sim6\times10^{61}$ erg 
has so heated all the intracluster gas over several Gyr that it
has been unable to cool onto the central cD galaxy to form stars.

In summary, both supermassive black-hole and starburst winds are driving outflows
that are currently extensive enough to impede mass buildup and to
regulate the SF rate over the past Gyr.
These and other energy exchange mechanisms in the ISM are sufficiently complex and 
poorly understood that our treatments of feedback in galaxy evolution remain
very uncertain.

\section{\label{sec:Interactions-with-Other}{R\^ole
of group/cluster environment in galaxy evolution}}

Section 6 outlined the evolutionary r\^oles of internally driven feedback.
However, evolution can also be stimulated by
events within the virialized, hence hot (${\sim10}^{8}$ K) environment of a
galaxy cluster. 
Ram pressure induced as a galaxy
transits this medium can erode the galaxy ISM \citep{Gunn72}, and it and stars
can be disrupted or merged by tides.
We therefore address how environment affects evolution,
first considering how cluster studies have constrained DM.

\subsection{Dark matter}

\subsubsection{Evidence from galaxy groups.}

Evidence for DM around galaxies in groups and bound pairs is slim
\citep{Persic96}. Mass comparable to that in the stars is estimated
from x-ray images \citep[see the review by][]{Mulchaey00} by assuming
that this gas is in hydrostatic equilibrium in the group's gravitational
potential. This is reasonable for the subset of groups with regular,
circular shape given the short sound-crossing time compared to local
cooling time. The intragroup medium has a significant abundance of
heavy elements, usually attributed to galaxy-scale winds (section \ref{sub:Galaxy-scale-winds}).
But the relative
effectiveness of supernova heating in the wind to that of the gravitational
potential of the group is uncertain because x-ray images even with
\textit{CXO} have not established the efficiency of energy transfer
from SNe to the gas because of uncertain gaseous filling factors \citep{Strickland04}.
The standard assumption is that ongoing galaxy winds are unimportant
heat sources and that the hot gas is spherically symmetric. 
assumptions on gas temperature and metallicity.
If the radial x-ray SB follows a \citet{King62} profile, 
temperature can also be determined at projected radius.
x-ray telescopes before \textit{XMM/Newton} and \textit{CXO} probably
missed much hot gas at larger radii, perhaps as much as in the galaxy stars. 
In fact, isothermal models fit data poorly,
arguing again for non-gravitational heating. 

\subsubsection{Evidence from the cores of galaxy clusters.}

The same x-ray analysis has been applied to galaxy clusters. \citet{Bautz03}
derive mass profiles and the central density slopes of 5 clusters
with \textit{CXO}; 4/5 are consistent with $\Lambda$-CDM
predictions. Only the cluster core may be in virial equilibrium. However,
in some well studied cases there is clear evidence from x-ray
images for transient core heating by shocks from sub-cluster
infall \citep[for example][]{Markevitch02}.
Such structures invalidate the isothermal assumption and may undermine
the assumption of hydrostatic equilibrium. A further complication
is a `cooling flow' in the centre of rich clusters and Egs
\citep{Fabian94} whereby the densest hot gas cools onto the giant
cD galaxies at the centres of clusters. In a few cases, central
gas is measured to have half the temperature of the cluster mean.
On the other hand, accretion rates up to 1000 M$_{\odot}$~yr$^{-1}$
are predicted yet are not seen in soft x-rays. Radio jets from the
AGN of the central galaxy appear to be reheating much of the gas and
quenching SF (section \ref{sub:Ubiquitous-AGN-and}). 

The deflection and focusing of distant background light into multiple images
(strong gravitational lensing) maps the distribution of
mass across the cores of rich clusters.
The pattern of images, for example cluster arcs, has been inverted to constrain
the core mass (hence $\Upsilon$, see section 5 of \citealt{Bartelmann01}).
DM can be mapped in sparser clusters by its
shear of background images (weak lensing) to give a lower
limit on the cluster core mass.
HST's ACS camera covered three
times the area of its predecessor with better spatial sampling, and
covered targets out to 1.5 Mpc radii without mosaicing (which complicates
derivation of a reliable shear pattern) on more distant clusters \citep[for example][]{Lombardi05}.
From such a map, \citet{Clowe06} find that the DM precedes by 100
kpc a distinct bow shock of hot gas falling through the `bullet
cluster' 1E0657-56. The collisionless DM continues unimpeded while
the x-ray gas is shocked hence delayed by the collision. DM in more
than a dozen clusters has been assessed by lensing and compared to virial and
x-ray measures. For example, the various mass estimates of cluster
Abell 2218 out to radius 0.4 Mpc are $M{}_{\rm{stars}}=7.2\pm4.3\times10^{12}$
\Ms, $M_{\rm{Xray}}=1.7\pm0.5\times10^{13}$ \Ms, and $M_{\rm{cluster}}=3.7\pm0.1\times10^{14}$ \Ms,
implying that $M_{\rm{dm}}=3.5\pm0.2\times10^{14}$ \Ms\  and $\Upsilon=300\pm60$
solar \citep{Squires96}; there is up to 100 times more mass than visible in stars \citep{Zwicky37}.
The lensing estimate for $M_{\rm{cluster}}$ agrees with the result
from virial analysis of member motions within the potential and from
x-ray measurements, although lensing gives the mass inside a cone
(along our l.o.s.) and the others give it in a sphere. The discrepancy
is largest when the mass filament that contains the cluster is projected
along our l.o.s. \citep{Squires96}.

Using accurate distances and a least-action infall into the Virgo cluster,
\citet{Tully04} obtain total
mass $1.2\times10^{15}$ \Ms, and $\Upsilon=850$, seven times
higher than that found by the Virial theorem for the core. This
analysis is unfeasible for other rich clusters.

\subsection{Interactions with the host cluster}

\subsubsection{Evidence for cluster effect on SF.}

The morphology--density relation ---
many more Eg/S0 galaxies than Sgs in rich clusters
and \textit{vice versa} in the low-density field \citep{Dressler80} ---
is striking.
Does the larger fraction of Egs in clusters arise from their birth
(nature) or from evolution induced by the environment (nurture)? A
strong clue is the `Butcher-Oemler effect' \citep{Butcher78}:
galaxy clusters at moderate redshift $(z\sim0.3)$ have a larger fraction
of photometrically blue galaxies than do clusters at low redshift.
This finding indicates substantial evolution
in cluster galaxies over the past $\sim5$ Gyr. Later studies have
shown that some blue galaxies have standard Sg emission-line spectra.
Others have spectra that are rare at present: very strong Balmer
absorption lines without emission. Such resemble the composite of
a main-sequence A and a K giant stellar spectrum, so are termed `E+A'
or `K+A' galaxies. By inference, their SF ended within the
past Gyr, leaving strong Balmer line relics. Their denotation as
`post-star burst galaxies' is controversial because
it implies that SF `ended with a bang', whereas some researchers argue that
SF simply declined suddenly, i.e., was truncated (see section 7.2.2). 
Balmer lines in some K+A galaxies are so strong
that they clearly resulted from a star burst. But only if you catch
it immediately after its burst can you distinguish unambiguously a
post-burst galaxy from one that is fading with truncated SF.

High-resolution images with HST and ground-based adaptive optics systems 
clarify that
many blue galaxies are normal Sgs, the spiral fraction in clusters
at $z>0.3$ much exceeds that today \citep[for example][]{Dressler97}, 
and there was a larger fraction
of interacting galaxies in $z>0.3$ clusters than at present
\citep[for example][]{Lavery88,vanDokkum99}. Another
clue is that the blue galaxies at higher-z tended to lie outside the
cluster core, whereas most galaxies within the core are red even at
high $z$. HST shows that the Eg fraction
remains constant at $\sim15\%$ from $z=0.5$ clusters to the
present; rather, it is the ratio of S0 to spiral galaxies that changes strongly
\citep[for example][]{Dressler97}.
Also, Egs at fairly high-z already seem to have a well-defined FP.
In summary, the favoured scenario has cluster Egs already concentrated
toward the cluster centre at early epochs $(z\sim2)$, whereas the
Sg population has evolved rapidly in the past $\sim5$ Gyr as they
converted to S0 galaxies while falling into the cluster.

More clues emerge from nearby clusters. \citet{Caldwell93} and \citet{Caldwell97}
find early-type galaxies in nearby clusters
with K+A `post-star burst' spectra that are generally too weak
to classify as true K+A galaxies as defined in higher-z clusters.
However, they \textit{do} show the same pattern of enhanced Balmer absorption lines
without emission. Also, these K+A galaxies in nearby clusters
tend to be rather low luminosity, i.e., they are much smaller galaxies
than those seen at higher z. From their major study of Coma cluster
galaxies, \citet{Poggianti04} argue that evolution in K+A galaxies
can be explained by a `downsizing' effect, namely that galaxies
terminating SF more recently are less massive. A possible explanation
for how downsizing might work is that more massive Sgs form stars
more rapidly, so later infalling massive `Sgs' no longer have
gas to make the K+A effect.

\subsubsection{Conversion of galaxy types.}

Two routes from spiral to S0 galaxies are proposed: tidal interaction/merger,
and interaction between the Sg ISM and the hot $(10^{8}\rm{\rm{K}})$,
diffuse intracluster medium (ICM). Both remove the ISM by stripping
or by inducing copious SF. Tidal interaction invokes both processes,
i.e., tides can remove significant gas and stars in a roughly equal
mass encounter,
but the interaction also drains much of the angular momentum
from the gas into the galaxy centres to trigger a star burst \citep{Mihos96}.
Tides tend to be strongest in \textit{low} velocity encounters so are
disfavoured in galaxy clusters with their very high velocity
dispersion. Thus, it has usually been assumed that tides and merging
instead make Egs from disc galaxies. Yet there is evidence from
high-resolution imaging of many merging/interacting Sgs in $z>0.3$
clusters \citep[for example][]{vanDokkum99}. 
So, an alternate scenario of `galaxy harassment' \citep{Moore96}
destabilizes cluster galaxies by the aggregate of many rapid encounters
with other galaxies and/or with the tidal field of the cluster. `Wet'
(gas-rich) and `dry' (-poor) mergers are distinguished by the
time required to organize the colliders into a spheroid: a dry merger
builds a discy bulge quickly by violent relaxation, whereas secular
processes (section \ref{sec:Bulges-of-Spiral}) can randomize the gas discs
in a wet merger into a boxy spheroid in $1-2$ Gyr. 

The alternative to tides is the `ram pressure stripping' hypothesis
of \citet{Gunn72}, who showed that an Sg falling through the hot
ICM has its ISM stripped when the ram pressure exceeds the local restoring
gravity of the disc.  Increasingly detailed
numerical models have simulated stripping
\citep[for example][]{Abadi99,Schulz01,Vollmer06}. Ram stripping
should be effective when the density of the ICM is $\gtrsim10^{-3}$
cm$^{-3}$ and the velocity dispersion of galaxies is $\gtrsim10^{3}$
\kms, typical of rich clusters. What is seen? First, single-dish
studies of many Sgs in nearby clusters show that their global HI
content is much depleted compared to field Sgs \citep{Gavazzi87,Giovanelli85,Solanes01}.
Radio interferometer mapping is more time consuming, so fewer such studies have been made.
But they do resolve the HI in cluster
Sgs and sometimes show a highly asymmetrical gas distribution,
as if the ISM on one side of the galaxy is being compacted by
ram pressure. Sometimes the radio continuum emission trails from the
compacted side, and one can see a rim of HII regions concentrated
along the squeezed edge \citep[or even ex-planar, as in][]{Kenney99},
as if SF is being induced along the leading shock where the galaxy
ISM plows into the cluster ICM \citep[for example][]{Gavazzi95,Crowl05}.

However, in the sparser Pegasus I cluster where the ICM density
and galaxy velocity dispersion are low, ram pressure
stripping should \textit{not} be occurring. Yet, \citet{Levy06} find
cases where HI is depleted globally and even concentrates on one
side of the galaxy where stars are forming. Other clues exist, but
gas stripping from Sgs remains obscure.

Advocating that S0 galaxies in rich clusters like Coma are
Sgs with stripped ISM raises questions.
Most problematic is that HI-deficient Sgs in clusters appear to
have systematically earlier morphological types and higher bulge/disc luminosity ratios
than those with normal HI content \citep[for example][]{Dressler86}.
How then can S0 galaxies with their apparently larger bulges be stripped Sgs? Perhaps,
as SF exhausts gas, discs fade to increase the bulge/disc flux ratio.
Too, \citet{Koopman98} have shown that the
apparent excess of early-type stripped Sgs in clusters is caused at least partially
by classification errors in morphology, due to the reduced SF rates.
In addition, \cite{Caldwell99}
have found examples of `bulges' of early-type galaxies in
nearby clusters that resolve into star-forming or post-starburst regions
(and thus are not `bulges' at all) when viewed
in high resolution HST images.

There is also the question of cluster S0 galaxy ages. If these are really
stripped Sgs, they would be younger than cluster Egs.
\citet{Kuntschner00} studied both S0 and elliptical galaxies
in the nearby Fornax cluster, and concluded that the S0s
are indeed younger. On the other hand, \citet{Jones00} claim no age
difference between these two types in clusters at $z\sim0.3$. Finally,
if S0s form in clusters by stripping Sgs, how do they form in sparse
environments?

A related question is: do most of the blue galaxies clustered
at $z\sim0.5$ `passivate' into the non-SF galaxies seen in
nearby clusters through a final major burst or by quenching/`strangling' their
SF as their ISM is stripped? Do data suggest that
their SF is simply truncated, or is there a last burst? The answer
is controversial. The Canadian Network for Observational Cosmology
consortium find no evidence for enhanced SF in $z\sim0.5$
clusters \citep{Balogh99}, so argue that all Sgs in clusters are `strangled'. Conversely,
the MORPHS consortium \citep[for example][]{Dressler04}, find strong
Balmer line absorption that cannot be reproduced by the sudden truncation
of SF; instead a star burst is required. In nearby clusters, \citet{Caldwell99}
find that some galaxies at much lower luminosity than the bright star-forming
galaxies at higher-z appear to be undergoing, or are seen right after,
a star burst. So, although these galaxies are much smaller than big
galaxies at higher-z with `K+A' signatures, they do seem to
have undergone a last burst of SF not truncation.

In summary, data show that galaxies
have evolved in both colour and morphology within clusters
since $z\sim0.5$. Tidal
damage and ram pressure stripping drive this evolution,
and are being modelled with growing sophistication. However, much
remains to be done to develop a coherent picture of rapidly evolving
galaxy properties within a cluster.

\section{New observational facilities}

A basic problem is the strong central
concentration of galaxy light, especially in early-types. Light
scattered from this point across the field of view of a spectrograph
dilutes absorption line strengths. Calibrating this effect is complicated
on a modern telescope because its altitude-azimuth mount (adopted
universally for rigidity) rotates the field of view during a target
track. The diffraction pattern of the mechanical supports of the secondary
mirror also rotates across the field, injecting time-variable structure
from the halos of bright sources within and even outside the field.
Dusty optics at well ventilated observing sites scatter light over
large field angles. Scattered light has an especially insidious effect
when photometric profiles are averaged azimuthally, as is often done when one
assumes axisymmetry. Diffraction gratings, including modern volume-phase
holographic gratings, also scatter light unintentionally as it is
dispersed.

Our atmosphere is scattering noise in many ground-based
instruments. Shortward of $\lambda2000$ nm, the incoming wavefront
can be `de-wrinkled' with an adaptive optics system to concentrate light. The ongoing
retrofit of telescopes with adaptive optics will serve to resolve more distant
stellar populations, allowing the powerful CMD discriminators to overcome
the degeneracies of spatially integrated spectra. A `laser guidestar'
allows an adaptive optics system to operate over almost the entire sky. Pixel SNR scales
as the square-root of the exposure time when the signal is Poisson-noise
limited. By reducing the size of the sky patch, an adaptive optics system reduces
noise twenty-fold or more. 
Complementing imagery is an IFS to obtain
spectra at the limit of the adaptive correction.
Hopefully, instruments on the very largest telescope(s) can be
optimized to study both faint, distant blobs in
the early universe \textit{and} bright, nearby puzzles.
It is encouraging that the European
Southern Observatory is building the Multi-Unit Spectroscopic Explorer (MUSE)
\footnote{\url{www.eso.org/instruments/muse}} instrument whose 
1 arc-min$^2$ field of view will be corrected by adaptive optics.
Such instruments could, for example, explore the DM content of an
Eg through the kinematics of its dynamically relaxed globular clusters.

Many galaxies are red objects whose dominant stellar population
emits mostly in the near-IR. The sky spectrum over $\lambda\lambda750-1500$
nm is covered with numerous and variable 
OH rotation-vibration band emission. The night
sky is dark between, but
light scatter within a spectrograph reduces this contrast.
Work is therefore underway to modify optical fibres to suppress OH
bands prior to wavelength dispersal. No technological barriers
have appeared, suggesting that 99\% of the bands will be suppressed
for another $4-8$-fold gain in SNR. 

After 2013, the James Webb Space Telescope (JWST) will attain high sensitivity
long-ward of $\lambda1000$ nm with background
set by dust emission and scattering (Zodiacal light). JWST
will probe dust-shrouded regions of SF in nearby galaxies including
the MWg. Missing within 5 years will be UV spectroscopy to assess the
full impact of massive stars, so a relatively inexpensive 1.7m aperture
World Space Observatory has been discussed.
JWST development has already commercialized large-format (4 megapixel), edge-buttable,
Presently, a single such detector cannot obtain simultaneous
J+H-band spectra at the $R=4000$ required to work in the dark between OH 
emission lines; a detector mosaic must be used. If one could use
OH suppressing fibres, the full near-IR spectrum could be projected at
$R=1500$ onto a single detector and a mosaic could be used to cover more 
of the sky instead
The ultimate such instrument in the next
decade will have a laser adaptive optics system feeding deployable integral-field modules
that are coupled by in-line OH sky-suppressing fibres to high-throughput
spectrographs. 

For brevity, we mention only a few ground
facilities under development for use by 2015.

The Large Synoptic Survey Telescope (\url{www.lsst.org}) is a 6.5m
effective aperture telescope for rapid surveys (15 sec exposures)
over 20,000 deg$^{2}$ and $\lambda\lambda320-1060$ nm in
six bands. For example, it will map the DM content
of galaxy clusters by weak lensing. Very cost-effective
liquid mirror telescopes are
also being developed for zenith strip surveys.

The 15-telescope Combined Array for Research in Millimeter-Wave Astronomy
(CARMA, \url{www.mmarray.org}) in the N. hemisphere is currently,
and the 64+ telescope Atacama Large Millimeter-Array (ALMA, \url{www.alma.nrao.edu})
in the S. hemisphere will by 2012, provide $0.1-1$ arcsec resolution
(to $0.01$ arcsec for ALMA) in millimetre and sub-millimetre
CO emission, 5-50 pc resolution at 10 Mpc. GMCs and the dependence
of the SF rate on environment can be studied throughout the local universe.

The Global Astrometric Interferometer for Astrophysics (GAIA,
\url{www.rssd.esa.int/GAIA}) is a 1-meter aperture space telescope optimized
for MWg kinematical studies. It will measure parallax distances
(up to 90 million stars with better than 5\% accuracy and 1 million with 
1\% accuracy) and proper
motions ($10^{9}$ stars, 1\% of the MWg population) out to a distance
of $\sim10$ kpc from the Sun for giants and 1 kpc for dwarfs. In
addition, it will obtain $R\sim10^{4}$ spectroscopy for radial velocities
and limited metallicity information near $\lambda870$ nm for $\sim10^{8}$
stars with $V<17.$ GAIA, combined with extensive ground-based followups,
will revolutionize our view of the MWg.

This impressive technology will allow us to probe deeper into
our nearest extragalactic neighbours, to understand how nature and nurture 
have combined to present them to us today.

\section{Conclusions}

The quest to study nearby galaxies across the electromagnetic spectrum has brought
these galaxies into sharp panchromatic focus to highlight their energy flows.  
Contemporary surveys are establishing both the build up of
mass/structure and the SF and chemical enrichment histories of
the universe.  However, while impressive progress has been made, this review
shows how precious little  we know about chemo-dynamical structures in
galaxies and even less about how they operate.  Making assumptions about
unobserved phase-space dimensions is always perilous.

We have focused on two themes related to nearby galaxies: 
First, through the behaviour of visible tracers, what can we say about
the distribution on their (presumably) dominant DM component?
Second, what can
we deduce about the internal structure, formation and evolution of their
baryonic component?

Despite decades of study, the DM distribution within galaxies remains
unclear. As we discussed, SB profiles in Egs are so steep that luminous
tracers are too faint for accurate kinematic measurements at the
radii where DM is presumed to dominate. Hence, except in the case
of Local Group dwarf spheroidal galaxies, the case for DM in 
individual Egs is still weak.
and its distribution is barely constrained. In disc galaxies,
DM is better established because RCs are often greatly extended in
cold H I.  But there remains enough ambiguity in population 
synthesis models of the visible matter (including uncertainty in 
the low mass end of the stellar IMF) that the relationship between 
mass and light is still much debated. Until these issues are resolved,
details of the DM will depend on the uncertain visible mass 
distribution. 

In regards to the internal chemo-dynamical structure and evolution of nearby
galaxies, we have seen that despite decades of study, how
elliptical and spiral galaxies are inter-related is still largely
unknown.  The most basic questions about the 3-D
structure of Egs remain unanswered, beyond the fact that
massive ones are primarily flattened by anisotropic velocity
dispersion while less massive ones tend to be flattened by
rotation.  After much controversy over, and gradual refinements in, population
synthesis models, we can now extract reliable \textit{light-weighted} mean
ages and chemical abundances (of Fe, Mg and perhaps Ca).  We are only just
beginning to progress to the far more difficult extractions of
real SF and chemical enrichment histories of Egs, because
these require accurate models of multiple spectral features at many wavelengths.

Our understanding of spiral galaxy bulges is even less certain
than Egs, largely because study of the bulge requires identifying
and removing the (thin and thick) disc component.  Information about bulge
ages still comes largely from photometric colours, not spectroscopic
line indices.  Recent understanding that pseudo bulges in late type spirals can constrain
the hierarchical galaxy formation picture will be capitalized once it is determined 
how often pseudo bulges arise from purely internal secular evolution and
from minor mergers.

The core conclusion of this review is that, if we are to
explore galaxy evolution scenarios such as $\Lambda$-CDM hierarchical
merging by establishing galaxy properties out to high redshift,
then we must improve considerably our understanding of nearby, present
epoch galaxies. These systems benchmark high-redshift evolutionary
studies.  The current generation of large ground-based
telescopes and space-based facilities contribute by detailing individual
galaxies.  Surface brightnesses of interest
extend deep into the sky-dominated regime, unfortunately yielding
only linear improvement in SNR with telescope diameter.
Perhaps for this reason, 8-10+ meter telescopes are being
used mostly to study high redshift galaxies at large lookback
times.  However, of equal importance to increased aperture are improved detectors
and more reliable control and removal of sky and galaxy light to
enable study of galaxies into new regimes of low SB.
Too, the development of integral field spectrometers has opened
new possibilities for studying galaxy inner regions.  In short,
there is much to be gained from glimpsing faint wisps within nearby ghosts.

\ack
GC thanks the Reynolds Foundation for sabbatical support and
Director Matthew Colless for hospitality at the Anglo-Australian Observatory
during the start and finish of this review.
We thank the referee for detailed and constructive comments.

\appendix
\section{\label{sec:What-(Astro-)Physical-Processes}Relevant Astrophysical
Processes}
\setcounter{section}{1}

Understanding galaxy evolution requires familiarity with two branches
of astrophysics: 1) the structure and evolution of stellar tracers
of the dynamical and evolutionary state of galaxies, and 2) dynamics
of self-gravitating systems. We first review basic stellar
evolution theory. We discuss how to simulate the composite light of
a coeval population, i.e. a cluster whose stars had the same birth.
We introduce methods to compare the integrated spectra of model
star clusters of unique age and chemical composition to the observed
spectra, aiming to decipher the SF and chemical evolution histories of galaxies.
Finally, we give a basic overview of self-gravitating systems, showing
how their internal structure can be constrained by their
observed light profiles and kinematics. Table \ref{tab:acronyms}
lists the acronyms used in this paper.

\begin{table}
\caption{\label{tab:acronyms}Acronyms used in this review.}
\begin{indented}
\item[]\begin{tabular}{ll}
\br
Acronym&Means\\
\mr
AGB&
(stellar) asymptotic giant branch\tabularnewline
AGN&
active galactic nucleus\tabularnewline
CBE&
(stellar) collisionless Boltzmann equation\tabularnewline
CDM&
cold, dark matter\tabularnewline
CMD&
colour-magnitude diagram\tabularnewline
CXO&
Chandra x-ray Observatory\tabularnewline
dIrr&
dwarf irregular galaxy\tabularnewline
dSph&
dwarf spheroidal galaxy\tabularnewline
DM&
dark matter\tabularnewline
Eg&
elliptical galaxy\tabularnewline
FP&
(galaxy) fundamental plane\tabularnewline
GMC&
giant molecular cloud\tabularnewline
HB&
(stellar) horizontal branch\tabularnewline
HI&
neutral hydrogen\tabularnewline
HII&
ionized hydrogen\tabularnewline
HST&
Hubble Space Telescope\tabularnewline
ICM&
inter-cluster medium\tabularnewline
IFS&
integral-field spectrometer\tabularnewline
IMF&
(stellar) initial mass function\tabularnewline
ISM&
inter-stellar medium\tabularnewline
IR&
infra-red\tabularnewline
LF&
luminosity function\tabularnewline
LSBg&
low surface-brightness galaxy\tabularnewline
MOND&
Modified Newtonian Dynamics\tabularnewline
Mpc&
million parsecs\tabularnewline
MS&
(stellar) main sequence\tabularnewline
MSTO&
(stellar) main sequence turn-off\tabularnewline
MWg&
Milky Way galaxy\tabularnewline
NSAR&
non-stellar (chemical) abundance ratios\tabularnewline
PA&
position angle (on the sky)\tabularnewline
RC&
(galaxy) rotation curve\tabularnewline
RGB&
(stellar) red-giant branch\tabularnewline
RGC&
(stellar) red-giant clump\tabularnewline
SB&
surface brightness (of unresolved stars)\tabularnewline
SDSS&
Sloan Digital Sky Survey\tabularnewline
SF&
star formation\tabularnewline
Sg&
spiral galaxy\tabularnewline
SNe&
supernovae\tabularnewline
SNR&
signal-to-noise ratio\tabularnewline
TFR&
Tully-Fischer relation\tabularnewline
URC&
(galaxy) universal rotation curve\tabularnewline
\br
\end{tabular}
\end{indented}
\end{table}

\subsection{\label{sub:Stellar-Evolution}Stellar Evolution}

\subsubsection{Basic equations of structure.}

These have been established for
decades, and are discussed in the elegant monograph by \citet{Schwarzschild58}.
Throughout the star they characterize local hydrostatic equilibrium,
mass conservation, energy production/conservation, and energy transport
(either by convection, or radiatively from a temperature gradient).
When coupled with an equation of state and after specifying
boundary conditions (at a `surface' in wavelength averaged optical depth because
the star has no sharp edge), one can integrate numerically the coupled
differential equations for internal opacity and energy generation
as functions of density, temperature, and chemical composition to
obtain a self-consistent model.

\subsubsection{Time dependence.}

No evolution equation depends explicitly on time. A star
evolves because its chemical composition alters gradually through
pressure confined fusions. While the ignition condition for each fusion
is reached only deep in the star, convection can dredge
up fusion products to alter the composition of the inert `envelope'.
The main driver for evolution is that, as lighter elements
fuse into heavier ones in the core, or later in a hot surrounding shell,
the mean molecular weight of the gas mix declines
to lower pressure. To maintain hydrostatic equilibrium, the core
must contract to higher density and temperature. Because fusion rates
are extremely sensitive to temperature, core energy release increases
thereby expanding and cooling the envelope to brighten and cool the
photospheric emission. Chemical composition usually changes very slowly
compared to the dynamical timescale. So, the star's evolution can
be traced numerically in time steps, each in hydrostatic equilibrium,
with composition altered by the fusion that occurred during the previous
step.
On a modern computer it is trivial to evolve a spherical, non-rotating,
unmagnetized star. While evolution models have become more sophisticated,
\citet{Iben74,Iben91} remain excellent introductions.

\subsubsection{The HR diagram.}

Before discussing how to derive galaxy ages from precise photometry
of their individual stars and/or colours and spectra of their integrated
starlight, a brief overview of stellar evolution is useful. Stellar
evolution is characterized by the Hertzsprung-Russell diagram,
wherein the star's effective temperature $T{}_{\rm{eff}}$, the
temperature of the blackbody that radiates the star's photospheric flux per area
from radius $R_{*}$,
is plotted against the bolometric (full spectral) luminosity\begin{equation}
L_{\rm{bol}}=4\pi R_{*}^{2}\sigma T_{\rm{eff}}^{4}\end{equation}
Numerical models evolve a star in this diagram. As hydrogen fusion
diminishes and the mean molecular weight of the core increases, the
star evolves first to higher $L_{\rm{bol}}$ and lower $T_{\rm{eff}}^{4}$
in the diagram. In practice it is impossible to measure either of
these, because both require absolute spectrophotometry from the UV
to the mid-IR. Instead, theory and observation intersect in the so-called
colour-magnitude diagram (CMD), wherein the logarithm of a visible-band
(or near-IR) flux is plotted vs. a visible-band (or near-IR) colour. Thus
one needs accurate transformations between $L_{\rm{bol}}$ and the
particular flux, and between $T_{\rm{eff}}^{4}$ and the particular
colour, almost invariably by building a model of the stellar atmosphere
to detail its photospheric spectrum.

Chemical composition, aside from providing input to determine
a star's age, supplies an independent chronometer for galaxy evolution.
Primordial gas from the Big Bang contains only hydrogen (mass fraction
X), helium (mass fraction Y), and traces of other light elements (Li,
Be, and B). Hence, details of the `heavy' element content (mass
fraction Z) of a star's photosphere gives the degree of prior SF,
subsequent evolution, and consequent heavy element production and
expulsion (in winds and nova/SNe) prior to the star's birth in the
enriched ISM.

Once photospheres of a sequence of stars have been so detailed, correlations
with heavy element content may be obtained for grosser spectral features
accessible to lower resolution spectroscopy or even filters selected
carefully to measure photometric colours. Such data demand less telescope
time, enabling study of fainter and more numerous targets in more
diverse environments. As datasets are built, the original correlations
are extended to less luminous stellar systems, to differing
abundances, and to galaxies that have led different lives.

\subsubsection{Colour-magnitude diagrams.}

Stellar evolution is complex as fusion shifts from core to shell hydrogen
burning and then to helium core and eventually shell burning. As the
dichotomy between an increasingly hot, dense core and an increasingly
extended, cool envelope widens, equations of stellar structure become
less certain. We give only the briefest overview of the important
phases of stellar evolution. We aim to 1) acquaint the reader with
how the various evolutionary phases can date a coeval population of
stars, which is crucial to determining SF histories of galaxies,
and 2) indicate which aspects of this program
are robust and well-understood, and which aspects are uncertain and
lead to an uncertain derived age. The underlying question is how accurately
can CMDs of Local Group galaxies and the integrated spectrum of starlight
from more distant galaxies constrain their SF histories?

\begin{figure}
\begin{centering}\includegraphics[scale=1.8]{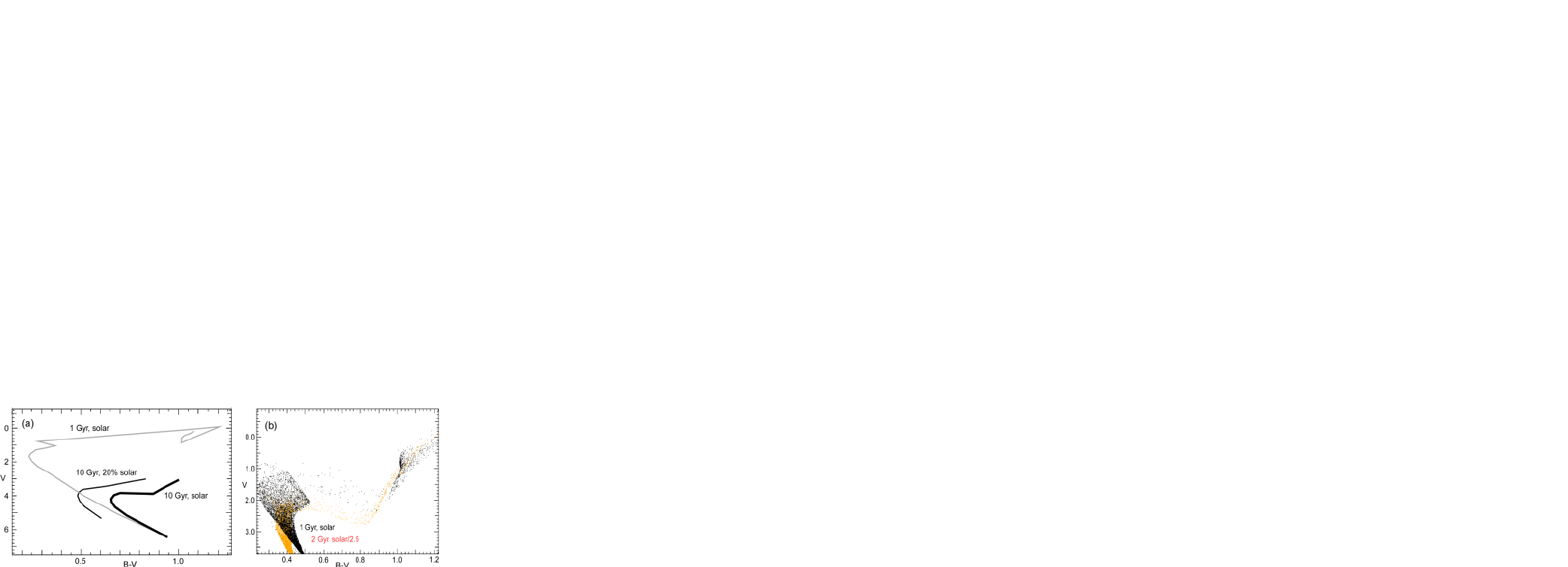}\par\end{centering}
\caption{\label{fig:Theoretical-distribution-of}(a) Isochrones from
the Geneva Observatory stellar evolution group
of 3 different stellar populations, all with identical Salpeter IMF, and plotted for
visible light filter bandpasses.
The vertical axis is related to log luminosity and the horizontal
to colour (hence temperature) with red (cooler) stars at right. The
grey isochrone is for a young (1 Gyr old) stellar population of solar
chemical composition. Black isochrones are for a 10 Gyr old population
of solar composition (thicker) and of heavy element abundance 20\%
of solar (thinner). (b) StarPop illustrates age-metallicity degeneracy.
Two stellar populations arising from a 1 Gyr-long episode of constant
SF are overlaid. Black plots a population of solar metallicity that
ceased SF 1 Gyr ago. Gray plots a population of roughly half solar
metallicity $(Z=0.008)$ that ceased SF 2 Gyr ago. RGB (right)
and sub-giant branches largely overlap.}
\end{figure}

The evolutionary track of a star --- the path over time in a CMD of a star of
given mass and chemical composition --- can be computed.
By combining tracks over
a range of mass but with fixed chemical composition, one can predict
the locus of all stars on a CMD at given age, an \textbf{isochrone}.
An excellent Java Applet, StarPop
\footnote{\url{http://astro.u-strasbg.fr/~koppen/starpop/StarPop.html}},
synthesizes the evolution of a stellar population by sampling isochrones
from a specified SF history. Figure \ref{fig:Theoretical-distribution-of}a
illustrates the effects of age and chemical composition on stellar sequences in a CMD.

\subsubsection{Phases of stellar evolution.}

In Figure \ref{fig:Theoretical-distribution-of}a, note the long diagonal
swath of young, solar chemical composition stars of different mass. Such `main
sequence' (MS) stars are fusing hydrogen in cores. In contrast,
the older solar-chemical composition population has a short MS because core
hydrogen in the most massive stars has exhausted; only
lower mass stars are still core fusing hydrogen.

Note the similar evolutionary sequences beyond the MS in both populations.
Specifically, atop the MS is the MS turnoff (MSTO), a sequence of
stars of increasingly higher mass and luminosity because the higher
mass stars are increasingly evolved by depletion of their core hydrogen.
In the CMD of a star cluster, the MSTO is our most effective chronometer.
Because more massive stars have higher central temperatures and pressures,
fusion exhausts fuel more rapidly. Hence, the mass
of a star of known chemical composition that is just exhausting core
hydrogen can yield a stellar age reliable to $\sim1$ Gyr on populations
older than $1-2$ Gyr. 

Subsequent, shorter phases of stellar evolution are:
\begin{itemize}
\item The red giant branch (RGB), again a sequence of increasing mass, now
increasingly evolved. It illustrates the increasing dichotomy between
hot core and cool low-density envelope as hydrogen fusion switches
to a shell. RGB stars are crude chronometers for populations older
than 1 Gyr because stars over a wide mass range funnel into the same
part of the CMD. The most important spectral feature for RGB
stars is the triplet of singly ionized calcium at $\sim\lambda800$
nm; it is useful even at lower spectral
$R\sim6000$ that washes out Fe lines. \citet{Gallart05} review the
uncertainties and conclude that, for a resolved population, the integrated
SF rate from its start until $\sim2$ Gyr ago can be estimated
to within a factor of four from counts of RGB and asymptotic giant
branch (AGB, see below)
stars, provided that temporarily over-luminous AGB long-period variables
are accounted for. 
\item Subgiant-branch stars, evolving relatively slowly toward the
RGB by burning H in a shell around an inert He core. All stars with
MS life longer than 2 Gyr experience this stage.
\item The red giant clump (RGC) is the helium core-fusion MS
for stars 1-10 Gyr old. Helium ignites suddenly because some electrons
in the core became quantum statistically degenerate when
the star ascended the RGB, augmenting the normal gas-pressure adjustments
for hydrostatic equilibrium
with a temperature-insensitive, incompressible `lattice' of
electrons. The result is a discontinuous loss of envelope mass and
a readjustment of the star's interior structure. As it ascends the
RGB, the star loses mass from its outer envelope at a rate that is
ill defined and intractable theoretically, rendering all evolutionary
calculations beyond the early part of the RGB uncertain.
\item Horizontal branch (HB) stars also burn He in their cores, are low
mass, older than 10 Gyr, metal-poor, and have a more extended range of
temperatures than RGC stars.
RR Lyrae stars are examples, recognized individually because of their
variability. The ratio of red to blue HB stars constrains the metallicity
and age. \citet{Gallart05} review the potential to date ages with
the RGC and HB, and conclude that they work reasonably well for $1-3$
Gyr ages. 
\item The asymptotic giant branch (AGB). After core helium burning at the
RGC, the star burns both helium and hydrogen in separate shells with helium fusion
unstable.
AGB stars are the most luminous known. They are chronometers
for populations older than 1 Gyr, but are less accurate when younger
because of uncertain wind-driven mass loss, thermal
pulses from helium shell fusion flashes, changes in photospheric abundances
from heavy element dredge-up (in particular, carbon), and mixing.
Uncertain contribution of AGB stars can alter the fluctuations
of the SB of galaxy starlight. If the RGB tip is reached below $\sim$8 \Ms,
the stellar envelope is gone and nuclear burning soon ceases.
\item A short, bright `planetary' nebula phase wherein the ejected
envelope of a $0.8-3$ \Ms\ star is photoionized by UV from
the exposed helium, carbon or oxygen core. Thereafter, it fades as a low luminosity
white dwarf whose
core is supported by a barely cooling degenerate electron lattice.
White dwarfs are too faint to study beyond the MWg.
\end{itemize}
The black sequences in Figure \ref{fig:Theoretical-distribution-of}a
show the importance of chemical composition in stellar evolution.
Both interior (opacities, hence radiative energy transfer, and mean molecular weight, hence
internal pressure) atmospheric
(spectral line strengths, hence the emergent radiation field) properties
depend on chemical abundances. As a result, the metal-poor (thinner line)
MS lies blueward (hotter) of the MS of the metal-rich isochrone (thicker
line). The message is that an accurate CMD (from photometric imagery),
an independent measure of its chemical composition (from high resolution
spectra of the absorption lines in individual stars), and a reliable
theoretical isochrone, can be combined to date the population. But,
isochrones are only as accurate as stellar evolution, which is uncertain
in several areas. There are uncertainties in
\begin{itemize}
\item key nuclear reaction rates $^{12}C(\alpha,\gamma)^{16}O$,
$^{14}N(p,\gamma)^{15}O$, and even the triple-$\alpha$ reactions
are either poorly constrained or have been revised recently to alter
evolution timescales \citep{Weiss05}. The problem for
determining astrophysical cross-sections is the difficulty in producing
sufficiently low-energy but high-luminosity accelerator beams.
\item opacities that determine radiative energy transport.
Widely used opacities are from the Opacity Project
\citep{Seaton94,Seaton96} and the OPAL \citep{Iglesias96} consortium.
A larger opacity lowers the H fusion rate and core temperature. 
\item the chemical composition of even the Sun, the basis of abundance analyses
of all other stars. Recent redetermination by \citet{Asplund05a}
indicates that the abundance of O and other major elements must be
reduced $\sim$1.5-fold, a controversial result because it worsens
agreement between theory and helioseismology spectra of the solar
interior \citep[for example][]{Bahcall05,Antia06}. Increasing helium
abundance Y increases $T_{\rm{eff }}$ and $L_{\rm{bol}}$ and
decreases the width of the MS.
\item convection, which does not seem to have even a mid-term solution.
We know so little about it that models parametrize it by a
convective scale height, the typical distance travelled by a convective
cell in units of the local pressure scale height. Because convection
transports energy in the outer layers of the Sun, its scale height sets the solar radius.
So, this scale is fixed \textit{ab
initio} in a 4.6 Gyr model to reproduce the solar radius, and its
value is assumed to apply universally to stars whatever their
internal structure, increasing the core mass. Our ignorance
is further compounded by convective overshoot: the scale is increased
to account for the inertia of bubbles that turn around not where their
acceleration reaches zero but beyond, at zero velocity. Overshoot
is clearly highly uncertain, but appears to be necessary to match
model isochrones with the CMDs of intermediate-age star clusters \citep[for example][]{Kozhurina97}.
Finally, convection plays a huge r\^ole by intermixing layers that
are otherwise stratified into different chemical composition. For
example, deep mixing transports fresh hydrogen into the core, prolonging
a star's life. Models for intermediate-mass stars are compared by
\citet{Dominguez99}, who find 10\% variations in H-burning lifetimes
and 30\% for He core-burning.
\item handling mass loss. As a star ascends the RGB, its envelope
distends and becomes less bound. An AGB star loses mass through either
a gradual wind or sudden thermal pulses. The empirical Reimer's Law \citep{Schroder05}
estimates gradual mass loss by a wind from an 
RGB star, but data are hardly reliable enough to predict the star's
mass when it ignites helium in its core. Thus, all evolutionary
calculations become more uncertain once the star ascends the RGB.
Specifically, the location of a star in the HR diagram during its
core helium burning, HB phase depends on the envelope mass (whereas
the mass of the helium core, hence the star's luminosity, is relatively
insensitive to initial mass): stars with lighter envelopes have hotter
$T_{\rm{eff}}$, so are bluer. Uncertain mass loss confounds prediction
of a star's position along the HB. Hence, isochrones that include
HB and post-HB phases have made basic assumptions about mass loss
that usually force the morphology of the HB to agree with that
observed in star clusters.
\item element diffusion from gravitational settling in the outer, radiative
layers that can mask the star's true chemical composition.
Helium diffusion in the core will alter the star's lifetime. An independent
check on its efficiency comes from other elements, \citet{Straniero97}
suggesting those not burning at H-burning temperatures, Fe or Ca,
as candidates and both measurable in Local Group galaxies down to the MSTO
with large telescopes.
\item the unobserved stellar rotation that augments hydrostatic equilibrium
and breaks the spherical symmetry of the star to greatly complicate models.
Rotation is ignored when many models are generated for an isochrone.
Rotation alters deep mixing hence chemical abundances.
\item Coulomb effects that soften the equation of state. 
They reduce MSTO ages by 10-15\%,
and ages derived from B- and V-band colour differences by 40-50\%.
\end{itemize}
These uncertainties grow at more advanced evolutionary stages as the
internal structure of the star becomes more extreme. Hence, the most
robust estimate of the age of a star cluster remains the $T_{\rm{eff}}$
and/or luminosity of the MSTO, as determined from a CMD. As we will
see, when individual stars cannot be resolved, one can still infer
MSTO characteristics by deciphering the integrated spectrum of galaxy
starlight.

\subsubsection{Sensitivity to basic parameters.}

Broad bandpasses bluer than V cannot find T$_{\rm{eff}}$ accurately
because the spectrum is blanketed by a forest of absorption
lines whose strengths vary with chemical abundances and surface gravity
$g=GM/R^{2}$. Larger telescopes and more sensitive detectors have
supplanted photometry with spectral determination of stellar
parameters. For 5000 K$\le$T$_{\rm{eff}}$$\le$8000 K, Balmer line strengths yield
accuracy $\pm50-80$ K. Surface gravity can
be established to a multiplicative accuracy of 1.5--2.5 by its pressure
modification of some line profiles (mainly Fe, Ca I,
Mg Ib triplet at $\lambda$520 nm, and CN bands at $\lambda$380 +
$\lambda$410 nm) \citep[for example][]{Fuhrmann97}. \citet{Lebreton01}
discusses how star diameters derived from interferometric measurements
have calibrated to internal accuracy $\sim1\%$ a combined empirical/model
atmosphere Infrared Flux method to deliver $T_{\rm{eff}}$ for A-F
dwarf and giant stars \citep{Blackwell98} and, an SB method with
comparable accuracy \citep{DiBenedetto98}. Recent work 
\citep[for example][]{Bessell98} improved
treatments of atomic/molecular line blanketing in O-K and
A-M stars.

\subsubsection{Sensitivities to variations in Y and Z.}

\begin{figure}
\begin{centering}\includegraphics[scale=0.3]{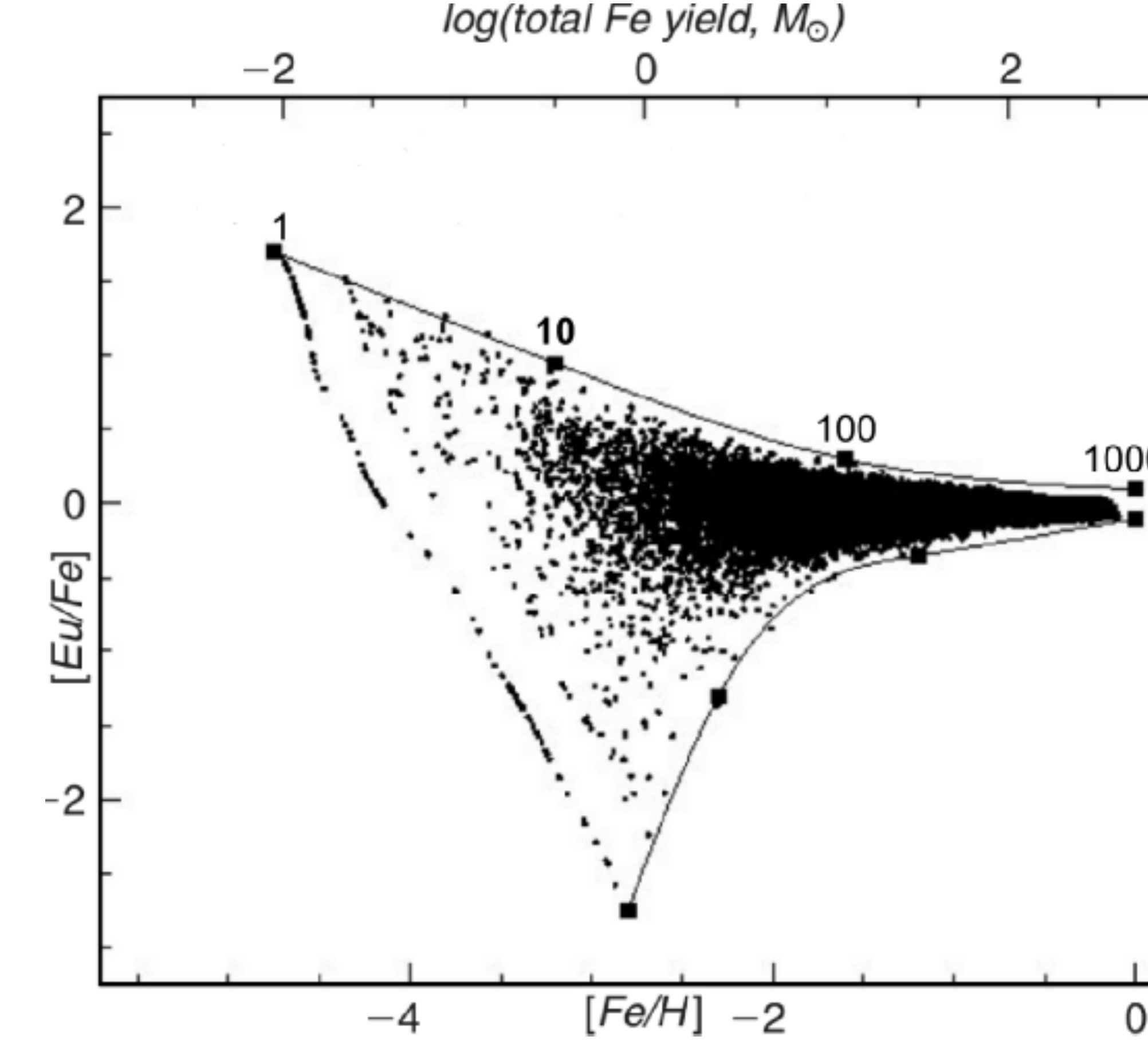}\par\end{centering}
\caption{\label{fig:SNyeildsim}
\citep[From][used with permission of Publications of the Astronomical Society of Australia, 
the authors and the publisher CSIRO.]{Bland-Hawthorn04}
How successive generations of low-mass stars
are enriched by just SNe II, as traced by europium production (r-process).
The initial wide scatter in abundances at left converges to a universal
value at right. The stars are formed with a Salpeter IMF and the element
yields are from \citet{Tsujimoto98}. The upper bound comes from lower-mass
SN, the lower from higher-mass. From left to right, \fullsquare\ indicates
the number of prior enrichments: 1, 10, 100, 1000. Just 10
enrichments from high-mass SNe suffice to enrich a cloud to $[Fe/H]=-2$.}
\end{figure}

\citet{Lebreton01} reviews measured chemical abundances; \citet{Wallerstein97}
review the derived patterns. Abundance estimates are less accurate for hot
and young stars because most of their ionization structure is in the
UV, and because spectral features are blurred by rapid stellar rotation.
Elements to study include
\begin{itemize}
\item O, Mg: products of SNe II that probe contributions from $>20$ \Ms\ 
and trace chemical mixing in globular clusters.
\item Si, Ca, Ti, Cr: also products of SNe II but less strongly weighted
toward more massive progenitor stars. Combined with O and Mg, these
`$\alpha-$elements' (even-proton number nuclei) probe the distribution
of initial masses (the so-called initial mass function, IMF) \citep{McWilliam97}.
\item Mn, Co: from SNe II, with yields depending on metallicity of
the progenitor star.
\item Eu: indicate rapid (r-process) neutron pickup (compared to the rate
of $\beta$-decay) nucleosynthesis in SNe II.
\item Y, Zr, Ba, La: indicate slow (s-process) neutron pickup nucleosynthesis
in lower-mass AGB stars.
\item Fe, Ni: track metallicity from SNe Ia. Fe also provides
parameters such as $T_{\rm{eff}}$ and $g$.
\end{itemize}
Detailed abundances constrain the SF history. $\alpha$-elements come
preferentially from hydrostatic burning in massive stars before their
SNe II make $r$-process elements. Figure \ref{fig:SNyeildsim} \citep[Fig. 1 of][]{Bland-Hawthorn04}
shows how successive generations of low-mass stars enrich by SNe II.
Roughly 1 Gyr after the star burst, white dwarf deflagration/detonation
SNe Ia (from mass
accretion in a binary star system) make more Fe-peak elements. An
observed overabundance of $\alpha$-elements compared to Fe-peak elements
implies a brief or recent SF episode \citep{Trager00,Trager04}; their
detailed abundances depend on the mass of the supernova progenitor. During
this time, AGB stars make odd-proton nuclei in the $s-$process. The
ratio of $\alpha$/Fe-peak element abundance decreases as a star burst
progresses. So the relative abundances of $r-$ to $s-$process elements
gives the relative importance of SNe Ia to AGB stars, hence the relative
yields of higher and lower mass stars.

\subsubsection{Uncertain isochrones.}

The pace of stellar evolution is exponential with mass. To combine tracks
to form an isochrone, one must interpolate in
the $(\log M,\log X)$ plane (X is the core hydrogen mass abundance)
by assuming that stellar evolution is continuous in this plane (it
is not), and that tracks of stars of slightly different mass are
always proximate in the CMD (also untrue in some mass ranges). Isochrones
may be interpolated physically only when homology relations can
scale stellar properties. Homology is broken by mass loss and
when core processes switch
drastically over a small mass range such as the dominance of CNO over
proton-proton fusion. Interpolation to intermediate
masses then yields invalid tracks through unphysical parts of the CMD.
Homology relations anyway predict only indirectly observable interior
--- not observable photospheric --- properties. Finally, all models
ignore important evolutionary influences such as rotation, binary
mass transfer, and magnetic fields.

Isochrones differ slightly depending on how RGB and HB convections
are handled \citep{Gallart05}. For example Geneva
(\url{http://obswww.unige.ch/~mowlavi/evol/stev\_database.html})
and Padova (\url{http://pleiadi.pd.astro.it/}) group zero-age isochrones span observations
of both halo and disc stars only if temperatures are set 130--250 K hotter than observed.

\subsubsection{Mass functions.}

Stellar $M/L\equiv\Upsilon$ is found empirically to vary from $4\times10^{-5}$
(O5 star) to 34 (M5). This mass-luminosity relation is fundamental,
yet its measurement is limited to those stars in suitable binary orbits
that are also near enough to us to have a precisely measurable parallax;
a 10\% parallax error produces a 5\% error in $T_{\rm{eff}}$. 

Models must specify a stellar birth-rate. The usual simplification
forms stars in one burst, so have a single age, metallicity, and initial
mass function (IMF) that together define a so-called simple stellar
population for the entire region observed. The
IMF is the probability of finding a star of given mass;
it thus also sets the mass fraction of the population that
is returned to the ISM by evolved stars. As a distribution, its use
implies an unavoidable Poissonian sampling variance
\citep[see][for a correction procedure]{Cervino02}; for example,
the predicted colours of a $10^{5}$ \Ms\  cluster (which is $\sim6$
times the mass of the most massive OB star association known in the
MWg, Cygnus OB2) disperse by 3-10\% depending on the assumed IMF and
on whether or not a few rare but luminous stars are included. 

IMF's are based on censuses of nearby SF regions. However, these regions
make predominantly low mass stars, are subject to variance, and are
complicated observationally by unresolved binary stars. The \citet{Salpeter55}
power-law IMF $\zeta(m)\propto m^{-\alpha}$ with $\alpha=1.7$ is
standard and has been assumed in most studies. However, \citet{Kroupa01}
summarizes evidence that this IMF predicts more low-mass stars than
are observed in MWg and LMC open star clusters; he provides an alternative,
multiple-part IMF from fits to those regions. The mean stellar mass
with this IMF is $0.36$ \Ms; half of the mass is in stars $0.01\le m\le1$ \Ms\  
and the other half is in $1\le m\le50$ \Ms. The variances
arise mostly from unresolved binaries.

\citet{Bruzual03} show that $\Upsilon$ derived from different
plausible IMF's varies by multiplicative factor $\sim1.5$ for
up to 1 Gyr after a star burst, increasing to multiplicative factor
$\sim2.5$ for ages up to 10 Gyr. Another uncertainty on the low
end of the IMF is that the current census with mass $<0.5$ \Ms\ 
is incomplete beyond 5 pc \citep{Henry97}. The high end of the IMF
(especially for Wolf-Rayet stars) has uncertain normalization and
highest variance from observations \citep[see for example Fig. 8 of][]{Bruzual03};
this is irrelevant for models of early-type galaxies that have not
formed many stars recently. In their high-SNR study of 25 such galaxies
(see section \ref{sub:Global-scaling-parameters}), \citet{Cappellari06}
compare the dynamically determined $\Upsilon$ with the values implied
from population synthesis \citep{Vazdekis96} to fit the observed
absorption strengths. They obtain better agreement between the two
methods across their sample with the \citet{Kroupa01} IMF; its fewer
low-mass stars compared to Salpeter reduces the derived $\Upsilon$
and hastens the subsequent evolution of the star cluster.

\subsubsection{Building a model galaxy.}

Having considered the evolution of a single star (evolutionary track
in the HR diagram) and of a stellar system (isochrone), we can
now model a simple stellar population, i.e., a stellar system
that is co-eval with unique chemical composition. A grid of SSP's,
ranging in age and chemical composition, forms the basis for comparison
with real galaxies. To create the spectrum of an SSP, one selects
an isochrone of desired age and metallicity and an appropriate IMF.
Next, at each point in the HR diagram, the number of stars (weighted
according to the isochrone and the IMF) is calculated, and a stellar
spectrum is selected with the appropriate $T_{\rm{eff}}$, $\log g$, and metallicity. Finally,
the spectrum is weighted by the number of stars at that point and
by luminosity, and combined with those at all other points in the
diagram to synthesize the spectrum of the particular SSP. \citet{Bruzual03}
detail this procedure. Naturally, its success depends on having a
spectral database that covers a wide range in $T_{\rm{eff}}$, $\log g$,
and chemical composition. Because of the difficulty of modelling the
emergent spectra of stars with millions of atomic and molecular transitions,
it has been traditional to use `libraries' of uniform-quality
spectra of many stars whose $T_{\rm{eff}}$, $\log g$, and metallicity
has been determined from a high dispersion abundance analysis. Recent
libraries, each comprising spectra of $\sim$1000 stars and ranging
widely over atmospheric parameters, are

\begin{itemize}
\item The Indo-US library of coude-feed stellar spectra \citep{Valdes04}
at \url{www.noao.edu/cflib}, covers $\lambda\lambda$346 - 946 nm
at resolution $\sim$$\lambda$0.1 nm FWHM for 885 stars.
\item ELODIE \citep{Moultaka04} spans $\lambda\lambda$400 - 680 nm at
high dispersion. Based on stars with well-determined atmospheric
parameters, version \url{http://atlas.obs-hp.fr/elodie} has $>10,000$
spectra.
\item MILES \citep{Sanchez06} contains spectra at resolution
$\lambda$0.23 nm FWHM and covers $\lambda\lambda$352 - 750 nm. 
\end{itemize}
Empirical libraries have limits. It is difficult
to find stars that range over sufficient $T_{\rm{eff}}$, $\log g$,
and metallicity whose atmospheric parameters are known from fundamental
high-dispersion spectral analysis. This is particularly true for hot
stars, because only metal-rich examples exist in the Solar Neighbourhood and
it is now clear that the mean
abundance ratios of stars in Egs are non-solar (see section 3.6).
Establishing atmospheric parameters for
stars in spectral libraries is also problematic, because the high dispersion analyses
were made by different investigators using different model atmospheres
and $T_{\rm{eff}}$ scales from different photometric colours.

Models of stellar atmospheres have increased in sophistication through
great increases in both computational power and the number of included
atomic and molecular transitions. Hence, using a library of synthetic
stellar spectra is increasingly attractive. Theoretical stellar spectra
can be calculated for any chemical prescription. For full consistency,
one should first model an atmosphere based on the assumed chemical
composition, and then calculate the detailed emergent spectrum. While
calculating the model is very time consuming, use of theoretical spectra
is now widespread especially for constructing the hot stars in young
stellar populations. An example is the library of 1654 theoretical
spectra, sampled at $\lambda$0.03 nm and covering $\lambda\lambda$300
- 700 nm, that is described in \citet{Martins05}\footnote{\url{www.astro.iag.usp.br/~lucimara/library.htm}}.

Two approaches compare composite spectra of SSP's, gridded by age
and metallicity, to the observed spectrum to determine which
combination of SSP's matches best.

\begin{enumerate}
\item Focus on absorption lines in the model grids with particular sensitivity
to age and/or metallicity, then find their best match. The preeminent
example is the Lick system of spectral indices \citep[for example][]{Faber85}.
This approach has several attractions:

\begin{itemize}
\item It is transparent to which specific feature(s) determine
age and metallicity.
\item It is simple to compute a two-index diagram that separates age and
metallicity effects (for example H$\beta$ versus the Fe5270 index),
plot the grids that connect the various SSP models, overplot galaxy
measurements, and then interpolate ages and metallicities for a large
galaxy sample.
\item With model atmospheres, one can probe precisely which transitions
of various elements form the index, either through the central passband
or through the two continuum bands that straddle it in wavelength.
\end{itemize}
\item Use $\chi^{2}$ minimization to fit the entire spectrum to models.
This approach is problematic because a typical spectrum contains more than 1000
independent pixels, so computation is intensive and impractical for
the numerous $(\sim10^{6})$ spectra from an extensive survey such
as the SDSS. Moreover, spectral information is highly redundant, with
many features duplicating the overall degenerate sensitivity to age
and metallicity. If the SNR is low, co-adding many redundant features
is advantageous, although populations would then not likely be isolated.
If the SNR is high, information in the spectrum can usually be reduced
to a few parameters that are chosen judiciously. Various strategies
are possible:

\begin{itemize}
\item Pre-select line strength indices, as in 1) above. 
\item Use principal component analysis (\citealt[for example][]{Madgwick03,Ferras06})
to find the minimum eigenvector set that contains most of the spectral
information. Solving for the eigenvalues that characterize a given
observed spectrum yields enormous compression. However, spectral
evolution of stellar populations is very non-linear in time. Thus,
for a set of emission-free spectra of different ages (to avoid the
vexing emission-line `contamination' of absorption features),
the galaxy spectrum is synthesized by combining a very
young with a very old population that may be irrelevant to that galaxy's
SF history.
\item Compress spectra through a set of weighting vectors to match a set of realistic
population parameters (for example a dozen ages spaced logarithmically, each with
a metallicity). \citet{Panter03} have used
such compression (acronym MOPED) to extract SF histories from huge
numbers of SDSS spectra. Of concern is that this approach fits data
with a highly compressed $\chi^{2}$ minimization that may fail to
duplicate faithfully the observed spectrum. While the same criticism
can be levelled at the spectral index approach, mismatches between
models and data are there more obvious. For example, plotting the
age-sensitive Lick H$\beta$ index against the Fe-sensitive Fe5270
index on a model grid with various ages and metallicities but using
solar abundance ratios, yields a single metallicity. For massive Egs,
plotting H$\beta$ against Lick Mg b indices (Mg b measures primarily
a Mg-sensitive feature) yields a larger metallicity. Such results
have indicated clear non-solar abundance ratios (NSAR) in Egs \citep[for example][]{Worthey04,Trager00,Peletier89},
requiring consistent NSAR models \citep{Thomas03}. Such transparent
disagreement between model and data led to important insights on the
SF/chemical evolution histories of galaxies; it is far less clear with the MOPED and
principal component techniques.
\end{itemize}
\end{enumerate}

\subsection{\label{sub:Gravitational-stellar-dynamics}Gravitational stellar
dynamics}

Goals of gravitational dynamics are to

\begin{enumerate}
\item Understand the phase-space distribution of stars in galaxies. The
observed 2D projected stellar motions and SB distribution constrain
the 3D shape and motions in the potential. The stars, with the inferred
DM, establish the gravitational potential of the galaxy. (Gas is generally
negligible in gravity, simply acting as a tracer that may not be in
equilibrium.)
\item Establish the stability and secular evolution of the stellar and DM
distributions in galaxies. How do these two interact over time in
`isolated' galaxies? 
\item Establish the consequences of gravitational interactions between galaxies. 
\end{enumerate}
One must recognize that a
self-gravitating system differs greatly in physics from a
charged plasma, perhaps making gravitational dynamics seem
obscure and counter-intuitive. 

First, unlike in a plasma with equal numbers of $\pm$ charges, gravity
only attracts; no equivalent to Debye shielding limits interactions
to roughly the inter-particle separation. Thus, interactions in a
gravitational system are very long-range; indeed, the sum of weak
ones from very distant stars dominates over nearby
two-body encounters. Surrounded by $\gtrsim10^{5}$ stars, each star
sees a smooth potential and would need longer than the
age of the Universe to exchange significant energy with others;
self-gravitating systems are nearly collisionless. This implies that,
if the potential has been stable since a galaxy formed, then a
star retains memory of its original orbit and its observer can be
a `galactic archaeologist'. Conversely, if evidence suggests
that certain orbit families have altered substantially, then by implication
the galactic potential was disrupted significantly at least once in
the past to scatter stars into new orbits.

Second, the stellar mean free path length exceeds greatly the system diameter,
so its equation of state, pressure, and temperature cannot be defined.
In contrast, charged plasma particles remain in a local equilibrium
characterized by the ideal gas law. 

As will soon be clear, solving the equations of a self-gravitating
system is daunting.
We touch only on key aspects, and
refer the reader to the classic monograph of \citet{Binney87} to
understand the full problem.
We first introduce the distribution function $f(\mathbf{x},\mathbf{v},t)$,
the fine-grained probability of locating stellar mass in six-dimensional
phase-space. Without collisions, the mass density within
this piece of phase space is invariant $df/dt=0.$ Hence, the coupled 
Boltzmann and Poisson equations in gravitational potential $\Phi$
\begin{eqnarray}
\begin{array}{c}
\frac{df}{dt}\equiv\frac{\partial f}{\partial t}+\mathbf{v}\bullet\boldsymbol{\bigtriangledown}f-\boldsymbol{\bigtriangledown}\Phi\bullet\frac{\partial f}{\partial\mathbf{v}}=0 \\
\\
\bigtriangledown^{2}\Phi(\textbf{x},t)=4\pi G\int f(\mathbf{x},\mathbf{v},t)d^{3}\textbf{v}
\end{array}
\end{eqnarray}\label{eqn:poisson}
describe the dynamics
of collisionless, self-gravitating galaxies.
Seven independent variables make $f$ hard to solve.
Instead, one often obtains insights from spatial
or kinematical moments, yielding three partial differential equations.
But, moments are \textit{global} averages over
the \textit{entire} distribution, and there certainly \textit{is} considerable
material (both DM and possibly unsettled baryons) in the faint
outer parts of galaxies, with unknown kinematics.
In fact, this material is very important to understand
because its slow rate of dynamical evolution (if undisturbed by tides)
means that it may retain a memory of its initial state.

\subsubsection{\label{sub:Velocity-moments}Velocity moments.}

Integrating the collisionless Boltzmann equation (CBE) (2) over velocity
yields the Jeans equations\begin{equation}
\frac{\partial\overline{v}_{j}}{\partial t}+\overline{v}_{i}\frac{\partial\overline{v}_{j}}{\partial x_{i}}=-\frac{\partial\Phi}{\partial x_{j}}-\frac{1}{\nu}\frac{\partial(\nu\sigma_{ij}^{2})}{\partial x_{i}}\end{equation}
where \begin{equation}
\nu\equiv\int fd^{3}v\end{equation}
is the mass density
 and \begin{equation}
v_{i}\equiv\frac{1}{v}\int f\, v_{i}d^{3}v\end{equation}
 are the three velocity moments, and\begin{equation}
-v\sigma_{ij}^{2}\equiv-v(\overline{v_{i}}\,\overline{v_{j}}-\overline{v_{i}v_{j}})\end{equation}
 is the velocity dispersion stress tensor with \begin{equation}
\overline{v_{i}v_{j}}\equiv\frac{1}{v}\int v_{i}v_{j}fd^{3}v\end{equation}
As mentioned, there is no equation of state
to relate motions (through tensor $\boldsymbol{\sigma}^{2}$)
to mass density $\nu$. Integrating higher moments $v_{i}v_{k}$ of
the CBE over velocity does not make this link, it simply introduces
an unspecified third-order tensor $\overline{v_{i}v_{j}v_{k}}$. The
only route forward is to truncate the moment sequence, for example
by making hopefully reasonable, but nonetheless non-unique, assumptions
on $\boldsymbol{\sigma}^{2}$. This approach was first applied to
galaxy formation by \citet{Larson69}, who truncated and then integrated
numerically the stellar and gas dynamical moment equations to model
observed properties of Egs.

For example, a non-rotating galaxy with rotationally invariant densities
and velocities has $\overline{v_{\theta}^{2}(r)}=\overline{v_{\phi}^{2}}(r)$.
The anisotropy of tensor $\boldsymbol{\sigma}^{2}$ is described by 

\begin{equation}
\beta(r)=1-\overline{v_{\theta}^{2}(r)}/\overline{v_{\rm{r}}^{2}}(r)\end{equation}
with $\beta=-\infty,\,0,$ and 1 for orbits that are circular, isotropic,
and radial, respectively. Even with this simplifying assumption,
we unfortunately have only five equations to solve for six variables:
$\nu$, $\overline{v_{\rm{r}}}$, $\overline{v_{\theta}}$, $\Phi$,
$\overline{v_{\rm{r}}^{2}}$, and $\beta$ (mass density, two independent
mean velocities, gravitational potential, velocity dispersion along
radii in the galaxy, and anisotropy of the velocity dispersion). Next,
we assume a spherically
symmetrical Eg in a static configuration, i.e., without streaming
motions. The Jeans equations then simplify to\begin{equation}
\frac{1}{\nu}\frac{d(\nu\overline{v_{\rm{r}}^{2}})}{dr}+2\frac{\beta\overline{v_{\rm{r}}^{2}}}{r}=-\frac{d\Phi}{dr}\end{equation}
Setting \begin{equation}
\frac{\partial\Phi}{\partial r}=\frac{GM(r)}{r^{2}}\end{equation}
recasts the radial Jeans equation into\begin{equation}
v_{\rm{c}}^{2}=\frac{GM(r)}{r}=-\overline{v_{\rm{r}}^{2}}(\frac{d\ln\nu}{d\ln r}+\frac{d\ln\overline{v_{\rm{r}}^{2}}}{d\ln r}+2\beta)\end{equation}
to solve for $M(r)$ \citep[see section 4.2 of][for details]{Binney87}.
If the radial variations of mass density, velocity
dispersion, and velocity anisotropy $\beta$ can be assumed or determined
empirically by deprojecting measurements of luminous tracers such
as planetary nebulae or globular clusters, we can recover the mass distribution. For example,
\citet{Sargent78} assumed $\beta=0$ (i.e., isotropic velocity dispersion)
in the central regions of Eg M87, and found from the inward rise in
both luminosity and velocity function dispersion that $\Upsilon$ must increase toward
the centre, indicating a supermassive black-hole. Subsequently, \citet{Binney82} relaxed
the assumption $\beta=0$, and found that rising $\Upsilon$ is unnecessary
if large ad hoc variations in $\beta(r)$ are allowed.

\subsubsection{Choice of velocity distribution function.}

Using only velocity moments does not ensure a viable solution of the
Jeans equations (positive distribution function everywhere). Nor is any solution based
on an assumed velocity dispersion tensor necessarily stable. An alternative
approach to solving the CBE is to use prior knowledge of the distribution function. For
a spherical galaxy, assume that the distribution function is both Maxwellian and `isothermal',
namely an isotropic velocity dispersion that is independent of radius.
Unfortunately, such a galaxy has infinite mass and radius.
When a dynamical system can be truncated tidally, a `lowered
Maxwellian' can be introduced. The distribution function of a lowered Maxwellian that truncates
at positive energy $\varepsilon$ is\begin{equation}
f_{\rm{K}}(\varepsilon)=\left\{ \begin{array}{cc}
\rho(2\pi\sigma^{2})^{-3/2}(\exp[\varepsilon/\sigma^{2}]-1) & \varepsilon>0\\
0 & \varepsilon\le0\end{array}\right.\end{equation}
\citet{Michie63} used (A.12) to model self-consistently the mass distributions
of globular clusters, and \citet{King66} derived a model family that fit their
SB profiles, although velocity function and proper motion measurements
later showed anisotropic motions outside the cores.
As discussed in section \ref{sub:Intrinsic-shape}, observations show
that Egs are actually triaxial, so have anisotropic velocity dispersions.
Egs also have structured centres, not the constant density core of
King models. Hence, King models are overly restrictive
and are not fully realized in Egs.

\subsubsection{Integrals of motion.}

The CBE can also be solved in terms of integrals of motion, namely
independent functions of phase-space $(\mathbf{x},\mathbf{v})$ that
have constant value along any orbit. In a static potential (no dissipation),
energy $E(\mathbf{x},\mathbf{v})=\frac{1}{2}v^{2}+\Phi(\mathbf{x})$
is one and, if potential $\Phi$ is axisymmetric, the component of
angular momentum around the rotation axis is too. In a spherical potential,
angular momentum $\mathbf{L}$ yields three integrals of motion. If
the potential is central, there is a fifth integral whose five-dimensional
surface in phase space $\psi(\mathbf{x},\mathbf{v})=$constant intersects
the 2D surface of constant $E$ and $\mathbf{L}$ to form a rosette-pattern
orbit that is restricted to a five-dimensional region of phase space.
In a $1/r$ potential, the rosette closes and the fifth integral is
said to be isolating. Otherwise, the entire 2D surface of constant
$E$ and $\mathbf{L}$ is covered eventually by the non-isolating
integral.

The dimensionality (hence complexity) of solutions to the CBE reduces
greatly with solutions that depend only on isolating
integrals. Orbits are regular if they have at least as many isolating
integrals as spatial dimensions. Realistic potentials have fewer isolating
integrals, hence have irregular orbits. The Strong Jeans Theorem states
that a distribution function composed of regular orbits is a function of just three isolating
integrals, $f(E,I_{2},I_{3})$. 

For example, assuming that $E$ is the only isolating integral in
a spherical galaxy is equivalent to assuming an isotropic velocity
dispersion tensor, which we noted is overly restrictive. To accord
fully with the Strong Jeans Theorem requires a distribution function that depends on
two further integrals without explicit forms. So, one next considers
distribution functions that depend on both $E$ and $\left|\mathbf{L}\right|$. Generally,
the third integral is far harder to identify.

\subsubsection{\label{sub:Stellar-orbits-and}Stellar orbits and the gravitational
potential.}

A galaxy can also be described by
the orbit families of an assumed potential that reproduce its inferred
distribution function and stellar motions.

An axisymmetric potential usually has a third integral with
$E$ and $L_{z}$. Some stars oscillate harmonically parallel
to the cylindrical coordinate axes to form box orbits (Lissajous figures,
which have no fixed sense of rotation and carry a star arbitrarily
close to the centre of mass). At larger radii one finds loop orbits
that are nearly circular and whose initial tangential velocity determines
the elliptical annulus that confines the orbit. This resembles
the annulus filled by an orbit in an axisymmetric potential as
$L_{\rm{z}}$ is varied. Box orbits fill the phase space
of more flattened axisymmetric systems. Orbital resonances,
central density cusps, or central point masses change box orbits to
boxlet orbits that tend to avoid small radii, or stochastic orbits
that cover phase space chaotically when the central mass grows to $\sim1\%$
of the system mass.

Realistic potentials and mass distributions thus have complex
orbits to integrate numerically, but restricted potentials can
be studied analytically. In particular, Stäckel (Eddington) potentials
are the most general that separate in curvilinear coordinates, hence
all orbits have three isolating integrals.
Realistic galactic potentials (including those for triaxial figures)
include both regular and irregular orbits, so Stäckel potentials are
too restrictive. But their solutions are far more tractable because
the moment equations now form a closed set. \citet{vandeVen03} discuss
general solutions obtained with Stäckel potentials.

\subsubsection{Spatial moments: tensor virial theorem.}

In \ref{sub:Velocity-moments} we obtained three Jeans velocity
moments by multiplying the CBE by each velocity component
and then integrating over all velocities. Now multiply
each velocity moment by a \textit{position}
coordinate and then integrate over all positions to form nine
tensor virial theorem (tensor Virial Theorem) equations\begin{equation}
\frac{1}{2}\frac{d^{2}I_{jk}}{dt^{2}}=2T_{jk}+\Pi_{jk}+W_{jk}\end{equation}
with the moment of inertia tensor \begin{equation}
I_{jk}=\int\rho x_{j}x_{k}d^{3}x\end{equation}
partitioned into the potential energy tensor \begin{equation}
W_{jk}=-\int\rho(x)x_{j}\frac{\partial\Phi}{\partial x_{k}}d^{3}x\end{equation}
 and the kinetic energy tensor \begin{equation}
K_{jk}=\frac{1}{2}\int\rho\overline{v_{j}v_{k}}d^{3}x=T_{jk}+\frac{1}{2}\Pi_{jk}\end{equation}
that sums ordered $T$ and random $\Pi$ motions, respectively. A `cold'
dynamical system is one where $W_{jk}$ and $T_{jk}$ balance, a `hot'
system balances $W_{jk}$ and $\Pi_{jk}$. 
The tensor Virial Theorem links the global spatio-kinematical properties of a galaxy.
While spatial data are straightforward to obtain in projection, kinematics
were limited until recently to samples along a PA's through
narrow slits. Complete spatio-kinematical coverage over the inner
part of an Eg is now attained in a single pointing of an integral-field
spectrograph (IFS).

The most important insight from the tensor Virial Theorem follows when gravity collects
matter from rest. As matter equilibrates, the
tensor Virial Theorem tells us that half of the gravitational energy released causes
motion within the potential well and the other half is dissipated
to achieve binding energy $E_{b}=-E$ equal to the kinetic energy;
the system has negative specific heat. For the MWg of mass $M_{\rm{g}}$
and average rotational velocity $\Theta_{0}\approx200$ \kms.
$E_{\rm{b}}=K\approx\frac{1}{2}M_{\rm{g}}\Theta_{0}^{2}$, hence
$\approx2.5\times10^{-7}$ of its rest mass energy must have dissipated
during its formation.

The tensor Virial Theorem can estimate the global $\Upsilon$ of
a nonrotating spherical galaxy by averaging stellar
velocity dispersion $\sigma(R)$ and SB $\sigma(R)$ maps 
over the projected radius $R$ from
IFS spectra\begin{equation}
\widehat{I}=3\pi\int_{0}^{\infty}\sigma(r)\sigma^{2}(r)rdr,\end{equation}
and by recognising that density follows from an Abel inversion integral
of the measured radial variation of the SB. 
With $\overline{I}$ an integral of $\sigma(R)$, the tensor Virial Theorem then gives
\begin{equation}
M/L\equiv\Upsilon=-\frac{2\widehat{I}}{\overline{I}}\end{equation}
In section \ref{sec:Spheroidal-Galaxies}
the tensor Virial Theorem constrains Eg 3D shapes by comparing radial
trends of mean rotational/random motion with SB ellipticity.
tensor Virial Theorem volume averaged properties require non-unique assumptions.

\subsubsection{Violent relaxation.}

Statistical mechanical quantities that depend on randomization such
as system temperature and entropy are seemingly undefinable in a collisionless
stellar system. Yet, virialization implies past dissipation. Violent
relaxation \citet{Lynden67} is the dissipator that erases most initial conditions by
widening the global range of stellar energies (but perhaps not the
energies of certain orbit families), independent of stellar mass.
It arises from rapid fluctuations in the gravitational field during
hierarchical mass buildup that scatters the stellar energies. 
Simulations \citep[for example][]{Dekel05}
show that the baryons approach virial equilibrium after a few dynamical
times, namely a Maxwell distribution that is nearly isotropic near
the centre\begin{equation}
f(\varepsilon)=\frac{\rho}{(2\pi\sigma^{2})^{3/2}}\exp(\frac{\Psi-v^{2}/2}{\sigma^{2}}),\end{equation}
with isothermal equation of state $p(\rho)=K\rho$ and with $\Psi$ the relative
potential. Meanwhile, stars outside remain anisotropic,
with many on radial orbits that originated in the mass buildup
(see section \ref{sub:Dark-matter-content}).

\begin{figure}
\begin{centering}\includegraphics[scale=1.8]{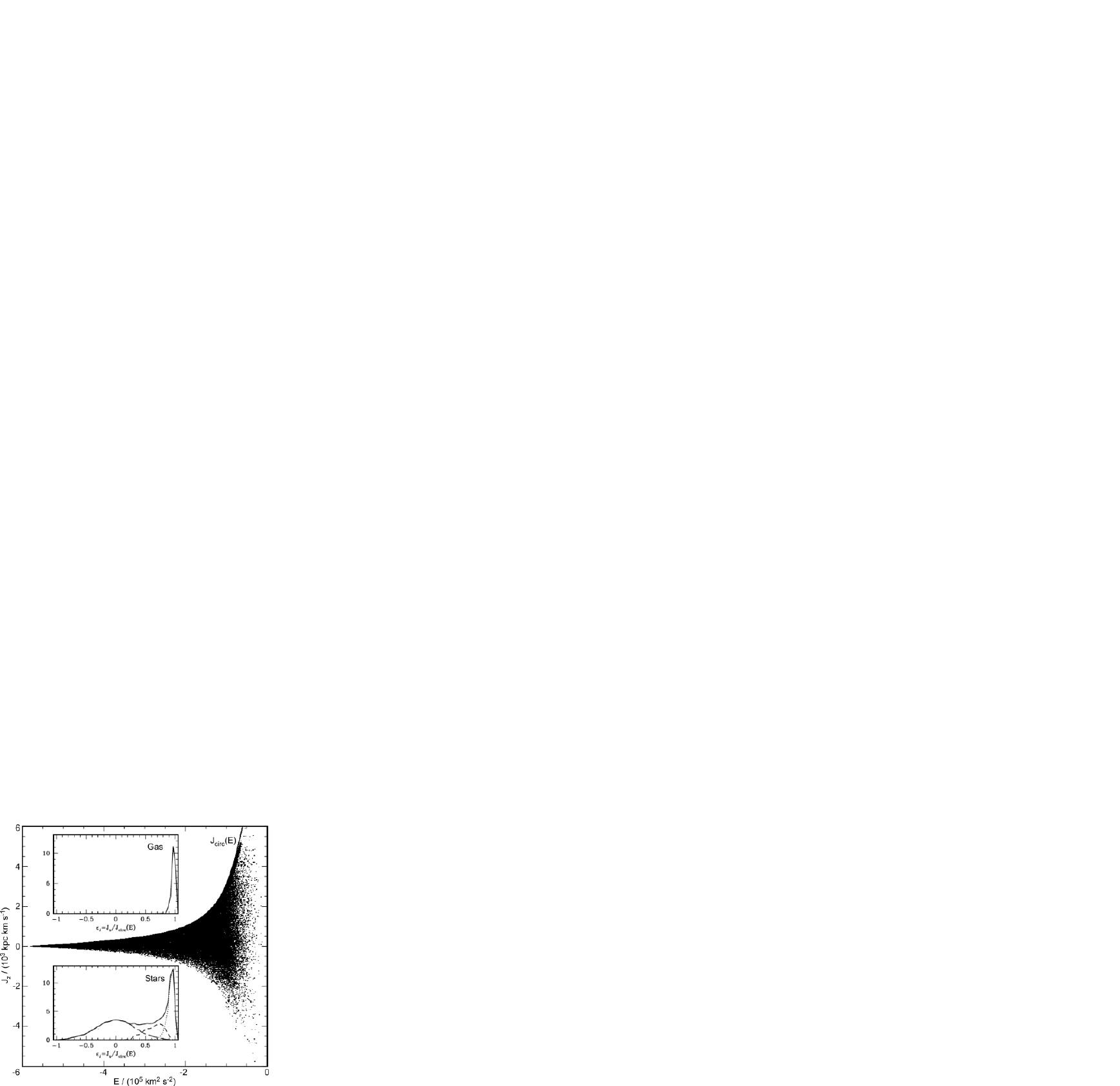}\par\end{centering}
\caption{\label{fig:Abadi-et-al}\citet{Abadi03}'s simulated formation of
an Sa galaxy. The z-component of specific angular momentum within $R_{\rm{e}}$
of the baryonic galaxy is plotted versus the specific binding
energy. Inserts plot the distribution of angular momentum
(normalized to a fraction of circular orbit value) of gas and stars,
together with a possible decomposition into different components
(labelled \dotted\ for thin disc, \broken\
for thick disc, \full\ for spheroid). Despite photometric
resemblance to Sa galaxy components, the simulated rotation curve
is too centrally peaked. }
\end{figure}

In the CDM paradigm, DM relaxes to form the potential well. The baryons
accrete into the well --- their dissipation characterized by parameter
\begin{equation}
\lambda=\widehat{J}\left|E\right|^{1/2}\frac{1}{GM^{3/2}}\end{equation}
with $\widehat{J}\equiv J/M$, which
sets \citep[for example][]{Dalcanton97} the SB of
the baryonic disc. Figure \ref{fig:Abadi-et-al} shows the distribution
of angular momentum from a $\Lambda$-CDM simulation \citep{Abadi03}, but
present data are too sparse to constrain such models.

As discussed in section \ref{sec:Bulges-of-Spiral}, $\widehat{J}$
of a disc spreads locally by collective
motions from dynamical resonances in quasi-periodic orbits. Some DM
distributions have minimal resonances hence maximal $\widehat{J}$
at the disc rim. We cannot yet measure stellar kinematics at
this SB, but can see if the stellar disc apears to truncate (section 5.1.2).

\subsubsection{\label{sub:Building-a-mass}Building a mass distribution and potential
self-consistently from stellar orbits.}

To model self-consistently a galaxy, one combines orbit
families of assumed $\Upsilon$ with a set of global weights
to reproduce the distribution function as derived from the SB profile and the velocity function including
asymmetries in absorption line profiles. Elucidating radial variations
of $\Upsilon$ would be an important insight from a successful model,
but current data are inadequate. At best, one posits a mass
distribution with a spherical DM halo and stars whose average $\Upsilon$
is consistent with the observed radial profiles of absorption line
strengths and photometric colours. The distribution function is assumed to be steady
state, but recall that galaxy shape may arise from both initial
conditions and external perturbations. 
To model an Eg one assumes inclination $i$ and then 
\begin{itemize}
\item 
use either a constraining technique such as
maximum entropy, or, fit to a functional form that has an analytical inversion, to
deproject the mapped SB $\sigma(R,z)$ with the Abel integral
\begin{equation}
\rho(r,z)=-\frac{\Upsilon}{\pi}\int_{r}^{\infty}\frac{d\sigma(R,z)}{dR}\frac{dR}{\sqrt{R^{2}-r^{2}}}\end{equation}
\item calculate the corresponding potential $\Phi(R,z)$ assuming constant
$\Upsilon$;
\item solve Jeans equations for mean velocities, then divides
random and streaming motions;
\item integrate space motions to get the velocity function and dispersion;
\item compare to data and iterate to refine $\Upsilon$, the velocity
division, and $i$. A valid model has positive distribution function throughout
the galaxy phase space, while satisfying (\ref{eqn:poisson}).
\end{itemize}
Velocity dispersions along the galaxy minor axis are always observed
to be smaller than predicted by this procedure, implying a third integral
and/or triaxial shape (section \ref{sub:Intrinsic-shape}). Surface photometry
finds non-elliptical isophotes for spheroidal galaxies. Some patterns
are consistent with embedded discs, which may include dust and gas
moving on inclined rings and counter-rotating compared to the stars.
The cold ISM settles into stable orbits by differential (with radius)
precession that collides clouds and dissipates their kinetic energy.
The ISM may form a warped disc in an oblate potential or polar ring
if prolate. An example is the dusty, but optically thin in near-IR, warped
disc in Centaurus A (NGC 5128), studied recently with \textit{SST} \citep{Quillen06};
based on transient tidal features (the warp and outer stellar shells),
it may have arisen as a small gas-rich satellite merged $\sim0.1$
Gyr ago.

Studies that assume a fixed, or slowly varying, potential from
a mass model $\rho(r)$ use the orbit summing method of \citet{Schwarzschild79}.
Many orbits are computed for many oscillations in the specified
potential; most are tubes or boxes without net rotation. By noting
regions within an equipotential surface that some orbits never traverse,
Schwarzschild found three integrals of motion for most orbits in his triaxial model potential.
Next he summed orbits with positive
weights to build the assumed $\rho(r)$. He found the weights by linear
programming, but now maximum entropy and the \citet{Lucy74}
or related stack-smooth-iterate \citep{Bak00} algorithms are used.
Methods such as `dithered' initial conditions for orbit integration
\citep{Cappellari06} ensure the smooth distribution functions required to fit modern
data; today over 400,000 orbits are summed with weights selected to
reproduce the stellar density and velocity function.

Different distribution functions are consistent with a given mass model because
the velocity function is degenerate with galaxy inclination unless additional features
such as a stellar disc, ionized gas \citep[for example][]{Krajonovic05},
neutral gas \citep{Morganti06}, or dust lanes \citep{Quillen06}
also constrain inclination. If the mass distribution and potential
are constrained by x-ray data, the orbital anisotropy is no longer
degenerate. Today Jeans models are often used because of their transparency,
uniqueness, and efficiency to generate an ensemble of noise-free distribution functions
across a grid of structural parameters for Monte Carlo error analysis
of the more sophisticated but non-unique Schwarzschild technique \citep[for example][]{Krajonovic05}.

\section{A Brief Review of Hierarchical Galaxy Formation}

Hierarchical buildup from small to large mass develops cosmic structure
from the early linear phase explored in the spectrum of temperature
fluctuations in the CMBR into the non-linear regime. In particular
\citep[for a comprehensive review, see][]{Baugh06}, $\Lambda$-CDM matches
the power spectrum of the fluctuations
as mapped by WMAP and terrestrial
instruments, the Hubble diagram for high redshift
SNe Ia, and the galaxy correlation function in recent wide-angle galaxy
surveys. \textit{Direct} study of underlying DM halos
is impossible. Instead, we study the baryonic component that dissipates
energy within the DM halo, forms stars, and leaves hot and cool gas.
Key for a researcher interested only in the halo is to determine to
what extent light traces DM. Appendix A introduces some of the galactic astrophysics that
intrudes.

Galaxy and cluster build up in the $\Lambda$-CDM
framework has been explored by simulations that
follow up to $10^{9}$ DM `particles' from post-recombination
density perturbations to the present epoch \citep[for example][]{Baugh06}.
Volumes $\sim500$ Mpc on a side have been
simulated with DM particles as small as $10^{8}$ \Ms.
Because DM is cold and non-interacting,
gravitational N-body simulations can track its concentration into halos.
In this way
galaxy masses as small as $10^{9}$ \Ms\  are followed,
unfortunately not the dwarf end of the galaxy mass function.
Alternatively, galaxy build up can be followed in a statistical sense
through semi-analytic prescriptions founded on the work of \citet{Press74}.
The consensus in comparing both methods is that the predictions
of $\Lambda$-CDM simulations are robust.
However, study of baryonic matter within CDM halos requires
accurate models of galaxy astrophysics over the mass range
encompassed locally by a box $\sim12$ Mpc on a side. For example,
almost all SF seen at high redshifts comes from a cooling flow into
CDM halos, yet its non-linear effects on
the surrounding ISM (`feedback', see section \ref{sub:Feedback})
is well beyond present simulations.

A successful model must also reproduce the fate of baryons as galaxies merge.
\citet{Helly03} compares the two tactics in use:
1) solve simultaneously the equations of hydrodynamics for
gaseous baryons during the N-body simulation of the DM halo \citep[for example][]{Katz92,Steinmetz99,Cen00,Pearce01}.
Naturally, at some point considerable sub-grid astrophysics must be
parametrized. 2) graft this parametrized astrophysics onto statistical
semi-analytic models \citep[for example][]{Cole00}.
We are still far from fundamental treatment of the gaseous baryons.

The starting point as the DM halo forms is to cool and settle baryonic
gas into a SF disc. An Eg forms when two near-equal mass disc galaxies
merge. It can later grow a new disc to form a large-bulge Sg only if hot
halo gas remaining from the merger can cool and settle. The scenario
reproduces the scale of early-type galaxies by forming galaxies in
DM halos where baryon cooling is efficient: objects lighter than
$10^{12}$ \Ms\  form galaxies, those heavier form clusters. A persistent
problem is that this scenario concentrates baryons too centrally,
the so-called `angular momentum catastrophe'. DM halos acquire
much of their angular momentum in mergers from the orbits of the progenitor
galaxies. The baryonic gas, in cooling early toward the centres of
the DM halos, does not gain this angular momentum \citep[for example][]{Steinmetz99,Navarro94}.
$\Lambda$-CDM simulations therefore predict very structured halos
and a central cusp of DM.

Despite these difficulties, simulations have provided robust results.
Specifically, larger DM halos tend to assemble at lower redshifts
(i.e., more recently) than smaller ones, and in general DM halos
form more rapidly in denser than sparser environments \citep{Baugh06}.

The situation for baryons, particularly baryons now in
stars, is murkier. Early simulations predicted that they would
indeed follow DM through a process of adiabatic compression. More massive
Egs are predicted to have younger luminosity-weighted mean ages than
lower mass Egs, and those in clusters were predicted to be older in
the mean than those in sparser regions. Egs of given mass
are predicted to scatter significantly in luminosity-weighted mean
age because of the stochastic nature of mergers \citep[for example][]{Kauffman98}.
However, as we discuss in section \ref{sec:Spheroidal-Galaxies} and section \ref{sec:Bulges-of-Spiral},
the inferred mean ages of Egs indicate that low mass Egs typically
contain younger stars than massive Egs. This anti-hierarchical
behaviour (called `down-sizing') led to new prescriptions for
cooling and feedback (see section \ref{sub:Feedback}) related to SF and
active galactic nuclei (AGN). For example, \citet{deLucia05} suppress
cooling of massive Egs that ends SF early on, while
\citet{Croton06} and \citet{Bower06}
use AGN for mass-dependent feedback to inhibit hierarchical mass
build up of baryons.

In summary, observations are now connecting with models but both
are still too uncertain to establish
firmly the viability of the $\Lambda$-CDM scenario in the non-linear
regime.

\end{document}